\documentclass[preprint,aps]{revtex4-1}
\usepackage{epsf,epsfig,subfigure,graphicx,amsmath,amssymb,amsfonts,amssymb}
\usepackage{graphicx,color}
\usepackage{hyperref}
\def\plb#1#2#3#4{#1, Phys. Lett. {\bf #2B}, #3 (#4)}
\def\npb#1#2#3#4{#1, Nucl. Phys. {\bf B#2}, #3 (#4)}
\def\prd#1#2#3#4{#1, Phys. Rev. {\bf D#2}, #3 (#4)}
\def\prl#1#2#3#4{#1, Phys. Rev. Lett. {\bf #2}, #3 (#4)}
\def\mpl#1#2#3#4{#1, Mod. Phys. Lett. {\bf A#2}, #3 (#4)}

\def\beq{\begin{equation}}
\def\eeq{\end{equation}}
\def\bea{\begin{eqnarray}}
\def\eea{\end{eqnarray}}
\def\ba{\begin{array}}
\def\ea{\end{array}}
\def\bec{\begin{center}}
\def\ec{\end{center}}

\newcommand{\dis}[1]{\begin{equation}\begin{split}#1\end{split}\end{equation}}

\newcommand{\tev}{\,\textrm{TeV}}
\newcommand{\gev}{\,\textrm{GeV}}

\begin{document}
%%%%%%%%%%%%%%%%%%%%%%%%%%%%%%%%%%%%%%%%%%%%%%%%%%

%\draft

%\preprint{SNU-CTP/12001}

%%%%%%%%%%%%%%%%%%%%%%%%%%%%%%%%%%%%%%%%%%%%%%%%%%
%% Title
%%%%%%%%%%%%%%%%%%%%%%%%%%%%%%%%%%%%%%%%%%%%%%%%%%
\title{\Large\bf Neutrino Assisted Gauge Mediation}
% See-Saw Mediation
% Gauge mediation with (RH) neutrino messenger
% S4 flavor symmetry and \theta_13
%%%%%%%%%%%%
%%%%%%%%%%%%%%%%%%%%%%%%%%%%%%%%%%%%%%%%%%%%%%%%%%
%% Author Name(s) & Address
%%%%%%%%%%%%%%%%%%%%%%%%%%%%%%%%%%%%%%%%%%%%%%%%%%

\author{Hyung Do Kim\email{hdkim@.snu.ac.kr}, Doh Young Mo, Min-Seok Seo}
\affiliation{ Department of Physics and Astronomy and Center for Theoretical Physics, Seoul National University, Seoul 151-747, Korea
 }

%%%%%%%%%%%%
%%%%%%%%%%%%%%%%%%%%%%%%%%%%%%%%%%%%%%%%%%%%%%%%%%
%% Contact Information(s)
%%%%%%%%%%%%%%%%%%%%%%%%%%%%%%%%%%%%%%%%%%%%%%%%%%
%\ead{hdkim@phya.snu.ac.kr}

% PACS codes: 05.45.-a, 45.50.-j
%\date{\today}
%\maketitle

%\tighten

%%%%%%%%%%%%%%%%%%%%%%%%%%%%%%%%%%%%%%%%%%%%%%%%%%%%%%%%
\begin{abstract}
%%%%%%%%%

Recent observation shows that the Higgs mass is at around 125 GeV
while the prediction of the minimal supersymmetric standard model is below 120 GeV
for stop mass lighter than 2 TeV
unless the top squark has a maximal mixing.
We consider the right-handed neutrino supermultiplets as messengers
in addition to the usual gauge mediation
to obtain sizeable tri-linear soft parameters $A_t$ needed for the maximal stop mixing.
Neutrino messengers can explain the observed Higgs mass for stop mass around 1 TeV.
Neutrino assistance can also generate charged lepton flavor violation including $\mu \to e \gamma$
as a possible signature of the neutrino messengers.
We consider $S_4$ discrete flavor model and show the relation of the charged lepton flavor violation, $\theta_{13}$ of neutrino oscillation and muon $g-2$.

% Key Words: Supersymmetry, 125GeV Higgs, gauge mediation, see-saw mechanism, S4 flavor symmetry, $\theta_{13}$ of neutrino oscillations, charged lepton flavor violation, $\mu \to e \gamma$

%%%%%%%%%
\end{abstract}
%%%%%%%%%%%%%%%%%%%%%%%%%%%%%%%%%%%%%%%%%%%%%%%%%%%%%

%\pacs{02.30.Mv, 05.45.-a}

% Put \maketitle here in IOP style.
\maketitle

% body of paper here

%\narrowtext
%\tighten
%%%%%%%%%%%%%%%%%%%%%%%%%%%

%\newpage

%%%%%%%%%%%%%%%%%%%%%%%%%%%%%%%%%%%%%%%%%%%%%%%%%%%%%%%%%%%%
\section{Introduction}\label{sec:Introduction}

The observation of the Standard Model Higgs-like new boson with mass at around 125 GeV \cite{:2012gk, :2012gu}
changes the current understanding of new physics at the weak scale.
The minimal supersymmetric standard model (MSSM) can explain 125 GeV
with a relatively light stop of 1 to 2 TeV
in the context of maximal stop mixing.
From the model building point of view, it is quite difficult
to realise the maximal stop mixing scenario starting from ultraviolet (UV) theory.
In minimal gauge mediation (MGM) \cite{Dine:1993yw, Dine:1994vc, Dine:1995ag, Giudice:1998bp}, soft tri-linear A term is not generated at the messenger scale
and the radiatively generated A term at the weak scale is not large enough to realise
the maximal stop mixing.
As a result, colored superpartners should be as heavy as 5 to 10 TeV
to explain 125 GeV mass of the Higgs boson \cite{Ajaib:2012vc, Feng:2012rn}.
Therefore, the explanation of 125 GeV Higgs boson mass needs an extra help in minimal gauge mediation.
Next to the minimal supersymmetric standard model (NMSSM) can use the extra
contribution from the Yukawa F-term of the singlet.
For instance, look at \cite{Bae:2012am}.
Extra vector-like fermions are added in minimal gauge mediation \cite{Martin:2009bg, Martin:2012dg, Bae:2012ir}.
Direct coupling of visible sector fields with messengers can help.
Higgs-messenger mixing \cite{Kang:2012ra, Craig:2012xp}
and matter-messenger mixing \cite{Shadmi:2011hs, Albaid:2012qk, Abdullah:2012tq,  Evans:2012hg,  Evans:2011bea}
can generate Yukawa mediated contribution
including A term at the messenger scale.
However, at the same time the virtue of gauge mediation is gone and they would spoil nice flavor preserving spectrum
and can possibly cause the flavor problem at the weak scale.
For the Higgs-messenger mixing, $A/m^2$ problem\cite{Kang:2012ra, Craig:2012xp}  which is analogous to $\mu/B\mu$ problem can arise and the electroweak symmetry breaking is difficult to achieve if the mixing coupling is large.
General gauge mediation \cite{Buican:2008ws} can avoid this problem by using the mechanism of radiatively generated maximal stop mixing \cite{Dermisek:2006ey, Dermisek:2006qj}.

In this paper we consider the right-handed neutrino supermultiplets as the messengers
of supersymmetry breaking in addition to the messengers charged under
the Standard Model (SM) gauge group, e.g. ${\bf 5}$
and ${\bf \bar{5}}$ of $SU(5)$.
The setup is motivated from \cite{Choi:2011rs} which provides a solution to $\mu$ problem
in gauge mediation (more precisely $\mu/B\mu$ problem) \cite{Dvali:1996cu, Giudice:2007ca}.
For the solution in \cite{Choi:2011rs} to work, the messenger scale
should be higher than the Peccei-Quinn breaking scale, $10^9 \sim 10^{11}$ GeV.
For the gauge mediation to be the dominant contribution compared to the Planck suppressed
higher dimensional contribution, the messenger scale should be lower than $10^{15}$ GeV.
Therefore, the See-Saw scale
with order one neutrino Dirac Yukawa couplings,
$10^{13} \sim 10^{14}$ GeV, is well motivated as the messenger scale
if we accept \cite{Choi:2011rs} as a solution of the $\mu$ problem in gauge mediation.
If the messenger scale is at around the See-Saw scale, the natural question is
why the right-handed neutrino supermultiplets do not serve as messengers
of supersymmetry breaking.
Apparently there is no harm to couple the right-handed neutrino superfields
directly to the messengers. Majorana mass of the right-handed neutrino
and the messenger mass of ordinary ${\bf 5}$ and ${\bf \bar{5}}$
might have the same origin in this case.
In summary, the minimal set of messengers are ${\bf 5}$, ${\bf \bar{5}}$ and ${\bf 1}$.
This is different from previous studies relating gauge mediation and See-Saw mechanism \cite{Joaquim:2006uz, Mohapatra:2008wx, FileviezPerez:2009im}. They employ particles relevant to See-Saw mechanism as messengers, and these particles are also charged under the SM gauge group, which can be seen in the Type-II or Type-III See-Saw. Therefore, gauge mediation and neutrino Dirac Yukawa mediation have a common messenger.  In our case, in contrast, neutrino Dirac Yukawa messenger is the right-handed neutrino, the SM singlet.

If the right-handed neutrinos couple to the supersymmetry breaking field,
neutrino Dirac Yukawa coupling generates the A term and soft scalar mass of lepton doublet
and up-type Higgs at the See-Saw scale after integrating out the neutrino messengers.
125 GeV Higgs mass can be explained with stop lighter than 2 TeV in this setup.
At the same time the stop mass gets an extra Yukawa mediation and maximal stop mixing can be easily realised.

As there is a neutrino Dirac Yukawa contribution to the soft parameters
in addition to the ordinary gauge mediation,
interesting new physics signature is expected.
The mechanism of the charged lepton flavor violation is different from that in mSUGRA \cite{Borzumati:1986qx}
or SUSY GUT \cite{Ciuchini:2007ha} in which the origin is the running of soft parameters
above the See-Saw scale.
Though the origin is different, the spectrum looks similar.
The crucial difference is that here the flavor violation appearing in lepton doublet soft scalar mass
is $16\pi^2$ bigger than the one in mSUGRA.
Therefore, the naive expectation is that order one neutrino Dirac Yukawa coupling
would be incompatible with the current bounds of various charged lepton flavor violation constraints including $\mu \to e \gamma$.

The computation of the charged lepton flavor violation needs a complete flavor model.
Current observation of the charged lepton mass and lepton mixing matrix (PMNS) can be
explained in a consistent way with the neutrino Dirac Yukawa matrix which is proportional to the identity matrix. This is not an ad hoc assumption but can be explained in the context of non-Abelian discrete flavor symmetry, e.g., tribimaximal PMNS\cite{Harrison:2002er} from $S_4$.
Therefore, order one neutrino Dirac Yukawa coupling can generate order one A term
at the messenger scale and at the same time can be consistent with the charged lepton flavor violation constraints as long as it is proportional to the identity matrix.

$S_4$ flavor symmetry is the most natural and/or simple if $\theta_{13} =0$
as the tribimaximal mixing can be nicely realised.
However, small but sizeable $\theta_{13}$ ($\sin \theta_{13}  \sim 0.15$) can be accommodated
with the extra complication\cite{Lin:2009bw, Ishimori:2012fg, Altarelli:2012bn, King:2012vj}.
If the origin of $\theta_{13}$ is the modification of Majorana mass of the right-handed neutrino,
there would be no off-diagonal element in the lepton doublet soft scalar masses
as the neutrino Dirac Yukawa would be still proportional to the identity.
In this case the model is free from the cLFV constraints. Nevertheless, the sparticle spectrum needed to explain the observed Higgs mass
is heavy enough such that it is hard to explain the muon anomalous magnetic moment at the same time.
If $\theta_{13}$ is due to the deviation of the neutrino Dirac Yukawa matrix
from the identity, sizeable charged lepton flavor violation is expected.
We compute the charged lepton flavor violating processes in both cases
and show that interesting parameter space exists if $\theta_{13}$ is a combination
of two contributions from neutrino Dirac Yukawa and Majorana mass matrix.

The contents of the paper is following.
In section 2, we explain the setup for neutrino assisted gauge mediation
in which the right-handed neutrino is added as messengers in addition to the ordinary SM charged messengers.
Also we discuss the implication for the Higgs mass.
In section 3, we explain our $S_4$ flavor model as a representative example to discuss possible phenomenological implication.
In section 4, we discuss charged lepton flavor violation in connection with muon anomalous magnetic moment, the neutrino mixing angle $\theta_{13}$ and the Higgs mass.
Then we conclude.

%%%%%%%%%%%%%%%%%%%%%%%%%%%%%%%%%%%%%%%%%%%%%%%%%%%%%%%%%%%%%%

\section{Neutrino Assisted Gauge Mediation and The Higgs Mass}

\subsection{Soft terms generated from right-handed neutrino messengers}\label{sec:softterms}

 The extremely small masses of neutrinos can be explained through the See-Saw mechanism\cite{Minkowski:1977sc, Yanagida:1979as, Yanagida:1980xy, GellMann:1980vs, Mohapatra:1979ia}, in which lepton number is violated at around the Grand Unified Theory (GUT) scale. In this paper, we consider the simplest model, type-I See-Saw. For this, we extend the MSSM superpotential by including right-handed Majorana neutrinos,
\dis{W&=\epsilon_{ab}\Big[(Y_U)_{ij}\bar{U}_iQ^a_jH_u^b-(Y_D)_{ij}\bar{D}_iQ^a_jH_d^b
 -(Y_E)_{ij}\bar{E}_iL^a_jH_d^b+(Y_\nu)_{ij} {N}_iL^a_jH_u^b
 \\
 &  + \mu H_u^a H_d^b\Big] + \frac{1}{2} M_N^{ij} {N}_i {N}_j ,}
 where $\epsilon_{ab}$ is a totally antisymmetric tensor with $\epsilon_{12}=1$.   The superfields in the superpotential represent right-handed neutrino-sneutrino pairs, in addition to the SM particles and their superpartners. They have the following SM gauge group SU(3)$_c \times$SU(2)$_L \times$U(1)$_Y$ quantum numbers:
  \dis{&Q: (3,2,\frac16),~~\bar{U}: (\bar{3}, 1, -\frac23),~~\bar{D}: (\bar{3}, 1, \frac13)
  \\
  & L : (1, 2, -\frac12),~~\bar{E}: (1,1,1),~~N: (1,0,0)
  \\
  &H_u: (1,2,\frac12),~~H_d: (1,2,-\frac12).}
Relative minus signs of Yukawa terms are given to make the sign of terms responsible for the fermion Dirac masses to be the same.

The relevant soft supersymmetry (SUSY) breaking terms are given by
  \dis{\mathcal{L}_{\mathrm{soft}} =& - (m_{N}^2)^i_j \tilde{N}^\dagger_i \tilde{N}_j -  (m_L^2)^j_i \tilde{L}^{\dagger i} \tilde{L}_j -  m_{H_u}^2 H_u^{\dagger}  H_u \\-& \Big[ \frac{1}{2} (B_N M)^{ij} \tilde{N}_i \tilde{N}_j  +  (\tilde{A_U})_{ij} \tilde{U}^{i} \tilde{Q}^{j} H_u -(\tilde{ A_D})_{ij} \tilde{D}^{i}\tilde{Q}^{j} H_d -(\tilde{A_E})_{ij} \tilde{E}^i\tilde{L}^j H_d  +B\mu H_u H_d + h.c.  \Big] .}

   We consider two origins of soft terms. The first one is gauge mediation. In the gauge mediation, sfermions obtain soft masses  given by \cite{Giudice:1998bp}
 \dis{m_{\tilde{f}}^2=4 \sum_a\Big(\frac{g_a^2}{16\pi^2}\Big)^2 C_a \sum_i \Big(\frac{F}{M_i}\Big)^2\ T_a({\cal R}_i) f(x_i)}
 at the messenger scales $M_i$, where $C_a$ is the quadratic Casmir $\sum_\alpha T^\alpha T^\alpha$ of the sfermion representation ${\cal R}_i$ under the corresponding gauge group labeled by $a$, which is given by $(N^2-1)/(2N)$ for SU(N) and $Y^2$ for U(1)$_Y$, $T_a$ is defined by ${\rm Tr}T^\alpha T^\beta = T_a({\cal R}_i)\delta^{\alpha \beta}$, and $f(x_i)$ is the loop function of $x_i=F/M_i^2$ which is close to one for small $x_i$. On the other hand, the 2-loop tri-linear A term is very small and can be neglected at the messenger scale.

  For the second origin of soft terms, we introduce a SUSY breaking spurion $X$ which couples to the right-handed neutrinos.
Majorana mass of the right-handed neutrino comes from the scalar vacuum expectation value (VEV) of the SUSY breaking spurion $X$,
\dis{ W \supset \lambda X N N.}
Then $N$ acts as the messengers of supersymmetry breaking, and the neutrino Dirac Yukawa coupling,
\dis {W \supset Y_\nu N L H_u,}
is interpreted as the direct mixing term among the messengers, Higgs and matter (leptons).

The SUSY breaking effects at the See-Saw scale $M_N=\lambda \langle X \rangle$ is studied in \cite{Giudice:2010zn}. When right-handed neutrinos couple to the SUSY breaking sector, Majorana mass matrix is analytically continued to be  $M_N \rightarrow (1+ \theta^2 B_N)M_N$, as in the case of gauge mediation\cite{Giudice:1997ni, ArkaniHamed:1998kj, Chacko:2001km}. Here, we assume that the flavor structure of the right-handed neutrinos is fully determined by $M_N$, so $B_N=F_X/X$ is a constant.

Then, SUSY breaking is transferred to the visible sector through the neutrino Dirac Yukawa interaction. Wave function renormalization from the interaction with right-handed neutrinos is given by
    \dis{ \delta Z_L = \frac{Y_{\nu}^{R \dagger} }{16 \pi^2} \Big( 1- \ln \frac{M^{R \dagger}M^R}{\Lambda^2} \Big) Y_{\nu}^{R}, ~~~~~ \delta Z_{H_u}= \mathrm{Tr} \delta Z_L \label{eq:wavere}}
  where
 \dis{ \lambda_N^R = [Z_N^{-1/2}]^T \lambda_N Z_L^{-1/2 } Z_{H_u}^{-1/2},~~~~~ M^R = [Z_N^{-1/2 }]^T M_N Z_N^{-1/2},}
 then analytically continued Majorana masses give the soft masses. From field redefinitions
  \dis{& L \rightarrow \big( 1- \frac{\delta Z_L \vert_0}{2} \big)(1- \theta^2 \delta Z_L \vert_{\theta^2})L
 \\
 & H_u \rightarrow \big( 1- \frac{\delta Z_{H_u} \vert_0}{2} \big)(1- \theta^2 \delta Z_{H_u} \vert_{\theta^2})H_u,}
 supersymmetric kinetic terms can be written in the simple form,
 \dis{\Phi^{\dagger}(1+ \delta Z_{\Phi}) \Phi \rightarrow \Phi^{\dagger}(1+ \theta^2 \bar{\theta}^2 \delta Z_{\Phi} \vert_{\theta^2 \bar{\theta}^2}) \Phi }
 then we can read off the one-loop corrections to the soft masses
 \dis{ \delta m_L^2 = - \delta Z_L \vert _{\theta^2 \bar{\theta}^2}~~~ \mathrm{and}~~~ \delta m_{H_u}^2 = - \delta Z_{H_u} \vert _{\theta^2 \bar{\theta}^2} \label{eq:smass}.}
 In the expression, $B_N$ is just a constant, not a matrix. So $\ln(M_N^\dagger M_N)$ in the wave function renormalization is separated into holomorphic and anti-holomorphic parts, respectively. Since $\theta^2\bar{\theta}^2$ term is not generated, we do not have one-loop soft masses.

Hence, as in minimal gauge mediation, soft masses are generated at two loop level. In \cite{Kang:2012ra}, it was shown that soft scalar masses of the fields which directly couple to messengers and those which do not are different. In our model, the slepton $\tilde{L}$ and the up-type Higgs $H_u$ couple to messengers $N$ directly to give soft terms,

   \dis{&\delta m_{L}^2 = \frac{B_N^2}{(4 \pi)^4} \Big[ \Big({\rm Tr} [Y_{\nu} Y_{\nu}^\dagger]+3{\rm Tr} [Y_U Y_U^\dagger] - 3 g_2^2 - \frac{1}{5} g_1^2 \Big) Y_{\nu}^\dagger Y_{\nu} + 3 Y_{\nu}^\dagger Y_{\nu} Y_{\nu}^\dagger Y_{\nu} \Big]
  \\
  &\delta m_{H_u}^2 = \frac{B_N^2}{(4 \pi)^4} \Big[ 4{\rm Tr} [Y_{\nu} Y_{\nu}^\dagger Y_{\nu}^\dagger Y_{\nu}] -\Big( 3 g_2^2 + \frac{1}{5} g_1^2 \Big){\rm Tr} [Y_\nu Y_\nu^\dagger] \Big]. \label{eq:mlsqaure}}

On the other hand, $\tilde{Q}$ and $\tilde{U}$ obtain two-loop soft scalar masses through the wave function renormalization of $H_u$ and the corrections are  given by
  \dis{&\delta m_{Q}^2=-\frac{B_N^2}{(4 \pi)^4} {\rm Tr}[Y_\nu Y_\nu^\dagger]Y_U^\dagger Y_U
  \\
  &\delta m_{U}^2=-\frac{B_N^2}{(4 \pi)^4} {\rm Tr}[Y_\nu Y_\nu^\dagger]Y_U Y_U^\dagger}
  while the soft masses of $\tilde{E}$ and $H_d$ come out of the wave function renormalization of $L$ and the corrections are given by
\dis{
  &\delta m_{E}^2=-\frac{B_N^2}{(4 \pi)^4} Y_EY_\nu^\dagger Y_\nu Y_E^\dagger
  \\
  &\delta m_{H_d}^2=-\frac{B_N^2}{(4 \pi)^4} {\rm Tr}[Y_EY_\nu^\dagger Y_\nu Y_E^\dagger].}

By replacing $Y_E \to Y_E(1+\delta A_E)$, $Y_U \to Y_U(1+\delta A_U)$, and $Y_D \to Y_D(1+\delta A_D)$, we have following soft terms at one loop level,
 \dis{&\delta A_E = -\delta Z_L \vert_{\theta^2},~~~\delta A_U = - \mathbb{I} \delta Z_{H_u} \vert_{\theta^2},
 \\
 &\delta A_D = 0,~~~\delta B = -  \delta Z_{H_u} \vert_{\theta^2}.\label{eq:msoft}}
 Unlike gauge mediation, right-handed neutrino mediation generates one-loop $A-$terms,
 \dis{&A_E = \frac{B_N}{16 \pi^2 } Y_{\nu}^\dagger Y_{\nu}
 \\
 &A_U = - {\rm Tr} A_E \times \mathbb{I}_{3 \times 3}
 \\
 &B = {\rm Tr} A_E.}

While gauge mediation contributions are flavor universal, See-Saw Yukawa mediation is flavor dependent
and one of the virtue of the gauge mediaion would disappear.
In the absence of See-Saw Yukawa mediation, cLFV can appear when the messenger scale is higher than the right-handed neutrino Majorana mass scale. See-Saw Yukawa contributes to the slepton soft mass through the renormalization group equation (RGE) ,
  \dis{\mu\frac{d}{d\mu}m_L^2=\mu\frac{d}{d\mu}m_L^2\Big|_{\rm MGM}+\frac{1}{16\pi^2}\Big[(m_L^2 Y_\nu^\dagger Y_\nu+Y_\nu^\dagger Y_\nu m_L^2)+2(Y_\nu^\dagger m_N^2 Y_\nu +m_{H_u}^2Y_\nu^\dagger Y_\nu +\tilde{A}_\nu^\dagger \tilde{A}_\nu)\Big]}
  which should be restricted by cLFV constraints \cite{Grossman:2011fz}. Here $\tilde{A}_\nu=A_\nu Y_\nu$ is used.  Since $m_L^2$ is two-loop generated, cLFV effects are further loop suppressed (at three loop level).
Unlike mSUGRA, this effect is known to be small in gauge mediation
as the messenger scale is at most comparable to the See-Saw scale
and the running can be made in a very short interval. This is not the cLFV that we are interested in.

In neutrino assisted gauge mediation, neutrino Dirac Yukawa couplings can introduce two-loop generated cLFV effects on $m_L^2$ as a result of gauge-Yukawa or Yukawa mediation,
\dis{\delta m_{L}^2 = \frac{B_N^2}{(4 \pi)^4} \Big[ \Big({\rm Tr} [Y_{\nu} Y_{\nu}^\dagger]+3{\rm Tr} [Y_U Y_U^\dagger] - 3 g_2^2 - \frac{1}{5} g_1^2 \Big) Y_{\nu}^\dagger Y_{\nu} + 3 Y_{\nu}^\dagger Y_{\nu} Y_{\nu}^\dagger Y_{\nu} \Big]}
in the charged lepton mass basis.
If the two loop generated slepton mass squared has a nonzero off-diagonal element,
it would generate cLFV.
Parametrically, this effect is much larger than the expected cLFV in mSUGRA or similar scenarios in which the effect comes from the running above the See-Saw scale.
We simply assume that both messengers ${\bf 5}, {\bf \bar{5}}$ and ${\bf 1}$ have the same masses
at the See-Saw scale. In principle these two masses can be different
and cLFV can arise if the singlet messenger is lighter than ${\bf 5}, {\bf \bar{5}}$.
However, this effect is loop suppressed compared to the Yukawa mediation we would not consider it in this paper.

Further discussion on cLFV is possible only when there is an explicit flavor model providing the neutrino Dirac Yukawa and charged lepton Yukawa matrices.
As a simple and illustrative example of the explicit model, we consider $S_4$ flavor symmetry
in Sec. \ref{sec:FlavorModel}. It will be shown that various types of See-Saw Yukawa $Y_\nu$ would predict different sizes of effects on cLFV.
Before moving onto the flavor discussion, let us consider the implication on the Higgs mass first.

\subsection{Higgs mass and superparticle spectrum}\label{sec:higgs}

%%%%%%%%%%%%%%%%%%%%%%%%%%%%%%%%%%%%%%%%%%%%%%%%%%%%%%%%%%%%%%%%%%%%%%%%%%%%%%%%%%%%%%%%
 \begin{figure}[!t]
  \begin{center}
   \includegraphics[width=0.45\textwidth]{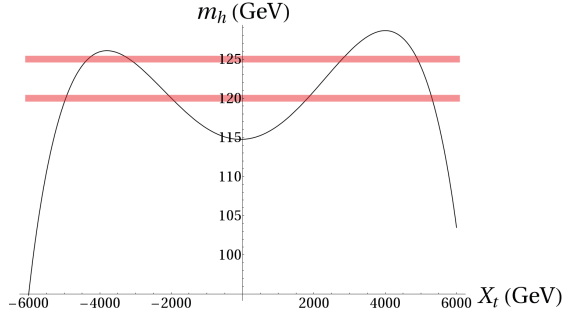}
  \end{center}
 \caption{Higgs mass with respect to $X_t$ for $\tan\beta=10$, stop mass $M_{\tilde{t}} \sim 2 \tev$.
  }
\label{fig:xHiggs}
\end{figure}
%%%%%%%%%%%%%%%%%%%%%%%%%%%%%%%%%%%%%%%%%%%%%%%%%%%%%%%%%%%%%%%%%%%%%%

Minimal gauge mediation does not generate $A_t$ at one loop
and the weak scale $A_t$ is radiatively generated by the gluino loop.
However, the same gluino contribution appears in stop soft scalar mass
and the relative ratio of $|A_t|$ and $m_{\tilde{t}}$ can not be large.
On the other hand, the physical light CP even Higgs mass in the MSSM
is affected by $\hat{X_t} \equiv (A_t -\mu /\tan \beta)/m_{\tilde{t}}$
and $\hat{X_t} \sim 2$ (or $\sqrt{6}$ more precisely) gives the maximum finite threshold correction
as shown in Fig. \ref{fig:xHiggs}.

One way to make $|\hat{X_t}| > 1$ at the weak scale is to start from tachyonic stop boundary condition
\cite{Dermisek:2006ey} which is explicitly realised in gauge messenger model \cite{Dermisek:2006qj}.
However, this option is not available in minimal gauge mediation.
The other possibility is to couple messengers directly to the visible sector fields
such that large negative $A$ term can be generated at the messenger scale.
If $A$ term is positive, the gluino contribution from the running cancels the $A$ term at the messenger scale.
Matter-messenger mixing
\cite{Shadmi:2011hs, Albaid:2012qk, Abdullah:2012tq, Evans:2012hg,  Evans:2011bea} also has been considered recently.
Messenger-matter-matter Yukawa coupling would generate the needed $A_t$ term
at the messenger scale. However, the full Yukawa couplings are written as $3\times 3$ matrices
and why all other dangerous Yukawa couplings between matters and messengers are absent except $33$ component remains to be a puzzle.
One way out is to consider Higgs-messenger mass mixing \cite{Chacko:2001km}
and to generate all the wanted Yukawa couplings between matter and messengers
from ordinary Yukawa couplings of matter with Higgs.
There would be a direct modification of squark spectrum if squark couples directly to the messenger.

%%%%%%%%%%%%%%%%%%%%%%%%%%%%%%%%%%%%%%%%%%%%%%%%%%%%%%%%%%%%%%%%%%%%%%%%%%%%%%%%%%%%%%%%
 \begin{figure}[ht]
  \begin{center}
      \includegraphics[width=0.70\textwidth]{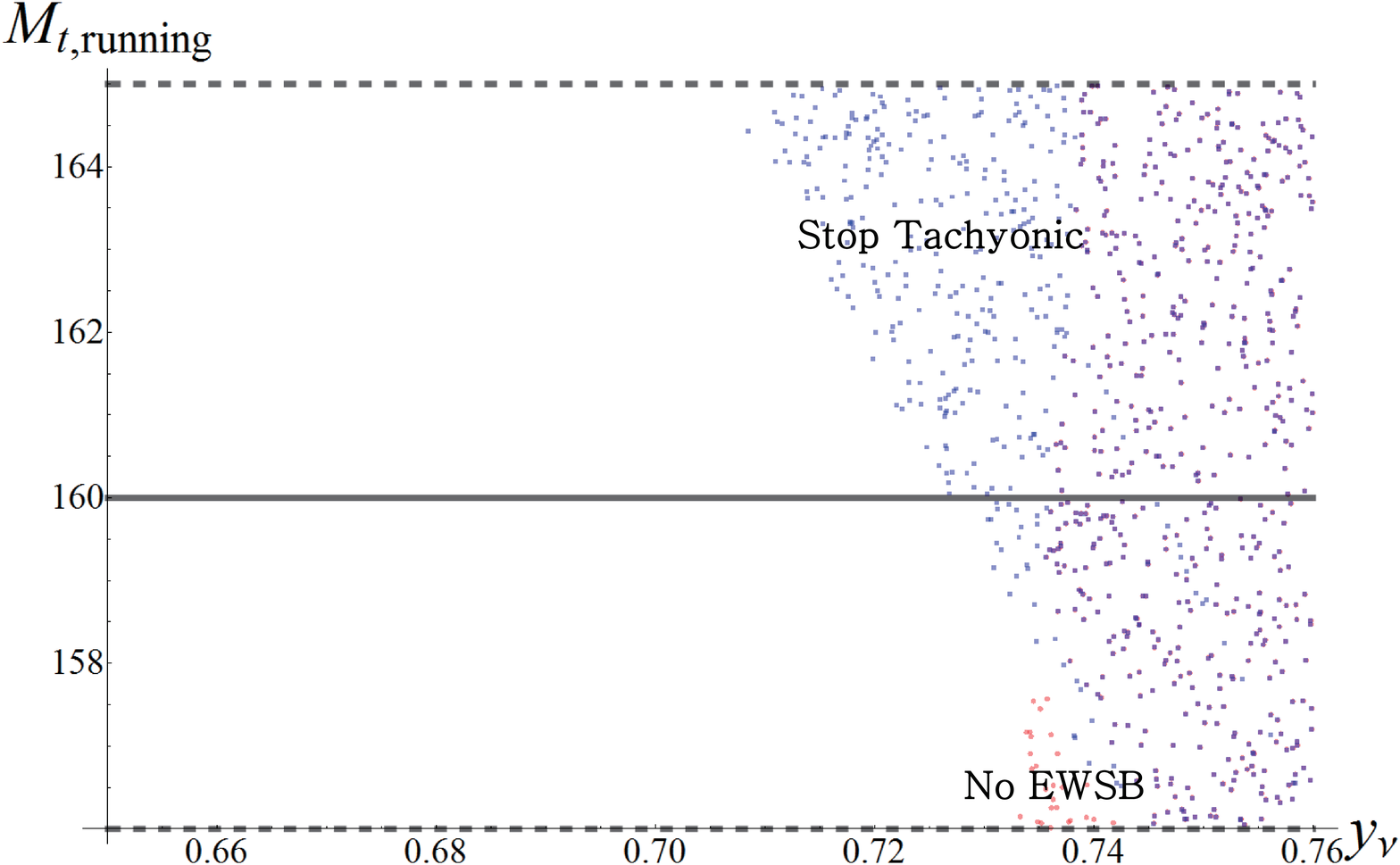}
      \includegraphics[width=0.70\textwidth]{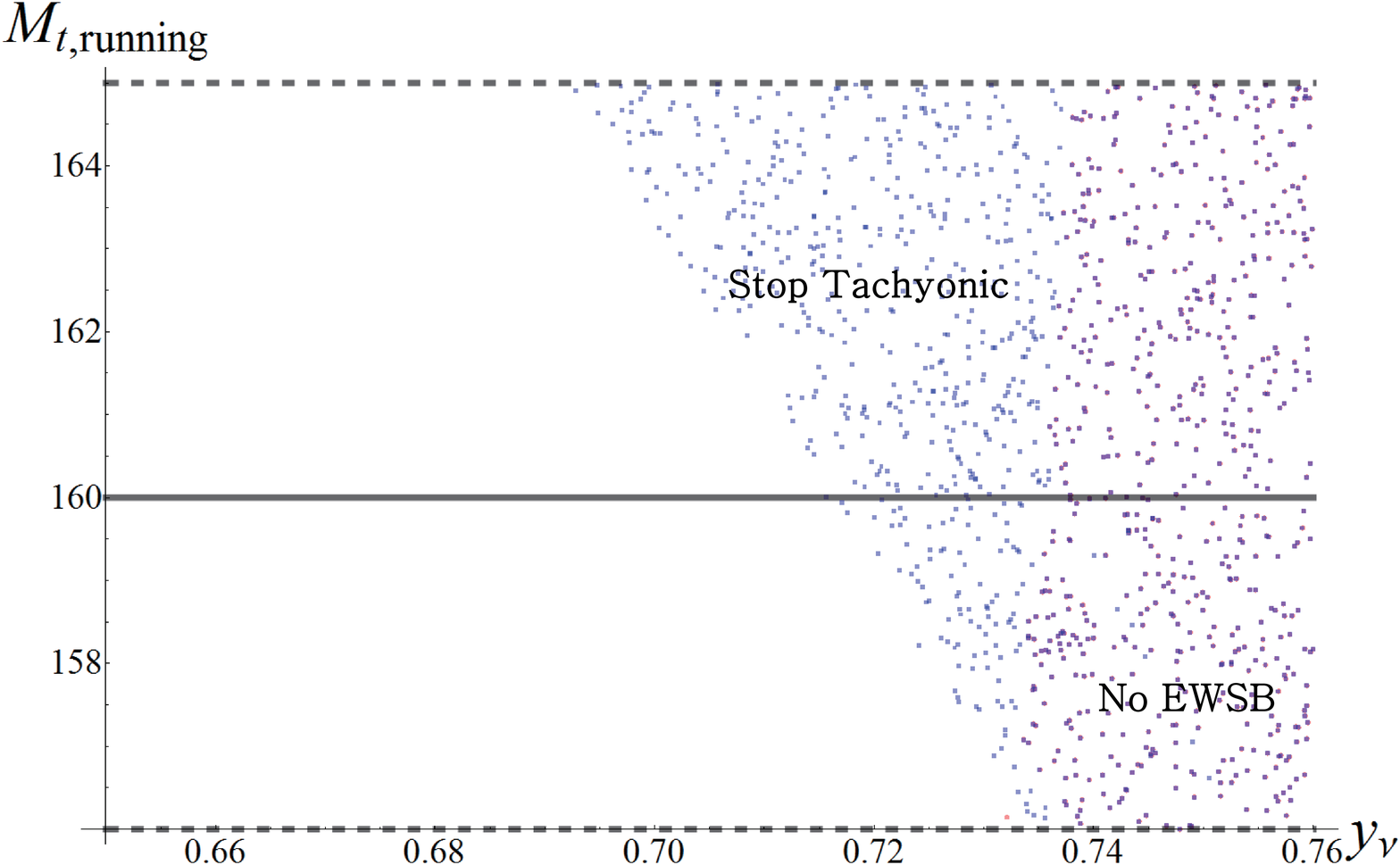}
  \end{center}
 \caption{ Phase diagrams indicating stop tachyonic and no EWSB region for $\tan \beta = 10$, $\tan \beta = 30$, respectively. $B_N$ is set to be $5 \times 10^5 \gev$.
  }
\label{fig:phase}
\end{figure}
%%%%%%%%%%%%%%%%%%%%%%%%%%%%%%%%%%%%%%%%%%%%%%%%%%%%%%%%%%%%%%%%%%%%%%

Higgs-messenger mixing through Higgs-messenger-messenger coupling or Higgs-Higgs-messenger coupling has been considered in \cite{Kang:2012ra, Craig:2012xp}. In this case, we often encounter $A/m^2$ problem.   To understand this, it is worth to emphasize that, the two-loop soft mass squared of the Higgs field $H_u$ which has direct coupling to messenger $\Phi$ has a structure of $m_{H_u}^2 \sim c \lambda^4-c^\prime \lambda^2 g^2$ where $\lambda$ is a coupling constant of Higgs and messenger fields and $g$ is the gauge coupling(s). On the other hand, the two-loop soft mass squared of fields $Q,\bar{U}$ which does not have a direct coupling with messenger has a form of $m_{Q_3,\bar{U}_3}^2 \sim -c_3\lambda^2 y_t^2$. This fact was extensively studied in \cite{Kang:2012ra}. For sufficiently large $\lambda$, large one-loop A terms are generated. At the same time, $m_{H_u}^2$ becomes positive so the soft mass of $H_u$ can be much larger than that in the pure gauge mediation. Moreover, the soft mass of $Q_3,\bar{U}_3$ can be much smaller. If the Higgs $H_u$ superfield directly couples to messengers whereas the top superfields do not, relatively light stop in natural SUSY can be easily obtained as we can have the small stop soft mass from the effect explained above and the large LR mixing from large A term. $A/m^2$ problem appears in $H_u$ soft terms such that large $A$ term at the same time generate large $m_{H_u}^2$ at the messenger scale
and it can make the electroweak symmetry breaking difficult. It is analogous to the famous $\mu/B\mu$ problem in gauge mediation.
 To avoid this but to allow the large $\lambda$ for maximal mixing, large $-c_2 \lambda^2 g^2$ part in $m_{H_u}^2$ is required. This can be achieved by introducing new gauge bosons or making strong interaction involved \cite{Kang:2012ra}.  On the other hand, one loop, negative contribution to $m_{H_u}^2$ can be considered if the messenger scale is low as analysed in detail in \cite{Craig:2012xp}.

%%%%%%%%%%%%%%%%%%%%%%%%%%%%%%%%%%%%%%%%%%%%%%%%%%%%%%%%%%%%%%%%%%%%%%%%%%%%%%%%%%%%%%%%
 \begin{figure}[!t]
  \begin{center}
      \includegraphics[width=0.70\textwidth]{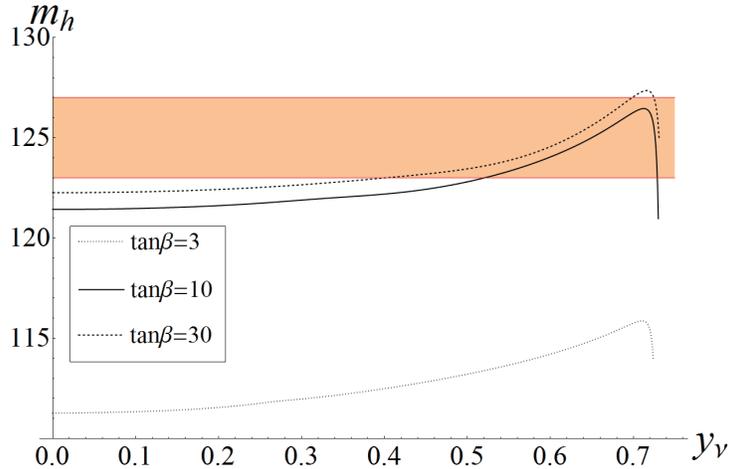}
  \end{center}
 \caption{ Higgs mass as a function of $y_{\nu}$ for  $B_N = 5  \times 10^{5} \gev$, $\rho = 0.1$. Higgs mass can be achieved with the help of Yukawa mediation for large $\tan \beta$ region. At $y_{\nu} \sim 0.7$, stop mass is approximately $1 \tev$.
  }
\label{fig:Higgsmass}
\end{figure}
%%%%%%%%%%%%%%%%%%%%%%%%%%%%%%%%%%%%%%%%%%%%%%%%%%%%%%%%%%%%%%%%%%%%%%

Neutrino assisted gauge mediation uses the Yukawa coupling among messengers (neutrinos), Higgs and lepton doublets.
As a result, Higgs and lepton doublet soft scalar masses get extra contribution from Yukawa mediation.
The same $A/m^2$ problem applies here and neutrino Dirac Yukawa coupling can not be taken to be a large value for successful electroweak symmetry breaking in principle. On the other hand, too large $m_{H_u}^2$ and too large A term may drive stop tachyonic through renormalization group running with top Yukawa.
The problem becomes worse as the stop soft scalar mass squared at the messenger scale gets a negative contribution from Yukawa mediation.
The situation is shown in Fig. \ref{fig:phase}. For the running mass of the top quark 160 GeV (the central value), the tachyonic stop appears before the real $A/m^2$
 problem prevents the successful electroweak symmetry breaking as we increase $y_\nu$. The crucial difference compared to the previous work in which $A/m^2$ problem is emphasized \cite{Kang:2012ra, Craig:2012xp} comes from the number of messengers. In neutrino assisted gauge mediation, the number of messengers is three, $N=3$. The $y^2$ contribution is effectively replaced by $N y_\nu^2$. Large $N$ effectively reduces the $A/m^2$ problem by $1/N$. At the same time smaller $y_\nu$ can provide the same impact with the aid of $N > 1$.
If tachyonic stop appears as $y_\nu$ gets larger, it would be easy to realise the maximal stop mixing by making the stop soft scalar mass sufficiently small.

%%%%%%%%%%%%%%%%%%%%%%%%%%%%%%%%%%%%%%%%%%%%%%%%%%%%%%%%%%%%%%%%%%%%%%%%%%%%%%%%%%%%%%%%
 \begin{figure}[!h]
  \begin{center}
      \includegraphics[width=0.50\textwidth]{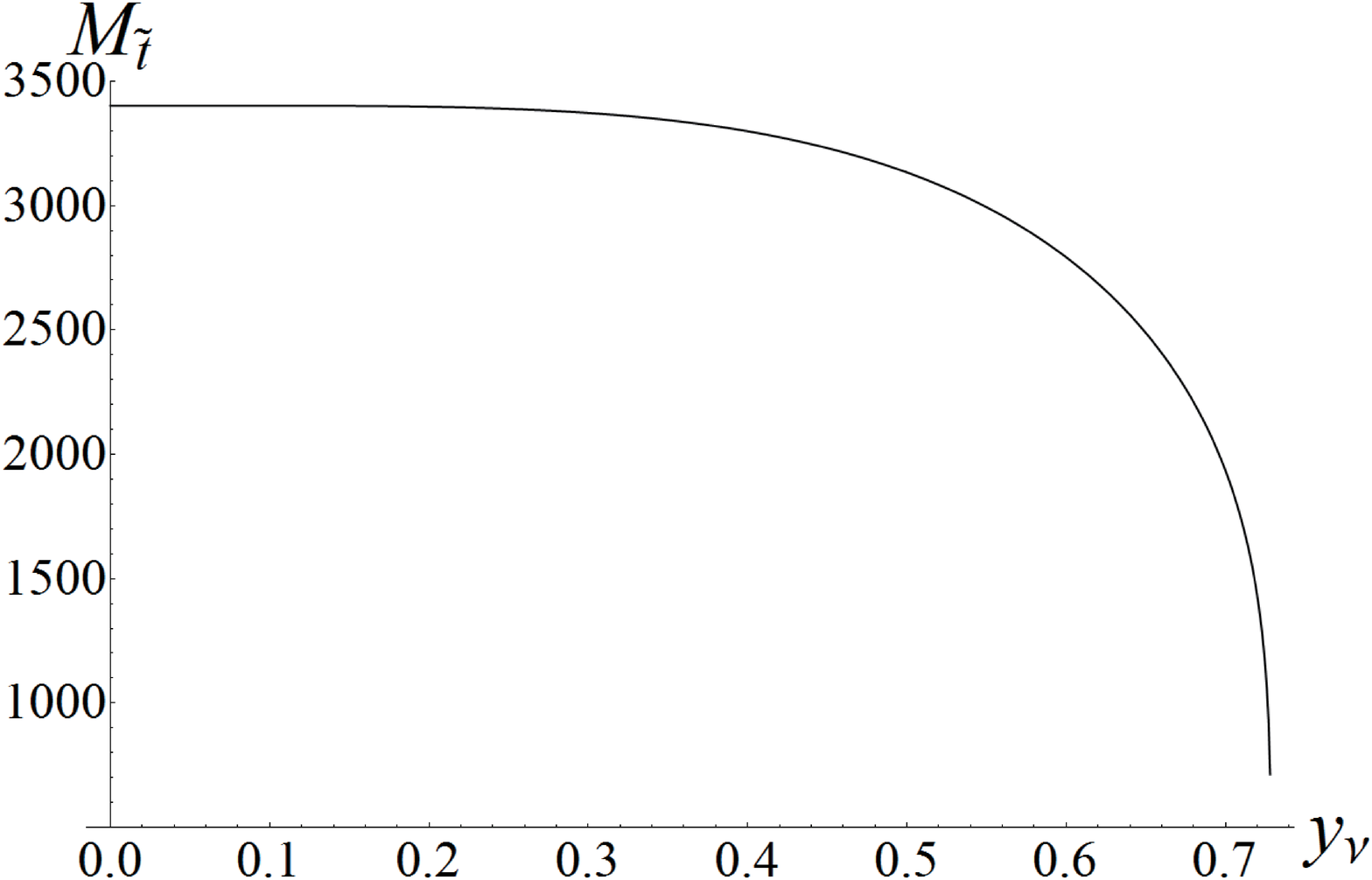}
%  \end{center}
 \caption{  $ M_{\tilde t}$ as a function of $y_{\nu}$ for $B_N = 5  \times 10^{5} \gev$, $\rho = 0.1$. $\tan \beta = 10$.
  }
\label{fig:stopmass}
%\end{figure}
%%%%%%%%%%%%%%%%%%%%%%%%%%%%%%%%%%%%%%%%%%%%%%%%%%%%%%%%%%%%%%%%%%%%%%
\vspace{10mm}

%%%%%%%%%%%%%%%%%%%%%%%%%%%%%%%%%%%%%%%%%%%%%%%%%%%%%%%%%%%%%%%%%%%%%%%%%%%%%%%%%%%%%%%%
% \begin{figure}[!h]
%  \begin{center}
      \includegraphics[width=0.50\textwidth]{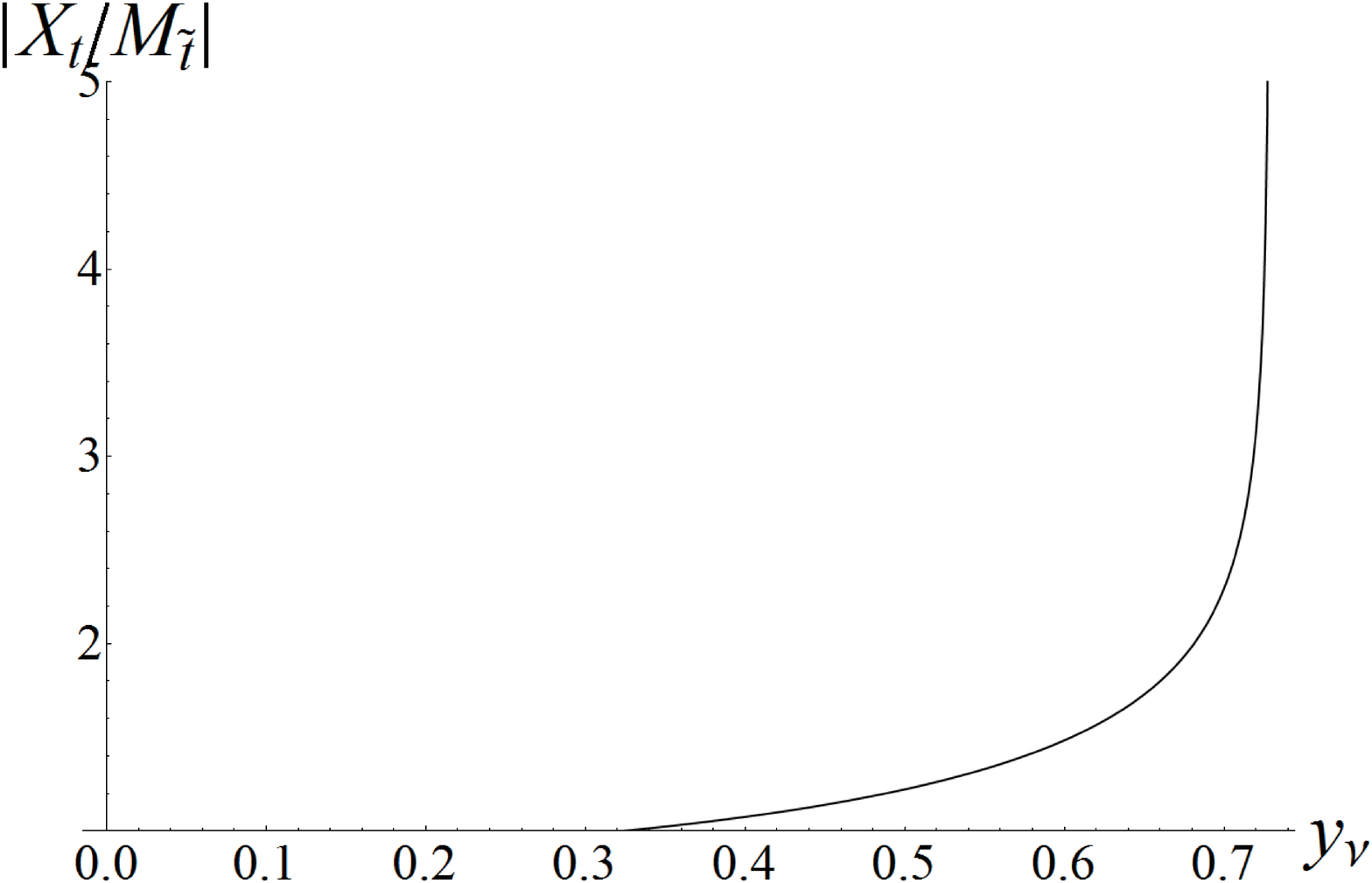}

  \end{center}
 \caption{ $X_t / M_{\tilde t}$ as a function of $y_{\nu}$ for $B_N = 5  \times 10^{5} \gev$, $\rho = 0.1$, $\tan \beta = 10$.
  }
\label{fig:XtMt}
\end{figure}
%%%%%%%%%%%%%%%%%%%%%%%%%%%%%%%%%%%%%%%%%%%%%%%%%%%%%%%%%%%%%%%%%%%%%%

Fig. \ref{fig:Higgsmass} shows the contribution assisted by neutrino messengers, compared to the minimal gauge mediation which corresponds to $y_\nu=0$
with stop mass at around 1 TeV. In the minimal gauge mediation,
the Higgs mass is computed to be at around $121 \sim 122$ GeV for $\tan \beta = 10 \sim 30$.
For $y_\nu = 0.7$, the Higgs mass can be as large as $125 \sim 126$ GeV.
4 to 5 GeV gain in the Higgs mass is obtained in neutrino assisted gauge mediation.
The gain does not look impressive but has an impact on allowed superparticle spectrum.
In the absence of $A_t$ at the messenger scale as is the case in minimal gauge mediation,
this extra 5 to 6 GeV can be achieved by making the logarithmic contribution large
and the stop mass should be as heavy as 5 to 10 TeV rather than 2 TeV.

Note that the plot stops at $y_\nu=0.72$.
Neutrino assisted gauge mediation is classified as Higgs-messenger mixing scenario
as the right-handed neutrino is the messenger and the neutrino Dirac Yukawa coupling connects Higgs, lepton doublet and the messenger (right-handed neutrino).
The stop soft scalar mass squared gets smaller and becomes tachyonic as the neutrino Dirac Yukawa coupling is increased as in Fig. \ref{fig:stopmass}.
The logarithmic correction to the Higgs mass also rapidly drops beyond $y_\nu \sim 0.7$ as the stop mass becomes too light (and becomes tachyonic)
as is shown in Fig. \ref{fig:Higgsmass}.
The maximal mixing is realised around this point, as shown in Fig. \ref{fig:XtMt}.
This also corresponds to the corner of the parameter space next to the critical point as in \cite{Giudice:2006sn}.

%%%%%%%%%%%%%%%%%%%%%%%%%%%%%%%%%%%%%%%%%%%%%%%%%%%%%%%%%%%%%%%%%%%%%%%%%%%%%%%%%%%%%%%%
 \begin{figure}[!t]
  \begin{center}
      \includegraphics[width=0.60\textwidth]{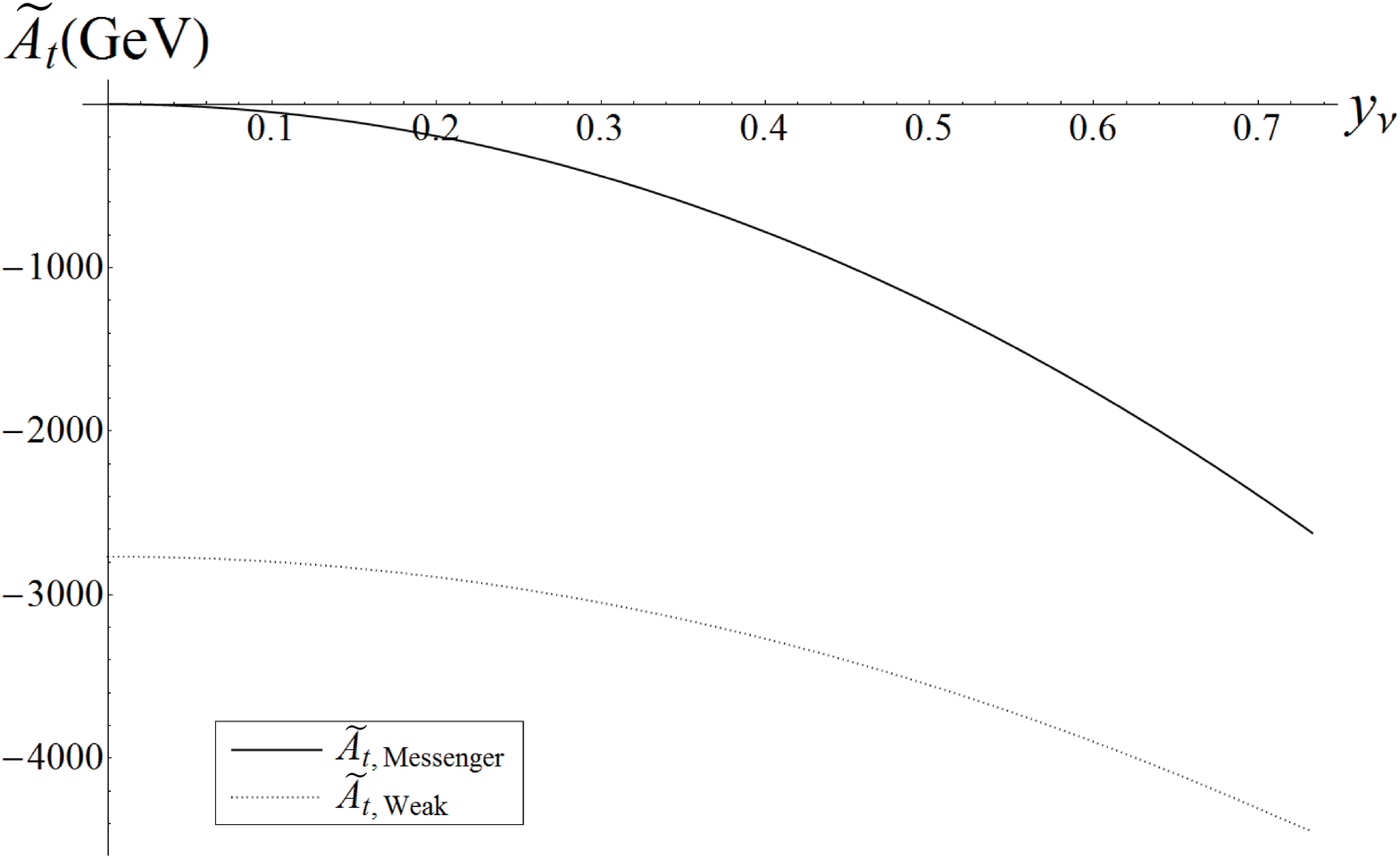}
  \end{center}
 \caption{ $\tilde{A}_t \equiv A_t Y_t$ as a function of $y_\nu$ for $\tan \beta = 10$, $B_N = 5 \times 10^5$GeV.  Without Yukawa mediation, one can obtain $\tilde{A}_{t} \sim -2700\gev$ at the weak scale by RG running effects. With help of neutrino mediation at the messenger scale,
 one can obtain $\tilde{A}_{t} \sim -4500\gev$ at weak scale. This drives more stop mixing, which helps $125 \gev$ Higgs mass.
  }
\label{fig:At}
\end{figure}
%%%%%%%%%%%%%%%%%%%%%%%%%%%%%%%%%%%%%%%%%%%%%%%%%%%%%%%%%%%%%%%%%%%%%%

Fig. \ref{fig:At} compares $A_t$ in the minimal gauge mediation
and the neutrino assisted gauge mediation both at the messenger scale and the weak scale.
Note that $A_t$ by itself is enhanced by 1.5 at the weak scale with the help of messenger scale $A_t$.

%%%%%%%%%%%%%%%%%%%%%%%%%%%%%%%%%%%%%%%%%%%%%%%%%%%%%%%%%%%%%%%%%%%%%%%%%%%%%%%%%%%%%%%%
 \begin{figure}[!t]
  \begin{center}
   \includegraphics[width=0.80\textwidth]{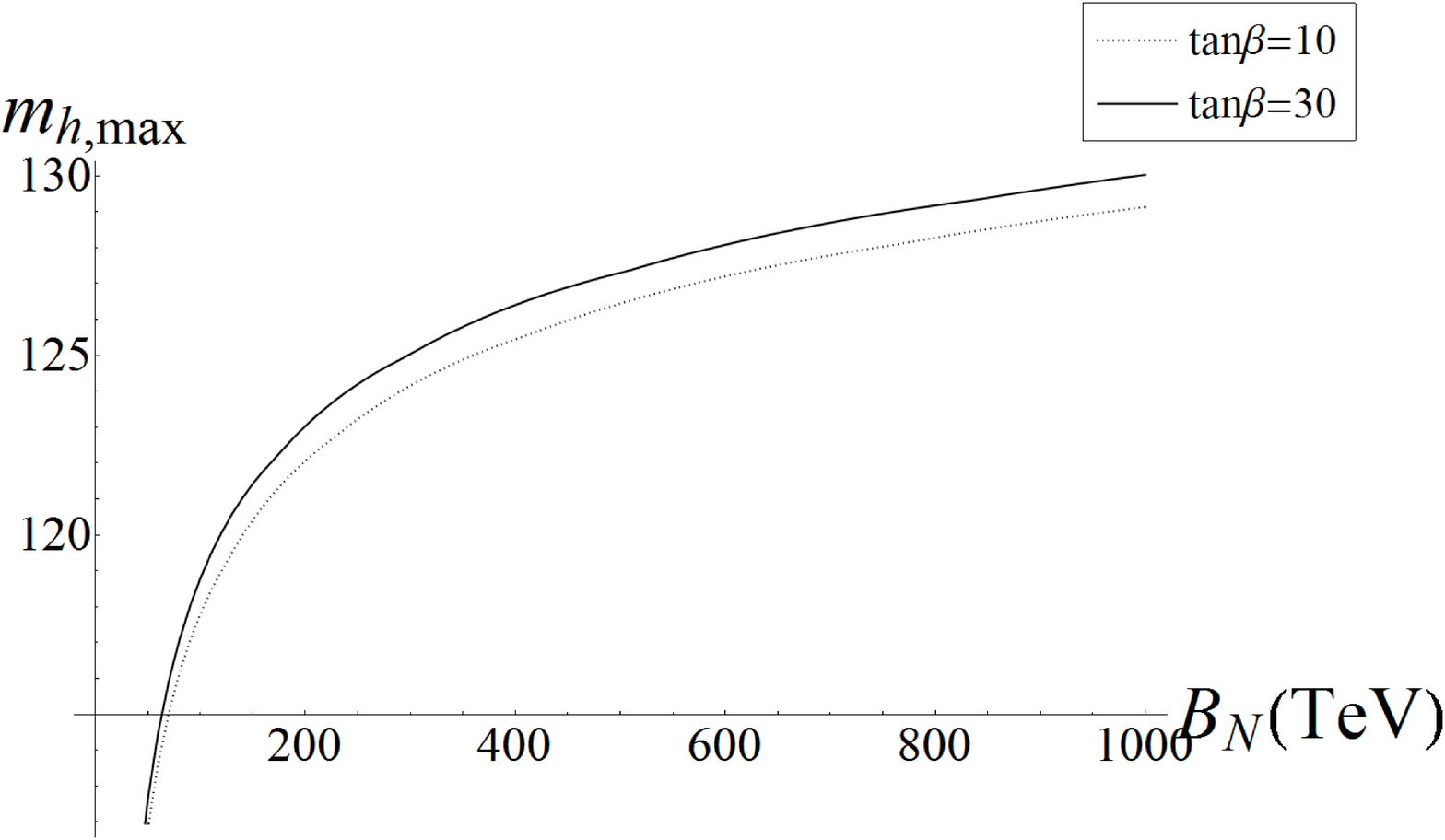}
  \end{center}
 \caption{Maximum values of Higgs mass as a function of $B_N$. For $\tan \beta = 10$, at least $B_N = 360 \tev$ is required to obtain $125\gev$ Higgs mass. For $\tan \beta = 30$,
  $B_N = 300 \tev$ is required. At two points $( \tan \beta = 10 , B_N = 360 \tev)$, $( \tan \beta = 30 , B_N = 300 \tev)$, we display sparticle spectrums in Table \ref{table:Spectrum125}. Also spectrums with $123 \gev$ Higgs mass are given for  $(\tan \beta = 10, B_N = 240 \tev)$, $(\tan \beta = 30 , B_N = 200 \tev)$. Sparticle spectrums are displayed in Table \ref{table:Spectrum123}.}
\label{fig:BNHiggs}
\end{figure}
%%%%%%%%%%%%%%%%%%%%%%%%%%%%%%%%%%%%%%%%%%%%%%%%%%%%%%%%%%%%%%%%%%%%%%

Fig. \ref{fig:BNHiggs} shows the relation between $B_N$ and the Higgs mass.
The neutrino Dirac Yukawa coupling $y_\nu$ is chosen to be close to $0.72$ which can maximize the Higgs mass for given $B_N$.

In summary, the minimal gauge mediation needs stop mass at around 5 to 10 TeV
to raise the Higgs mass up to 125 GeV. If the right-handed neutrinos are the messengers
of the supersymmetry breaking, so called `neutrino assisted gauge mediation', we can explain 125 GeV Higgs mass with lighter than 2 TeV stop mass.

%%%%%%%%%%%%%%%%%%%%%%%%%%%%%%%%%%%%%%%%%%%%%%%%%%%%%%%%%%%%%%%%%%%%%%%%%%%%%%%%
%%%%%%%%%%%%%%%%%%%%%%%%%%%%%%%%%%%%%%%%%%%%%%%%%%%%%%%%%%%%%%%%%%%%%%%%%%%%%%%%
\section{Flavor Model }\label{sec:FlavorModel}
%%%%%%%%%%%%%%%%%%%%%%%%%%%%%%%

 In this section, we consider models which can explain neutrino oscillations successfully. Since the SUSY breaking mediation through the neutrino Dirac Yukawa coupling is flavor dependent in general, sizable cLFV could be generated. To avoid this, the neutrino Dirac Yukawa coupling is set to be proportional to the idenitity. In the right-handed neutrino mass basis, it would be proportional to the unitary matrix so soft mass $m_L^2$, which depends on the combination $Y_\nu^\dagger Y_\nu$ is flavor universal.  It is easily achieved by employing the non-abelian discrete symmetry for the tri-bi maximal mixing of the PMNS matrix. Since the tri-bi maximal mixing should be modified to make $\theta_{13}$ nonzero, as reported by several observations\cite{Abe:2011fz, Hartz:2012np, Adamson:2012rm, An:2012eh, Ahn:2012nd}, small corrections should be added. When the neutrino Dirac Yukawa coupling has such corrections, such that it has a deviation from identity, cLFV is generated. We look for several ways to suppress cLFV, at least under the experimental bound.

%%%%%%%%%%%%%%%%%%%%%%%%%%%%%%%%%%%%%%%%
   \begin{table}[!h]
\begin{center}
\begin{tabular}{|c|c|c|c|c|}
\hline    Superfield  & $S_4$ & $Z_4$ & U(1)$_L$ & U(1)$_R$
\\[0.2em]
\hline  &&&&\\ [-1.1em]
$L$ &  ${\bf 3}$& 1& 1& 1
\\[0.4em]
$\bar{E}$ &${\bf 2+1}$ & 2 & -1& 0
\\[0.4em]
$N$ & ${\bf 3}$ & 3 & -1& 0
\\[0.4em]
$\Phi$ &  ${\bf 3+3^\prime}$ & 1& 0& 0
\\[0.4em]
$\chi$  & ${\bf 1+2+3}$ & 2 & 2 & 0
\\[0.4em]
$H_u$  &${\bf 1}$ & 0 &0 & 1
\\[0.4em]
$H_d$  & ${\bf 1}$ & 0 &0 & 1
\\[0.4em]
$X$ & ${\bf 1}$& 0&0& 2

\\[0.4em]
\hline
\end{tabular}
\end{center}
\caption{Charge assignments under
$S_4 \times Z_4 \times {\rm U(1)}_L \times {\rm U(1)}_R $  for leptons, flavons, Higgs, and SUSY breaking spurions. } \label{table:charges}
\end{table}
%%%%%%%%%%%%%%%%%%%%%%%%%%%%%%%%%%%%

  To make the PMNS matrix tri-bi maximal, we use $S_4$ discrete symmetry, since it is closely related to the permutation structure of Yukawa couplings. Other discrete symmetries, such as $A_4$, the even permutation of the $S_4$ could be used. The main difference is that the first and the second generation of the right-handed leptons belong to ${\bf 2}$ dimensional representation in $S_4$
  while they correspond to different one dimensional representations, ${\bf 1^\prime}$, ${\bf 1^{\prime\prime}}$ in $A_4$. In \cite{He:2006dk, He:2006qd, He:2011gb, BenTov:2012tg}, the structure we use is obtained from $A_4$ symmetry and discussion on the deviation from the tri-bi maximal mixing is in parallel. The $S_4$ symmetry model building is reviewed in \cite{Bazzocchi:2012st}. In Appendix A, we summarised representations and tensor products of $S_4$ group.

  For quark sector, the CKM matrix is close to the identity. Deviation from the identity has a hierarchy structure parametrized by some powers of the Cabibbo angle, $\lambda=\sin\theta_C$.  On the other hand, the PMNS matrix, mixing matrix in the lepton sector has large mixing angles. Even the smallest mixing angle, $\theta_{13}$ is in the order of $\lambda$. To explain this, it is natural to assume that $u-$ and $d-$ quark sectors have almost the same structure under the discrete flavor symmetry whereas the charged lepton and the right-handed neutrino sectors do not. This picture can be realised by introducing appropriate `flavons' charged under discrete symmetry group and more symmetries can be introduced to forbid useless couplings. Here, we consider the symmetry group $S_4 \times Z_4 \times {\rm U(1)}_L$, where U(1)$_L$ represents a lepton number, which may be discretized.
  In this paper, we consider superpotential for See-Saw mechanism with flavons $\Phi$ and $\chi$,
%\dis{W=\epsilon_{ab} [-\bar{E} \Phi L^aH_d^b +NL^aH_u^b]+\frac12 N\chi N.}
 \dis{W= -\l_{1ij} \bar{E}_i \Phi L_j H_d + \l_{2ij} N_i L_j H_u+\frac12 \l_{3ij} X N_i \chi N_j,}
 where $i,j=1,2,3$ are the generation indices and $X$ is a SUSY breaking spurion.
  For this, $S_4$, $Z_4$, U(1)$_L$ and U(1) R-symmetry quantum numbers are given in Table \ref{table:charges}.
  %as follows:
% \dis{& L : ({\bf 3}, 1, 1, 1),~~~\bar{E}: ({\bf 2+1}, 2, -1, 0),~~~N: ({\bf 3}, 3, -1, 0)
% \\
% &  \Phi : ({\bf 3+3^\prime}, 1, 0, 0),~~~\chi : ({\bf 1+2+3}, 2, 2, 0)
% \\
 %& H_u:  ({\bf 1}, 0,0, 1),~~~H_d : ({\bf 1}, 0,0, 1),~~~ X : ({\bf 1}, 0,0, 2)
 %}

 The charged lepton Yukawa couplings can be constructed from $\bar{E}\Phi L H_d$, the neutrino Dirac Yukawa coupling can be constructed from $NL$, and the Majorana mass of the heavy neutrinos can be constructed from $X N\chi N$. On the other hand, $\Phi^2$, $\chi^2$, and $\Phi\chi$ cannot couple to the combinations $\bar{E} L H_d$, $NL$, and $XNN$ to make singlets.  Note that U(1)$_R$ is introduced to forbid unwanted coupling $ N \chi N$, which makes $B_N$ in a matrix form, not a constant.

The discrete symmetry quantum number can be extended to the quark sector, such as $Q: ({\bf 3},1,1, 1, 1)$, $\bar{U} : ({\bf 2+1},2,0, 0)$, and $\bar{D} : ({\bf 2+1},2,0, 0)$ under $S_4\times Z_4\times {\rm U(1)}_L \times$U(1)$_R$. The flavons $\Phi : ({\bf 3+3^\prime}, 1,0)$ make the singlet combinations ${\bar U}\Phi Q H_u+{\bar D}\Phi Q H_d$ and Yukawa couplings $Y_U$ and $Y_D$ have the same form as the charged lepton Yukawa coupling. They are diagonalized by the same unitary matrix so CKM matrix is the identity in the leading order. If another type of flavon couples to either of up and down quark sectors to give subleading corrections of order $\lambda$, it would explain the Cabibbo angle.

Lepton $L_i$ is in the {\bf 3} and $\bar{E}_j$ is in the {\bf 1+2} representations, in which $(\bar{E}_1)_{\bf 1}+(\bar{E}_2,\bar{E}_3)_{\bf 2}$. Also there are the SM singlet flavons $\Phi_{\bf 3}$, and $\Phi_{\bf 3^\prime}$ in the ${\bf 3}$, and ${\bf 3^\prime}$ representations.
We do not provide a complete vacuum alignment in this setup.
Instead in Appendix B, we show a few simple examples in which the aligned vacuum is realised.
If, for instance, VEVs are arranged to be  $\langle\Phi_{\bf 3}\rangle =v_2 (1,1,1)$, and $\langle \Phi_{\bf 3^\prime}\rangle =v_3 (1,1,1)$, we have the following Yukawa structure
   \dis{ Y_E=\lambda_E \frac{1}{\sqrt3}
\left(
\begin{array}{ccc}
 c &c& c\\
 a & a\omega & a\omega^2 \\
 b & b\omega^2 & b\omega
  \end{array}\right) \label{eq:leptonyukawa}
  }
    where $a=(\lambda v_2+ \lambda^\prime v_3)/\Lambda$, $b=(\lambda v_2-\lambda^\prime v_3)/\Lambda$, $c=\lambda^{\prime \prime}v_2/\Lambda$,  and $\lambda, \lambda^\prime, \lambda^{\prime \prime}$ are coupling constants of $\bar{E}_{\bf 2} L_{\bf 3}\Phi_{\bf 3}$, $\bar{E}_{\bf 2} L_{\bf 3}\Phi_{\bf 3^\prime}$, and $\bar{E}_{\bf 1} L_{\bf 3}\Phi_{\bf 3}$, respectively. In this case, $Y_E^\dagger Y_E$ has the form of
  \dis{ Y_E^\dagger Y_E=|\lambda_E|^2
\left(
\begin{array}{ccc}
 a^2+b^2+c^2 & c^2+a^2\omega+b^2\omega^2 & c^2+b^2\omega+a^2\omega^2 \\
  c^2+a^2\omega^2+b^2\omega & a^2+b^2+c^2 & c^2+a^2\omega+ b^2\omega^2 \\
 c^2+b^2\omega^2+a^2\omega & c^2+a^2\omega^2+b^2\omega & a^2+b^2+c^2
  \end{array}\right)
  ,}
  which will be diagonalized to $|\lambda_E|^2((\epsilon^3)^2, (\epsilon)^2, 1)$ by the unitary matrix,
     \dis{ V_L^l=\frac{1}{\sqrt3}
\left(
\begin{array}{ccc}
1&1&1\\
1& \omega^2 & \omega \\
 1&\omega &\omega^2
  \end{array}\right). \label{leptonunit}
  }
 Here we use $\epsilon \simeq m_\mu/m_\tau$ as the order parameter. Then,  $c=\epsilon^3$, $a=\epsilon$ and $b=1$.

      On the other hand, let heavy neutrinos $N_i$ be in the triplet ${\bf 3}$. $\Phi_{\bf 3}$ and $\Phi_{\bf 3^\prime}$ cannot couple to the combination $L_i N_j$ by $Z_4$ and U(1)$_L$ symmetries as well as the SM gauge symmetry. Since the combination $L_1N_1+L_2N_2+L_3N_3$ is a singlet, we naturally have the neutrino Dirac Yukawa coupling $Y_\nu$ proportional to the identity. Finally, $XN_iN_j$ has again  the form of ${\bf 3} + {\bf 3^\prime}+{\bf 1}+{\bf 2}$. $\Phi$'s cannot couple to it while singlet $\chi_{\bf 1}$ and triplet $\chi_{\bf 3}$ in the singlet and triplet representation can do, so we have the following Majorana mass term:
          \dis{ M_N=
\left(
\begin{array}{ccc}
 w_1 & 0&w_2 \\
  0 &w_1 & 0 \\
w_2&0& w_1
  \end{array}\right)
  }
 where $\langle \chi_{\bf 1} \rangle=w_1$ and $\langle \chi_{\bf 3} \rangle=w_2(0,1,0)$, respectively. Therefore, the neutrino mass matrix $M_\nu=-v_u^2 Y_\nu^T M_N^{-1} Y_\nu$ is diagonalized by
         \dis{ V_L^\nu=
\left(
\begin{array}{ccc}
 \frac{1}{\sqrt2} & 0&-\frac{1}{\sqrt2}\\
  0 &1 & 0 \\
\frac{1}{\sqrt2}&0& \frac{1}{\sqrt2}
  \end{array}\right)
  }
  so we obtain the PMNS matrix in the tri-bi maximal mixing,
           \dis{ V_{\rm PMNS}\equiv (V_L^l)^\dagger V_L^\nu=
\left(
\begin{array}{ccc}
 \sqrt{\frac{2}{3}} & \frac{1}{\sqrt3}&0\\
  -\omega\frac{1}{\sqrt6} &\omega\frac{1}{\sqrt3} & e^{-i5\pi/6}\frac{1}{\sqrt2}\\
 -\omega^2\frac{1}{\sqrt6}&\omega^2\frac{1}{\sqrt3}& e^{i5\pi/6}\frac{1}{\sqrt2}
  \end{array}\right).
  }
  In this construction, $S_4$ triplet flavons have VEVs in the direction of $(1,1,1)$ or $(0,1,0)$. These directions are easily stabilized compared to other directions, such as $(1,1,0)$, as argued in Appendix B.
 Note that $Y_\nu$ proportional to the identity does not give rise to LFV. In Eq. (\ref{eq:mlsqaure}), we see $m_L^2$ from the neutrino Dirac Yukawa mediation is flavor universal. In the right-handed neutrino and the charged lepton mass basis,  $Y_\nu$ moves to $ V_L^\nu Y_\nu$ and $m_L^2$ moves to $(V_L^l)^\dagger m_L^2 V_L^l$. As a result, PMNS matrix is mulitplied and will change $m_L^2$ matrix. However, if the neutrino Dirac Yukawa matrix is proportional to the identity matrix, the property of $V \mathbb{I} V^\dagger = \mathbb{I} V V^\dagger = \mathbb{I}$ cancels out  such effects.

There are various ways to put corrections to make non-zero $\theta_{13}$. Moreover, corrected neutrino mass matrix should be consistent with the measurements of $\theta_{12}$, $\theta_{23}$ as well as neutrino mass squared differences, $\Delta m^2_{\rm sol} \equiv m_2^2-m_1^2$ and $|\Delta m^2_{\rm atm}| \equiv |m_3^2 -m_2^2|$. Since the overall neutrino mass scale is not known, the important quantity is the ratio of neutrino mass squared differences, as described in \cite{BenTov:2012tg},
         \dis{\sqrt{|R|} \equiv \sqrt{\frac{\Delta m^2_{\rm atm}}{\Delta m^2_{\rm sol}}}.}
The measured values adopted in \cite{Beringer:1900zz} are given by
 \dis{&\Delta m^2_{\rm sol}=(7.50\pm 0.20)\times 10^{-5} {\rm eV}^2
  \\
  &\Delta m^2_{\rm atm}=(0.00232)^{+0.00012}_{-0.00008}{\rm eV}^2 \nonumber}
  \dis{&\sin^2(2\theta_{12})=0.857 \pm 0.024
  \\
  &\sin^2(2\theta_{23})>0.95
  \\
  &\sin^2(2\theta_{13})=0.098 \pm 0.013}
in the 90\% C. L. The global analysis for such quantities can be found in \cite{Fogli:2012ua, GonzalezGarcia:2012sz}.

        Suppose, for simplicity, we leave the charged lepton sector untouched and correct neutrino sector only. Moreover, we keep the mixings of $\nu_2$ with $\nu_{1, 3}$ forbidden, so that $V_L^\nu$ is modified to
    \dis{ V_L^\nu=
\left(
\begin{array}{ccc}
 \cos(\frac{\pi}{4}+\delta) & 0&-\sin(\frac{\pi}{4}+\delta) \\
  0 &1 & 0 \\
\sin(\frac{\pi}{4}+\delta)&0& \cos(\frac{\pi}{4}+\delta)
  \end{array}\right).
  }
   For small $\delta$, $\cos(\frac{\pi}{4}+\delta) \simeq (1/\sqrt2)(1-\delta)$ and $\sin(\frac{\pi}{4}+\delta) \simeq (1/\sqrt2)(1+\delta)$. From
     \dis{&V_{\rm PMNS}=(V_L^l)^\dagger V_L^\nu
  \\
  &=\frac{1}{\sqrt3}
\left(
\begin{array}{ccc}
1&1&1\\
1& \omega & \omega^2 \\
 1&\omega^2 &\omega
  \end{array}\right)
  \frac{1}{\sqrt2}
\left(
\begin{array}{ccc}
(1-\delta)&0&-(1+\delta)\\
0& 1 & 0 \\
 (1+\delta)&0 &(1-\delta)
  \end{array}\right),}
  we see (13) element of the PMNS matrix is given by
  \dis{|V_{e3}|=\Big|\frac{2\delta}{\sqrt6}\Big|.}
         If such corrections are entirely present in the right-handed neutrino Majorana mass term while $Y_\nu$ is untouched, there would be no observable charged lepton flavor violating process.
For example, let us introduce a doublet flavon $\chi_{\bf 2}$. Then, its VEV modifies the diagonal elements of the Majorana mass matrix. With $\langle \chi_2 \rangle =x^2 (1,1)$, diagonal term has a correction $x^2[2N_1N_1-N_2N_2-N_3N_3]$.
%On the other hand, with $\langle \chi_2 \rangle =x^2 (\omega ,\omega^2)$, diagonal term is given by $x^2[-N_1N_1+2N_2N_2-N_3N_3]$ and with $\langle \chi_2 \rangle =x^2 (\omega^2 ,\omega)$, diagonal term is given by $x^2[-N_1N_1-N_2N_2+2N_3N_3]$.
In principle, by introducing several doublets with different VEVs, each diagonal term can be different.

%%%%%%%%%%%%%%%%%%%%%%%%%%%%%%%%%%%%%%%%%%%%
\subsection{Model I}\label{Model I}
 %%%%%%%%%%%%%%%%%%%%%%%%%%%%%%%%%%%%%%%%%%%%

  Besides putting correction to $M_N$, one can find $S_4$ doublet VEVs giving corrections to $Y_\nu$ to make a sizable $\theta_{13}$ while dangerous charged lepton flavor violation is suppressed. To see this, consider the general $S_4$ doublet VEV, $(a, b)$ where $a$ and $b$ are complex numbers. With this VEV and coupling $\lambda_1$, $Y_\nu$ can be modified as
   \dis{\left(
\begin{array}{ccc}
 1+\lambda_1(a+b)& 0&0 \\
  0 &1+\lambda_1(b\omega+a\omega^2)& 0 \\
0&0& 1+\lambda_1(b\omega^2+a\omega)
  \end{array}\right)
  }
  In this case, $Y_\nu^\dagger Y_\nu$ in the charged lepton mass basis is given by
     \dis{\left(
\begin{array}{ccc}
 1+\lambda_1^2(|a|^2+|b|^2)& \lambda_1(a^*+b)+\lambda_1^2ab^*&\lambda_1(a+b^*)+\lambda_1^2a^*b \\
  \lambda_1(a+b^*)+\lambda_1^2a^*b &1+\lambda_1^2(|a|^2+|b|^2)&  \lambda_1(a^*+b)+\lambda_1^2ab^* \\
 \lambda_1(a^*+b)+\lambda_1^2ab^*&\lambda_1(a+b^*)+\lambda_1^2a^*b& 1+\lambda_1^2(|a|^2+|b|^2)
  \end{array}\right).
  }
If $\lambda_1(a^*+b)+\lambda_1^2ab^*=0$, all the off diagonal elements vanish. For example, $\lambda_1=1$ and $a=b=\omega$ is the case. This condition also implies that off diagonal terms of $Y_\nu^\dagger Y_\nu Y_\nu^\dagger Y_\nu$ vanish so we do not expect any sizeable cLFV.
However, this condition requires a cancellation of two different flavon contributions
and is considered as a serious fine tuning different from vacuum alignment.
We do not pursue this possibility any longer in this paper.

If $a^*=-b$ and both $|a|$ and $|b|$ are smaller than one, the (12) element of $Y_\nu$ is given by $-\lambda_1^2(a^*)^2$.
The (23) element is the same and the (13) element is its complex conjugate, $-\lambda_1^2a^2$.
In this way, LFV is suppressed quadratically even though it does not vanish.
For $Y_\nu^\dagger Y_\nu Y_\nu^\dagger Y_\nu$ term, the (12) element is $2[1+\lambda_1^2(|a|^2+|b|^2)][\lambda_1(a^*+b)+\lambda_1^2ab^*]+[\lambda_1(a+b^*)+\lambda_1^2a^*b]^2$.
For $Y_\nu^\dagger Y_\nu$ term, the (23) element is the same and the (13) element is its complex conjugate.
When $a^*=-b$, it is $-2\lambda_1^2a^2(1+2\lambda_1^2|a|^2)+\lambda_1^4(a^*)^4$, which is quadratically suppressed for small $a$.
For illustration, suppose $\lambda_1a=\lambda_1b=i\rho$. The stabilization of such doublet VEV is discussed in Appendix B. The neutrino Dirac Yukawa has the form of
      \dis{Y_\nu=y_\nu\left(
\begin{array}{ccc}
 1+2i\rho& 0&0 \\
  0 &1-i\rho& 0 \\
0&0& 1-i\rho
  \end{array}\right)
  \label{eq:modYnu}}
  and off-diagonal terms of $Y_\nu^\dagger Y_\nu$ in the charged lepton mass basis is suppressed to ${\cal O}(\rho^2)$,  as expected,
         \dis{(V_L^l)^\dagger (Y_\nu^\dagger Y_\nu) V_L^l=|y_\nu|^2\left(
\begin{array}{ccc}
 1+2\rho^2& \rho^2&\rho^2 \\
  \rho^2 &1+2\rho^2& \rho^2 \\
\rho^2&\rho^2& 1+2\rho^2
  \end{array}\right).
  }
   With this $Y_\nu$, neutrino mass matrix is given by
          \dis{M_\nu=-|y_\nu|^2 \frac{v^2\sin^2\beta}{2w_1}\frac{1}{1-x^2}\left(
\begin{array}{ccc}
 1+4i\rho-4\rho^2& 0&-x(1+i\rho+2\rho^2)\\
  0 &(1-x^2)(1-2i\rho-\rho^2)& 0 \\
-x(1+i\rho+2\rho^2)&0& 1-2i\rho-\rho^2
  \end{array}\right)
  }
 and the deviation of mixing from $\pi/4$ is given by
  \dis{\delta=\Big|\frac{-6i\rho+3\rho^2}{4x(1
+i\rho+2\rho^2)}\Big|\simeq \frac{3\rho}{2x}}
  such that
  \dis{|V_{e3}|\simeq \frac{3\rho}{\sqrt6 x}.}

To the first order in $\rho$, mass eigenvalues are given by
  \dis{-|y_\nu|^2 \frac{v^2\sin^2 \beta}{2w_1}\Big(\frac{1+i\rho}{1+x}, 1-2i\rho, \frac{1+i\rho}{1-x}\Big).}
  Taking absolute values of these eigenvalues, we obtain neutrino masses $-[|y_\nu|^2 v^2\sin ^2 \beta/(2w_1)](1/(1+x), 1, 1/(1-x))+{\cal O}(\rho^2)$.

\begin{figure}[!t]
  \begin{center}
      \includegraphics[width=0.65\textwidth]{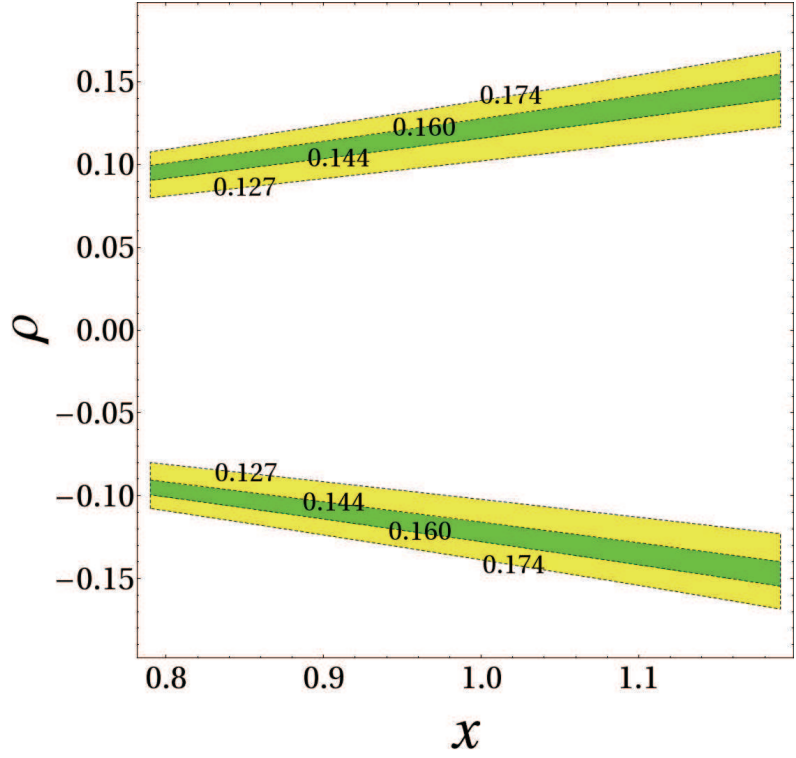}
  \end{center}
 \caption{ $\theta_{13}$ with respect to $\rho$ and $x$ parameters. All points in the colored region satisfy neutrino oscillation experiments. Neutrino $\theta_{13}$, indicated on contour label in radian, is measured as $0.144 < \theta_{13}  < 0.160$ in $1 \sigma$ level, $0.127 < \theta_{13}  < 0.174$ in $3 \sigma$ level.
  }
\label{fig:theta13}
\end{figure}

In summary, we expect that even though the charged lepton flavor violating effects are generated in the $A_E$ term at one loop and in the $m_L^2$ term at two loop, they can be suppressed by extra small expansion parameter $\rho$ proportional to $\theta_{13}$.
With the vacuum alignment of the doublet flavon $i(v,v)$, it is possible to cancel the first order correction of $\rho$ and the off-diagonal elements of the slepton mass squared would have $\rho^2$ suppression as a result.   Fig. \ref{fig:theta13} shows how measured $\theta_{13}$ can be explained for the choices of $\rho$ and $x$ parameters satisfying observed neutrino mass squared ratio, $\sqrt{R}$. The observed $\theta_{13} \sim 0.15$ can be accommodated for $\rho \sim 0.1$.

   %%%%%%%%%%%%%%%%%%%%%%%%%%%%%%%%%%%%%%%%%%%%%%%%%%%%%%%%%%%%%%%%%%%%%%%%%%%%%%%%%%%%%%%%
%%%%%%%%%%%%%%%%%%%%%%%%%%%%%%%%%%%%%%%%%%%%%%%%%%%%%%%%%%%%%%%%%%%%%%

%%%%%%%%%%%%%%%%%%%%%%%%%%%%%%%%%%%%%%%%%%%
\subsection{Model II}\label{Model II}
%%%%%%%%%%%%%%%%%%%%%%%%%%%%%%%%%%%%%%%%%%%%

Of course, $\theta_{13}$ can come from both Majorana mass correction and neutrino Dirac Yukawa correction. Only the neutrino Dirac Yukawa coupling can affect the cLFV. To see the two-parameter case, consider
     \dis{Y_\nu=y_\nu \left(
\begin{array}{ccc}
 1+2i\rho & 0&0\\
  0 &1 -i\rho& 0 \\
0&0& 1-i\rho
  \end{array}\right)}
    and
          \dis{ M_N=
\left(
\begin{array}{ccc}
 w_1 & 0&w_2 \\
  0 &w_1 & 0 \\
w_2&0& w_1(1-\zeta)
  \end{array}\right)
  .}

The neutrino mass is given by
          \dis{ M_\nu=-|y_\nu|^2 \frac{v^2\sin^2\beta}{2w_1}\frac{1}{1-x^2-\zeta}
\left(
\begin{array}{ccc}
 1+4i\rho-\zeta& 0&-x(1+i\rho) \\
  0 &(1-x^2)(1-2i\rho)-\zeta & 0 \\
-x(1+i\rho)&0& 1-2i\rho
  \end{array}\right) +{\cal O}(\rho^2, \zeta^2)
  }
  where $x=w_2/w_1$ again. Then, three neutrino mass eigenvalues are given by
  \dis{-|y_\nu|^2 \frac{v^2\sin^2\beta}{2w_1}\Big( \frac{1+i\rho}{1+x}+\frac{\zeta}{2(1+x)^2}, 1-2i\rho, \frac{1+i\rho}{1-x}+\frac{\zeta}{2(1-x)^2}\Big)}
  and
  \dis{\delta = \frac{\sqrt{36 \rho^2+\zeta^2}}{4x}.}
  Hence, we see the (13) element of the PMNS matrix is given by
  \dis{|V_{e3}|=\Big|\frac{2\delta}{\sqrt6}\Big|=\Big|\frac{\sqrt{36 \rho^2+\zeta^2}}{2\sqrt6 x}\Big|.}
  Moreover, $m_L^2$ from Yukawa mediation is controlled by the parameter $\rho$ only and $(V_L^l)^\dagger (Y_\nu^\dagger Y_\nu) V_L^l$ is the same as the previous case,
         \dis{(V_L^l)^\dagger (Y_\nu^\dagger Y_\nu) V_L^l=|y_\nu|^2\left(
\begin{array}{ccc}
 1+2\rho^2& \rho^2&\rho^2 \\
  \rho^2 &1+2\rho^2& \rho^2 \\
\rho^2&\rho^2& 1+2\rho^2
  \end{array}\right).
  }

In the limit of $\zeta \to 0$, both $\theta_{13}$ and cLFV come from the neutrino Dirac Yukawa which corresponds to the Model I.  In the opposite limit, $\rho \to 0$, $\theta_{13}$ is entirely obtained from Majorana mass term and cLFV does not appear.
% More discussion later

In addition, we can also constrain absolute mass scale of light neutrinos. The most stringent constraint on neutrino absolute mass is given by CMB data of WMAP experiment,
combined with supernovae data and data on galaxy clustering, $\Sigma_{j} m_{j} \lesssim 0.68$eV, 95\% C.L. Conservatively, we set the bound $2.6 \times 10^{14 } \gev \lesssim M_N$. Throughout paper, we use $M_N = 5 \times 10^{14}$GeV, the heaviest right-handed neutrino mass.

%%%%%%%%%%%%%%%%%%%%%%%%%%%%%%%%%%%%%%%%%%%%%%%%%%%%%%%%%%%%%%%%%%%%%%%%%%%%%%%%
%%%%%%%%%%%%%%%%%%%%%%%%%%%%%%%%%%%%%%%%%%%%%%%%%%%%%%%%%%%%%%%%%%%%%%%%%%%%%%%%
\section{Charged Lepton Flavor Violation}\label{sec:LFV}
%%%%%%%%%%%%%%%%%%%%%%%%%%%%%%%%%%%%%%%%%%%%%%%%%%%%%%%%%%%%%%%%%%%%%%%%%%%%%%%%
%%%%%%%%%%%%%%%%%%%%%%%%%%%%%%%%%%%%%%%%%%%%%%%%%%%%%%%%%%%%%%%%%%%%%%%%%%%%%%%%

Since flavor structures of supersymmetric particles can be different from those of SM partners, flavor number is easily violated in SUSY. In general, the structure of the slepton mass matrix raises dangerous cLFV. Such cLFV in SUSY is studied in  \cite{Hisano:1995cp, Arganda:2005ji}.
In our model, once the identity structure of the neutrino Yukawa coupling $Y_\nu$ is broken, cLFV is produced. As a possible modification, one may put off-diagonal terms into $Y_\nu$. On the other hand, when the degeneracy of $Y_\nu$ is broken, the combination of $Y_\nu$s in the charged lepton mass basis,  $(V_L^l)^\dagger(Y_\nu^\dagger Y_\nu)V_L^l$ has off-diagonal terms as shown in Sec. \ref{sec:FlavorModel}. The slepton mass squared gets extra contribution from neutrino Dirac Yukawa interactions,
\dis{\delta m_{L}^2 = \frac{B_N^2}{(4 \pi)^4} \Big[ \Big({\rm Tr} [Y_{\nu} Y_{\nu}^\dagger]+3{\rm Tr} [Y_U Y_U^\dagger] - 3 g_2^2 - \frac{1}{5} g_1^2 \Big) Y_{\nu}^\dagger Y_{\nu} + 3 Y_{\nu}^\dagger Y_{\nu} Y_{\nu}^\dagger Y_{\nu} \Big].}
In the charged lepton mass basis, $(V_L^l)^\dagger m_L^2 V_L^l$ has off-diagonal elements and cLFV appears. Even though this is a general feature, it is also possible to find some parameter space in which charged lepton number is conserved.
For example, in Sec. \ref{Model I}, off diagonal terms of the slepton soft  mass squared, $(m_L^2)_{12}$ can vanish for specific value of $y_\nu$.
Corresponding condition would be
 \dis{(\delta m_L^2)_{12}&\propto \Big[{\rm Tr}[Y_\nu^\dagger Y_\nu]+3{\rm Tr} [Y_U Y_U^\dagger] -3g_2^2-\frac15 g_1^2\Big](Y_\nu^\dagger Y_\nu)_{12}+3(Y_\nu^\dagger Y_\nu Y_\nu^\dagger Y_\nu)_{12}=0,}
 which is equivalent to
  \dis{(\delta m_L^2)_{12}&\propto \Big[3(1+2\rho^2)y_\nu^2+3y_t^2-3g_2^2-\frac15 g_1^2\Big]y_\nu^2 \rho^2+3y_\nu^4 2\rho^2
 \\
& \simeq y_\nu^2 \Big[3y_t^2-3g_2^2-\frac15 g_1^2\Big]\rho^2+9y_\nu^4 \rho^2+{\cal O}(\rho^4)=0.}
  Near the GUT scale, $g_1^2 \simeq g_2^2 \simeq 4\pi/28$, and $y_t \simeq 0.5$ so off diagonal term vanishes for $y_\nu \simeq 0.28$.  For this value of $Y_\nu$, there would be no unwanted cLFV.
This is different from the condition that diagonal contribution involving $Y_\nu$ vanishes,
 \dis{(\delta m_L^2)_{ii}&\propto\Big[3(1+2\rho^2)y_\nu^2+3y_t^2-3g_2^2-\frac15 g_1^2\Big]y_\nu^2(1+2\rho^2)+3y_\nu^4(1+8\rho^2)
 \\
& \simeq y_\nu^2 \Big[3y_t^2-3g_2^2-\frac15 g_1^2\Big]+6y_\nu^4 +{\cal O}(\rho^2)=0}
which is satisfied for $y_\nu \simeq 0.34$.

Of course, it does not mean that $y_\nu$ should take the lepton number conserving value. We have many constraints on $y_\nu$ from various observations. In this paper, we try to explain the $125$GeV Higgs mass with large A term generated from $y_\nu$. On the other hand, one may try to explain deviation of muon $g-2$ from the SM prediction. Moreover, degeneracy breaking parameter $\rho$ is used to explain sizable $\theta_{13}$. However, it is also difficult to find an appropriate value of $y_\nu$ which satisfies all of them. In this section, we present the cLFVs for parameters explaining the $125$GeV Higgs mass with large A term and $\theta_{13}$. Thereafter, we visit the muon $g-2$ constraints and the relation among $\theta_{13}$, cLFV, and the Higgs mass.

%%%%%%%%%%%%%%%%%%%%%%%%%%%%%%%%%%%%%%%%%%%%%%%%%%%%%%%%%%%%%%%%%%%%%%%%%%%%%%%%
%%%%%%%%%%%%%%%%%%%%%%%%%%%%%%%%%%%%%%%%%%%%%%%%%%%%%%%%%%%%%%%%%%%%%%%%%%%%%%%%
\subsection{Experimental status}\label{subsec:status}
%%%%%%%%%%%%%%%%%%%%%%%%%%%%%%%%%%%%%%%%%%%%%
 \begin{table}[t]
\begin{center}
\begin{tabular}{|c|c|c|}
\hline  Observables & Experimental bound & Future sensitivity
\\[0.2em]
\hline  &&\\ [-1.1em]
${\rm Br}(\mu \to e \gamma)$ & $2.4 \times 10^{-12} [63]$  & ${\cal O}(10^{-13})$ [63]
\\[0.4em]
${\rm Br}(\tau \to \mu \gamma)$&$4.4 \times 10^{-8}$[64]& $2.4\times10^{-9}$[69]
\\[0.4em]
${\rm Br}(\tau \to e \gamma)$& $3.3 \times 10^{-8}$[64]& $3.0\times10^{-9}$ [69]
\\[0.4em]
${\rm Br}(\mu \to 3e)$&$1.0 \times 10^{-12}$ [65]& ${\cal O}(10^{-16})$ [70]
\\[0.4em]
${\rm Br}(\tau \to 3e)$&$2.7 \times 10^{-8}$[66]& $2.3\times10^{-10}$ [69]
\\[0.4cm]
${\rm Br}(\tau \to 3 \mu)$ &$2.1 \times 10^{-8}$[66]& $8.2\times10^{-10}$[69]
\\[0.4em]
$\frac{\Gamma(\mu{\rm Ti}\to e{\rm Ti})}{\Gamma(\mu{\rm Ti}\to {\rm capture})}$&$4.3 \times 10^{-12}$[67]& ${\cal O}(10^{-18})$[71]
\\[0.4em]
$\frac{\Gamma(\mu{\rm Au}\to e{\rm Au})}{\Gamma(\mu{\rm Au}\to {\rm capture})}$&$7.0 \times 10^{-13}$[68]&
\\[0.4em]
\hline
\end{tabular}
\end{center}
\caption{Various LFV experimental bounds and future sensitivities. The table is adopted from \cite{Abada:2012cq}.} \label{table:LFVexp}
\end{table}

The current experimental bounds and future sensitivities for various cLFV processes in the 90\% C. L. are summarised in Table \ref{table:LFVexp}\cite{Beringer:1900zz, Hewett:2012ns}.

%%%%%%%%%%%%%%%%%%%%%%%%%%%%%%%%%%%%%%%%%%%%%%%%%%%%%%%%%%%%%%%%%%%%%%%%%%%%%%%%%%%%%%%%
 \begin{figure}[!t]
  \begin{center}
   \includegraphics[width=0.45\textwidth]{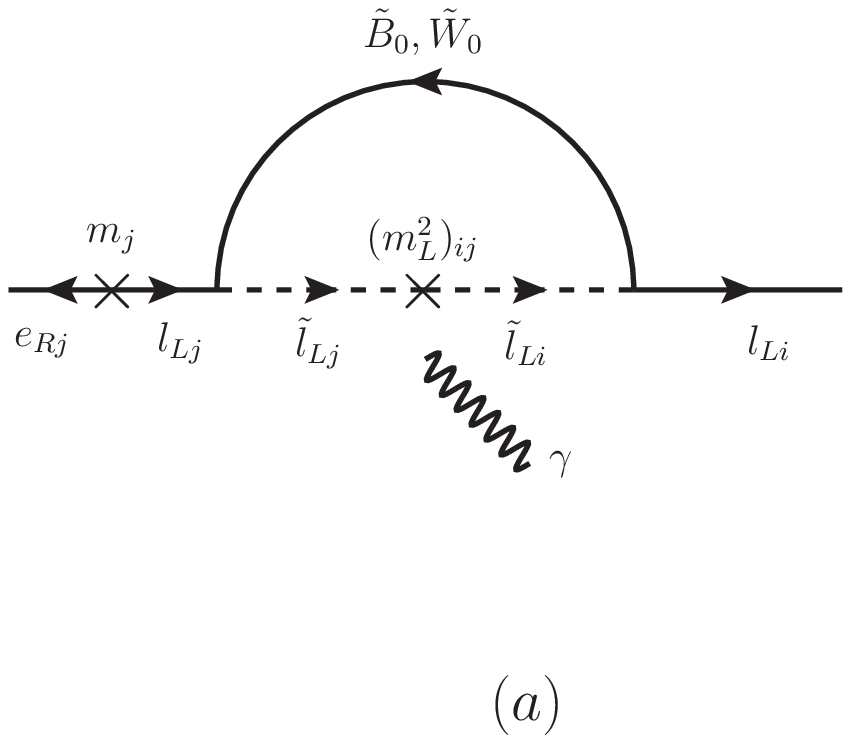}
  {\includegraphics[width=0.45\textwidth]{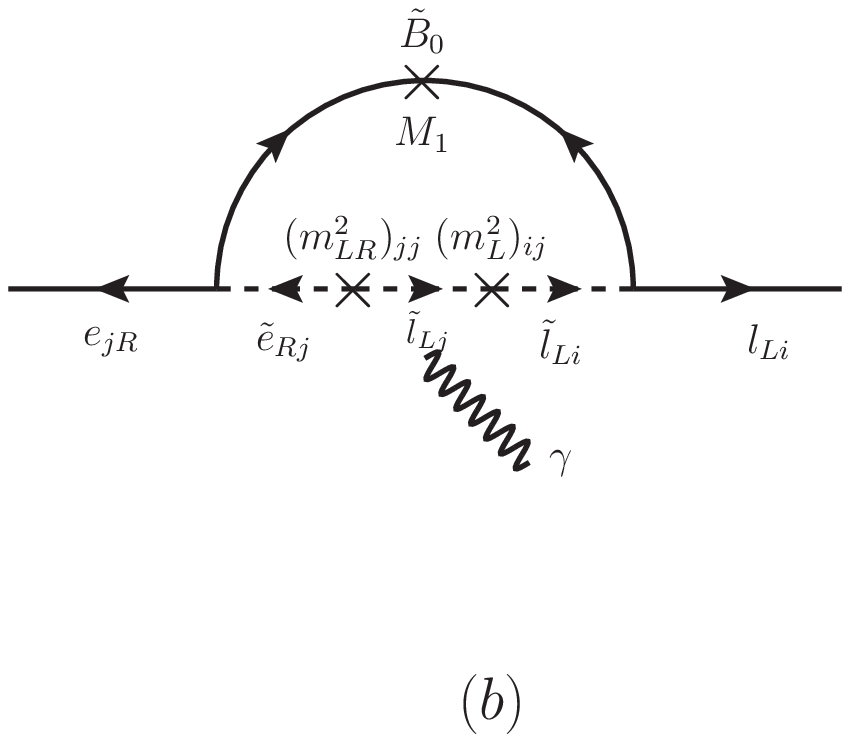}}
     \vskip 1.0cm
     {\includegraphics[width=0.45\textwidth]{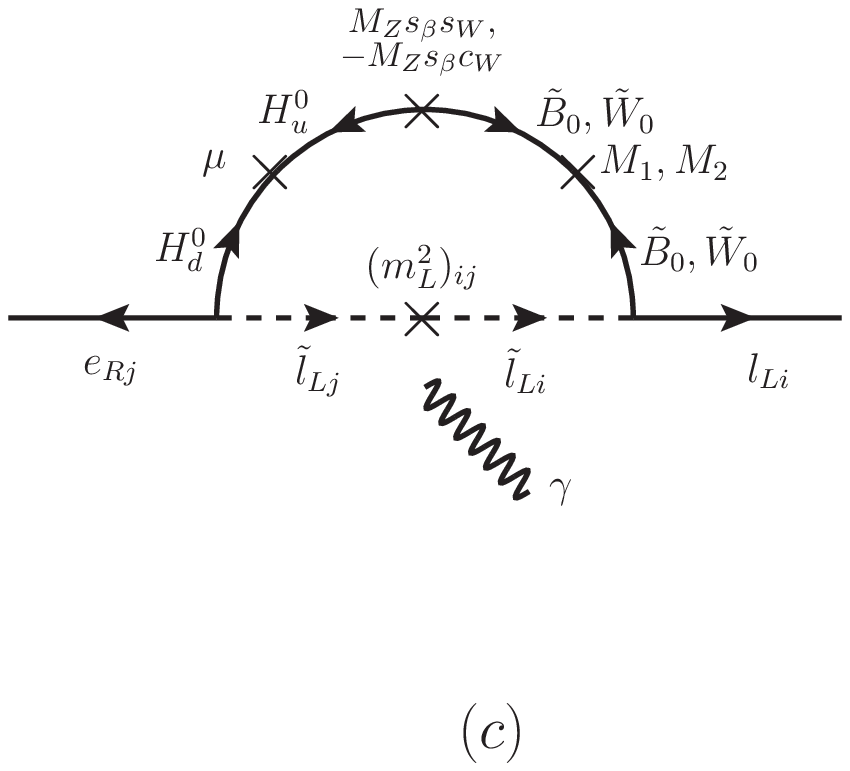}}
  \end{center}
 \caption{Feynman diagrams for $l_j \to l_i \gamma$ process with neutralino-charged slepton internal lines in the mass insertion scheme.
  }
\label{fig:LFVn}
\end{figure}
%%%%%%%%%%%%%%%%%%%%%%%%%%%%%%%%%%%%%%%%%%%%%%%%%%%%%%%%%%%%%%%%%%%%%%

%%%%%%%%%%%%%%%%%%%%%%%%%%%%%%%%%%%%%%%%%%%%%%%%%%%%%%%%%%%%%%%%%%%%%%%%%%%%%%%%%%%%%%%%
 \begin{figure}[!t]
  \begin{center}
   \includegraphics[width=0.45\textwidth]{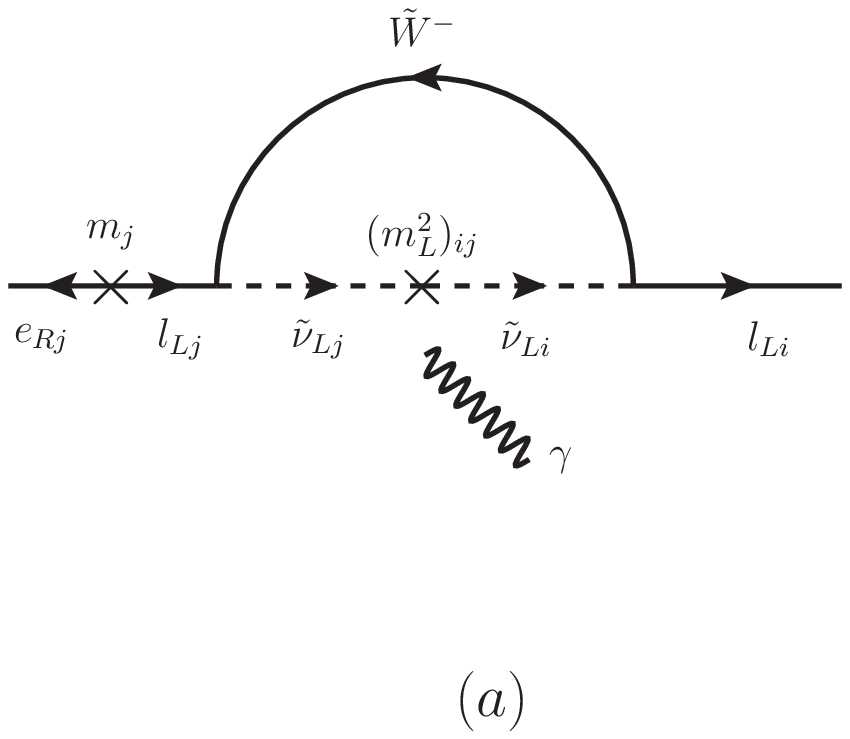}
  {\includegraphics[width=0.45\textwidth]{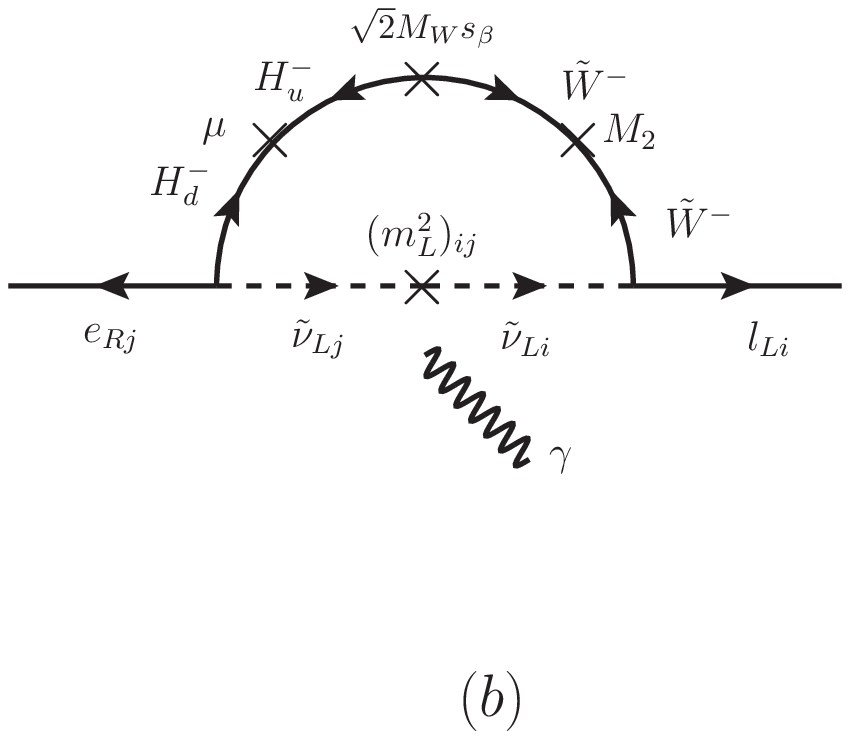}}
  \end{center}
 \caption{Feynman diagrams for $l_j \to l_i \gamma$ process with chargino-sneutrino internal lines in the mass insertion scheme.
  }
\label{fig:LFVc}
\end{figure}
%%%%%%%%%%%%%%%%%%%%%%%%%%%%%%%%%%%%%%%%%%%%%%%%%%%%%%%%%%%%%%%%%%%%%%

  %%%%%%%%%%%%%%%%%%%%%%%%%%%%%%%%%%%%%%%%%%%%%%%%%%%%%%%%%%%%%%%%%%%%%%%%%%%%%%%%%%%%%%%%
 \begin{figure}[!t]
  \begin{center}
         \includegraphics[width=0.70\textwidth]{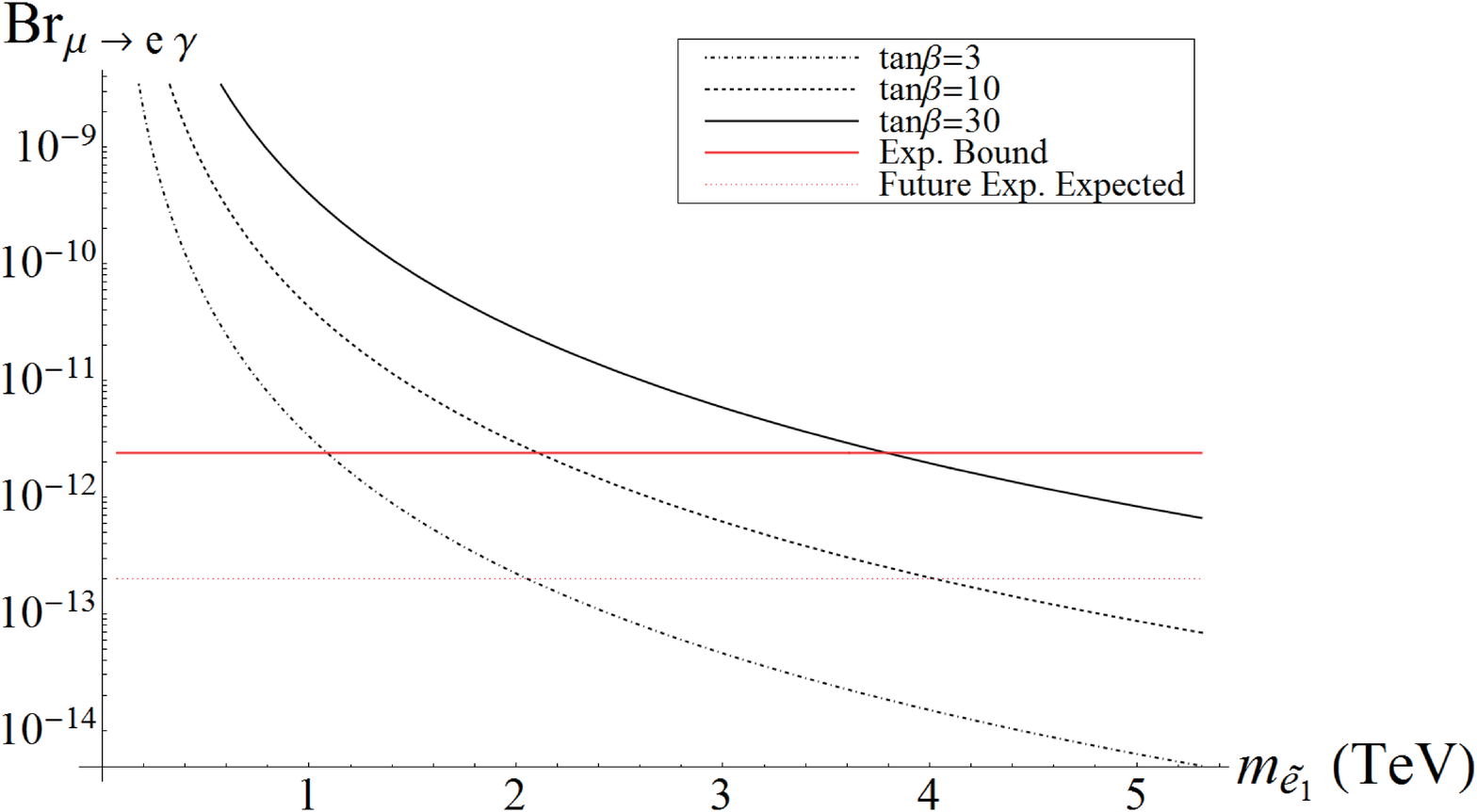}
             \vskip 0.5cm
      \includegraphics[width=0.50\textwidth]{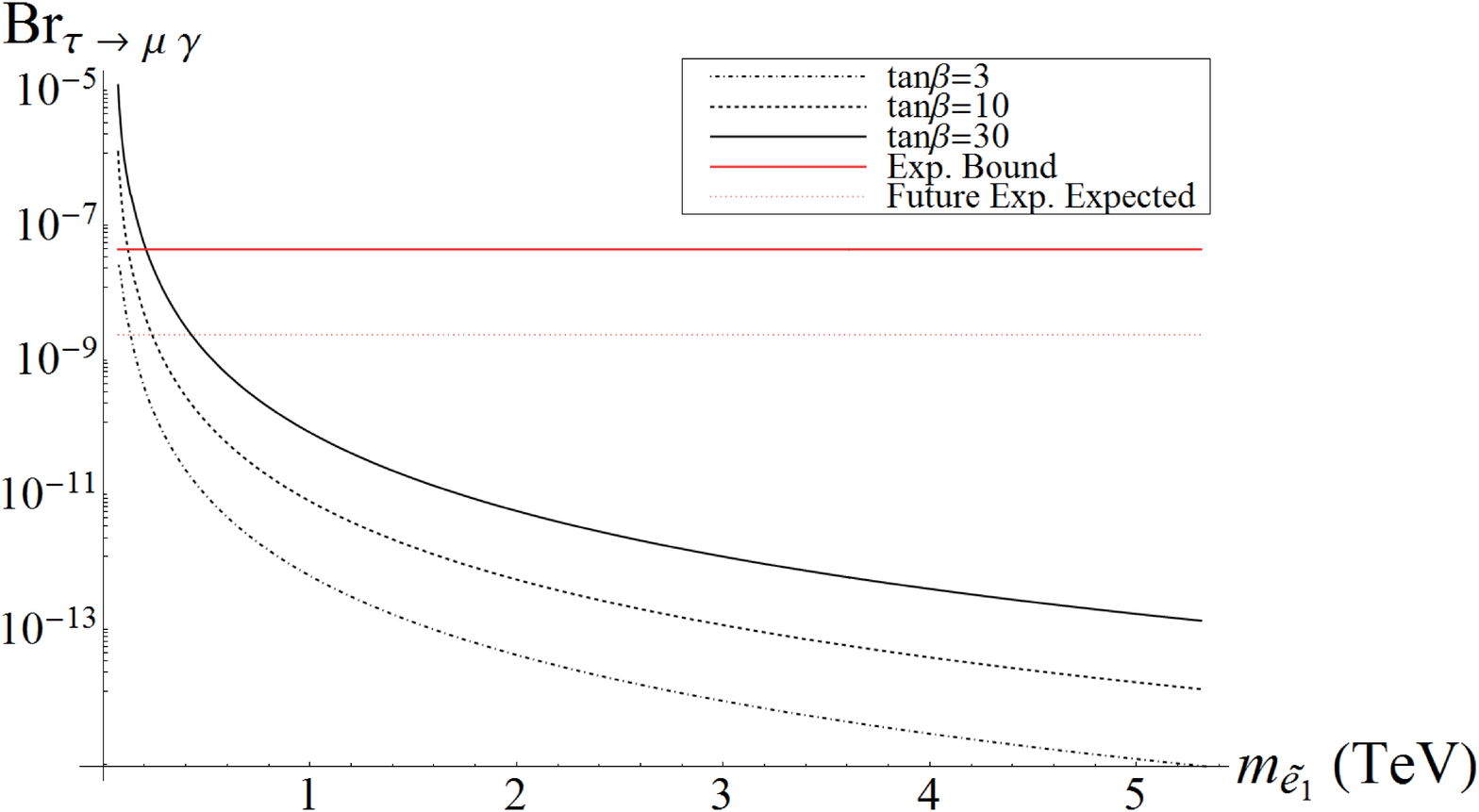}
            \vskip 0.5cm
   \includegraphics[width=0.50\textwidth]{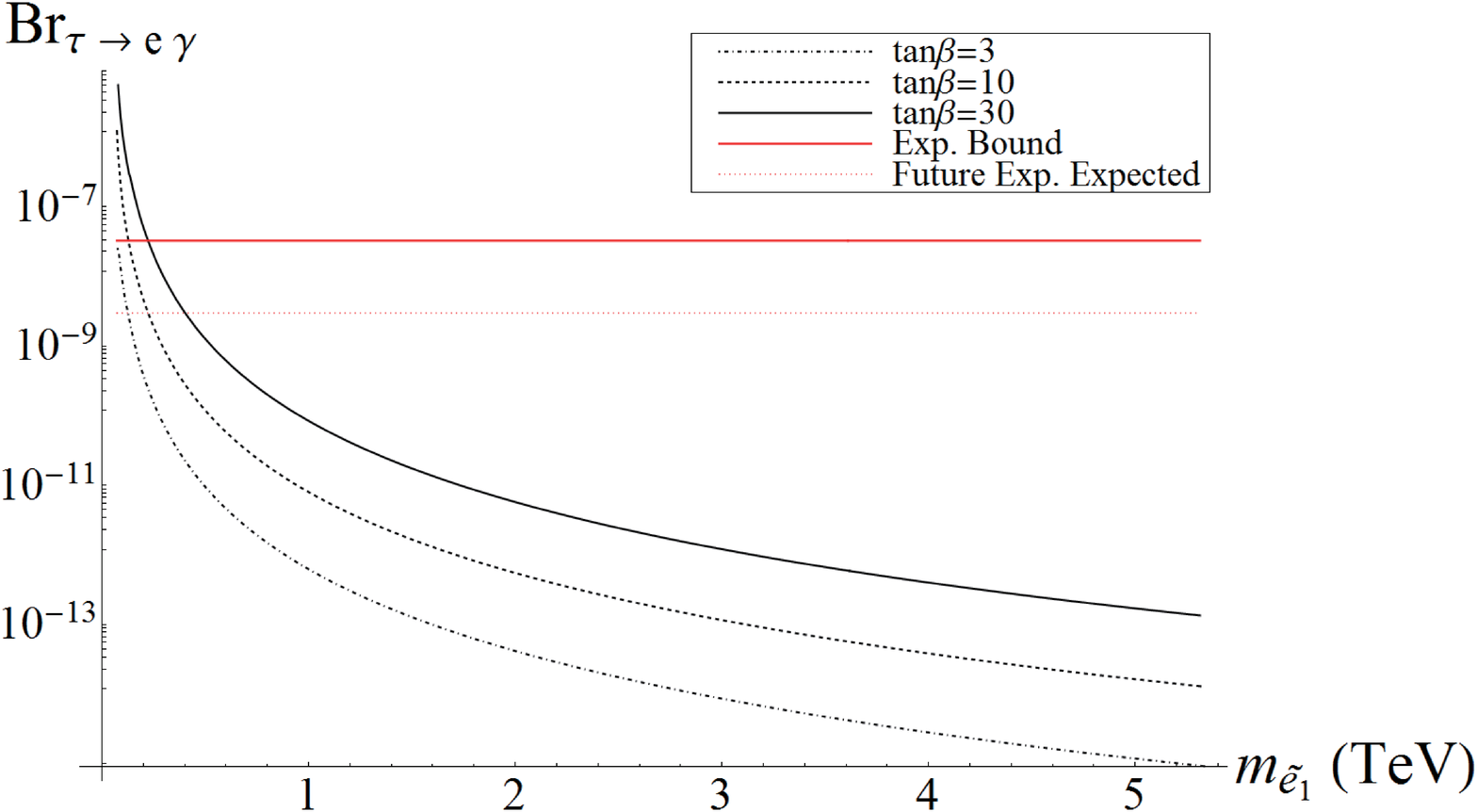}
  \end{center}
 \caption{ Branching ratios of $\mu \to e\gamma$, $\tau \to \mu\gamma$ and $\tau \to e\gamma $ with respect to the lightest selectron mass for $\tan\beta=3, 10, 30$, $y_\nu=0.65$ and $\rho=0.1$.
  }
\label{fig:ljligamma}
\end{figure}
%%%%%%%%%%%%%%%%%%%%%%%%%%%%%%%%%%%%%%%%%%%%%%%%%%%%%%%%%%%%%%%%%%%%%%
\subsection{$l_j \to l_i \gamma$}

The amplitude for $l_j\to l_i\gamma$ is written as
\dis{T=e\epsilon^{\mu *}\overline{u_i}(p-q)\Big[q^2\gamma_\mu(A^L_1P_L+A_1^RP_R)
+m_{l_j}i\sigma_{\mu\nu}q^\nu(A_2^LP_L+A_2^RP_R)\Big]u_j(p).}
On the mass shell ($q^2\to0$), gauge invariance imposes that the chirality preserving part does not contribute to the $l_j\to l_i\gamma$ process. Hence, chirality flipping should take place in the on-shell $l_j\to l_i\gamma$ process. The decay rate is given by
\dis{\Gamma(l_j\to l_i\gamma) =\frac{e^2}{16\pi}m_{l_j}^5(|A^L_2|^2+|A^R_2|^2)}
and the branching ratio yields approximately
\dis{{\rm Br}(l_j \to l_i \gamma)\sim \frac{\alpha^3}{G_F^2}\frac{1}{m_{\rm SUSY}^4} \Big(\frac{(m_L^2)_{ij}}{m_{\rm SUSY}^2}\Big)^2.}

 In the mass insertion scheme, the chirality flipping can be easily analyzed. Consider first the case of the neutralino-charged slepton internal loop, as shown in Fig. \ref{fig:LFVn}. Fig. \ref{fig:LFVn} (a) shows the chirality flipping from a fermion mass insertion in the external lepton line. In Fig. \ref{fig:LFVn} (b), chirality flipping takes place in the slepton internal line through the LR mixing insertion, $m_j(A-\mu\tan\beta)$. This term consists of flavor universal part $-m_j\mu\tan\beta$, which can be enhanced in the limit of large $\tan\beta$ and large $\mu$. The chirality flipping in Fig. \ref{fig:LFVn} (c) is given by the Yukawa coupling of the lepton-slepton-Higgsino vertex. This vertex contains $1/\cos\beta$ factor which combines with a $\sin\beta$ in the Higgsino-gaugino mixing to give a $\tan\beta$ dependence to the diagram. Therefore, this diagram is enhanced in the large $\tan\beta$ limit.  Note that it is inversely proportional to the $\mu$, the Higgsino mass. Since this diagram contains SUSY mass scale only, unlike other diagrams proportional to the Higgs VEV $v$ through $m_j$, it is dominant over all other diagrams with the neutralino-charged slepton internal loop in many cases. However, since the Higgsino-bino mass insertion $M_Z\sin\beta\sin\theta_W$ and Higgsino-wino mass insertion $-M_Z\sin\beta\cos\theta_W$ have the opposite signs, slight destructive interference occurs.

Next, the case of the chargino-sneutrino internal loop is shown in Fig. \ref{fig:LFVc}. Diagrams are similar to those of the neutralino-charged slepton internal loop, except the absence of the slepton LR mixing, since the right handed neutrinos are already integrated out. Chirality flipping can occur either in the external lepton line (Fig. \ref{fig:LFVc} (a)) or in the lepton-sneutrino-Higgsino vertex(Fig. \ref{fig:LFVc} (b)). The latter diagram dominates over the former one, and since it does not have a destructive interference, it becomes the leading contribution over all other diagrams in many cases. The similar argument also applies to the discussion of muon $g-2$, whose SUSY contribution comes from the same diagram with flavor conservation. Following this diagram, SUSY enhances the muon $g-2$ for positive $\mu$\cite{Lopez:1993vi, Chattopadhyay:1995ae}.

%WRITE DOWN THIS PART LATER
%
%
%
%
%
In Fig. \ref{fig:ljligamma}, we show branching ratios of various $l_j\to l_i \gamma$ processes for Sec. \ref{Model I}.  In  the graph, neutrino Dirac Yukawa couplings are fixed to be $y_\nu=0.65$ and $\rho=0.1$, while $\tan\beta$ and SUSY breaking scale are varied. Since off-diagonal terms of  $m_L^2$ in the charged lepton mass basis are the same, normalized branching ratio, $\Gamma(l_j \to l_i\gamma)/m_j^5$ are almost identical. Therefore, branching ratios are closely related to the total decay rate of mother particle. For example, since total decay rate of tau is about $5.3$ times larger than that of muon, branching ratio of Br$(\mu \to e \gamma)$ is about $5.3$ times larger than Br$(\tau \to e \gamma)$ and Br$(\tau \to \mu \gamma)$ which are almost the same.

\subsection{$l_j^- \to l_i^-l_i^-l_i^+$}

 In many cases, dominant contribution comes from the photon penguin. $Z$ boson penguin is suppressed in general because of the accidental cancellation when the neutralino or chargino is pure gaugino or pure Higgsino\cite{Hirsch:2012ax}. Such accidental cancellation is broken by introducing TeV scale physics which couples to the sneutrino with the sizable coupling. This can be realised in the R-parity violating model or in the TeV inverse seesaw, for example\cite{Hirsch:2012ax,  Abada:2012cq}.

  In our case, photon penguin is  a leading contribution, so we have a simple relation between Br$(l_j \to l_i \gamma)$,
  \dis{\frac{{\rm Br}(l_j\to3l_i)}{{\rm Br}(l_j \to l_i \gamma)}=\frac{\alpha}{3\pi}\Big(\ln\frac{m_{l_j}^2}{m_{l_i}^2}-\frac{11}{4}\Big).\label{eq:l3lllg}}
The box diagram is suppressed in general, except for some special cases, such as in SUSY with Dirac gauginos\cite{Fok:2010vk}.

\begin{figure}[!b]
  \begin{center}
      \includegraphics[width=0.70\textwidth]{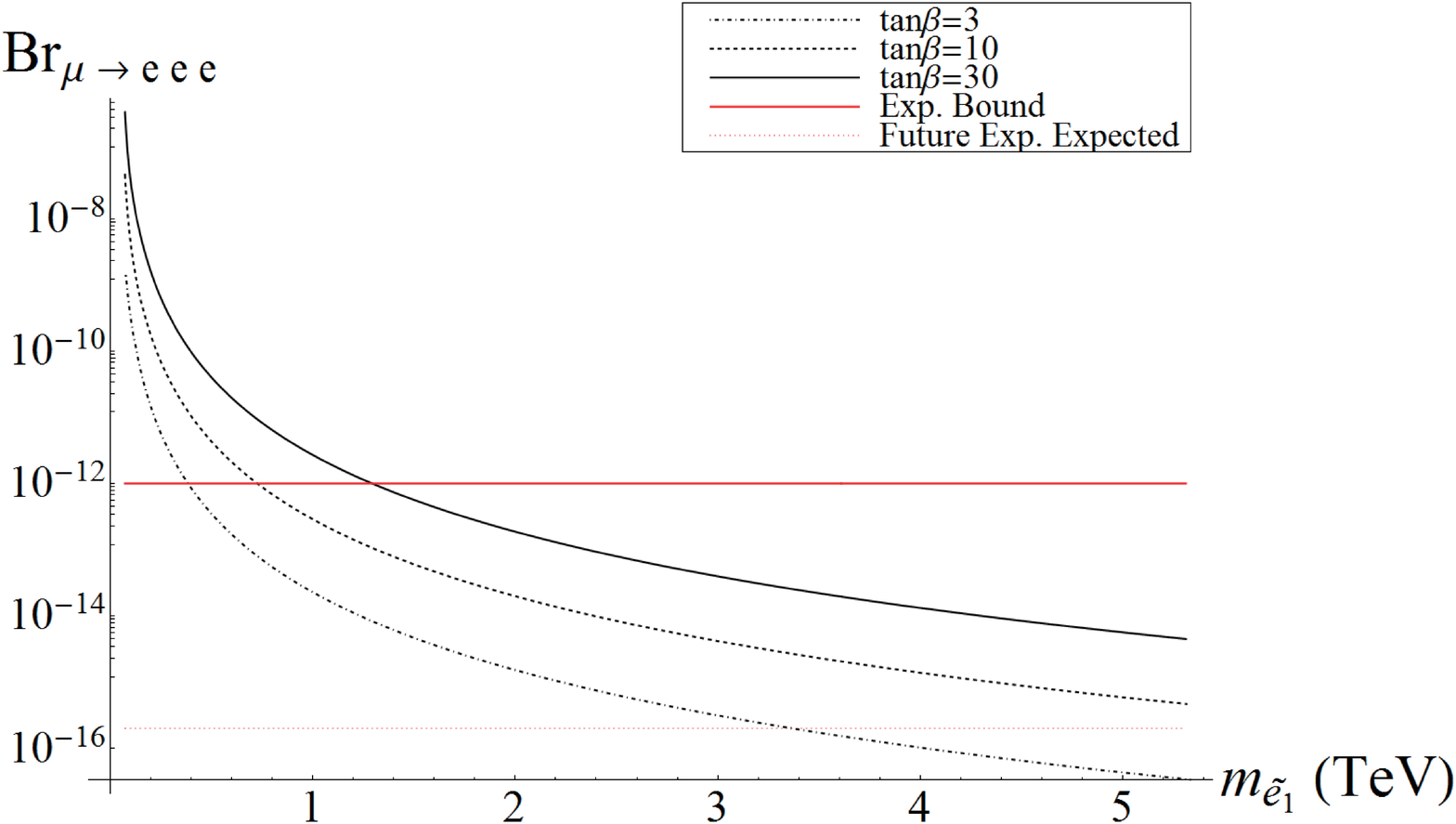}
            \vskip 0.5cm
   \includegraphics[width=0.50\textwidth]{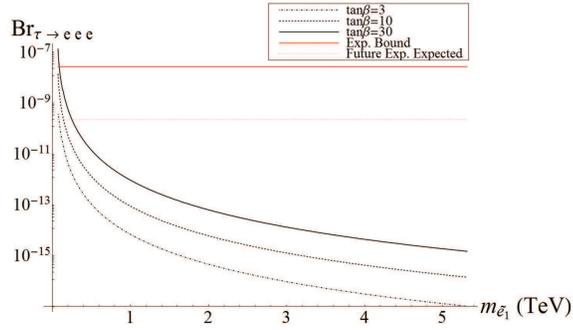}
               \vskip 0.5cm
   \includegraphics[width=0.50\textwidth]{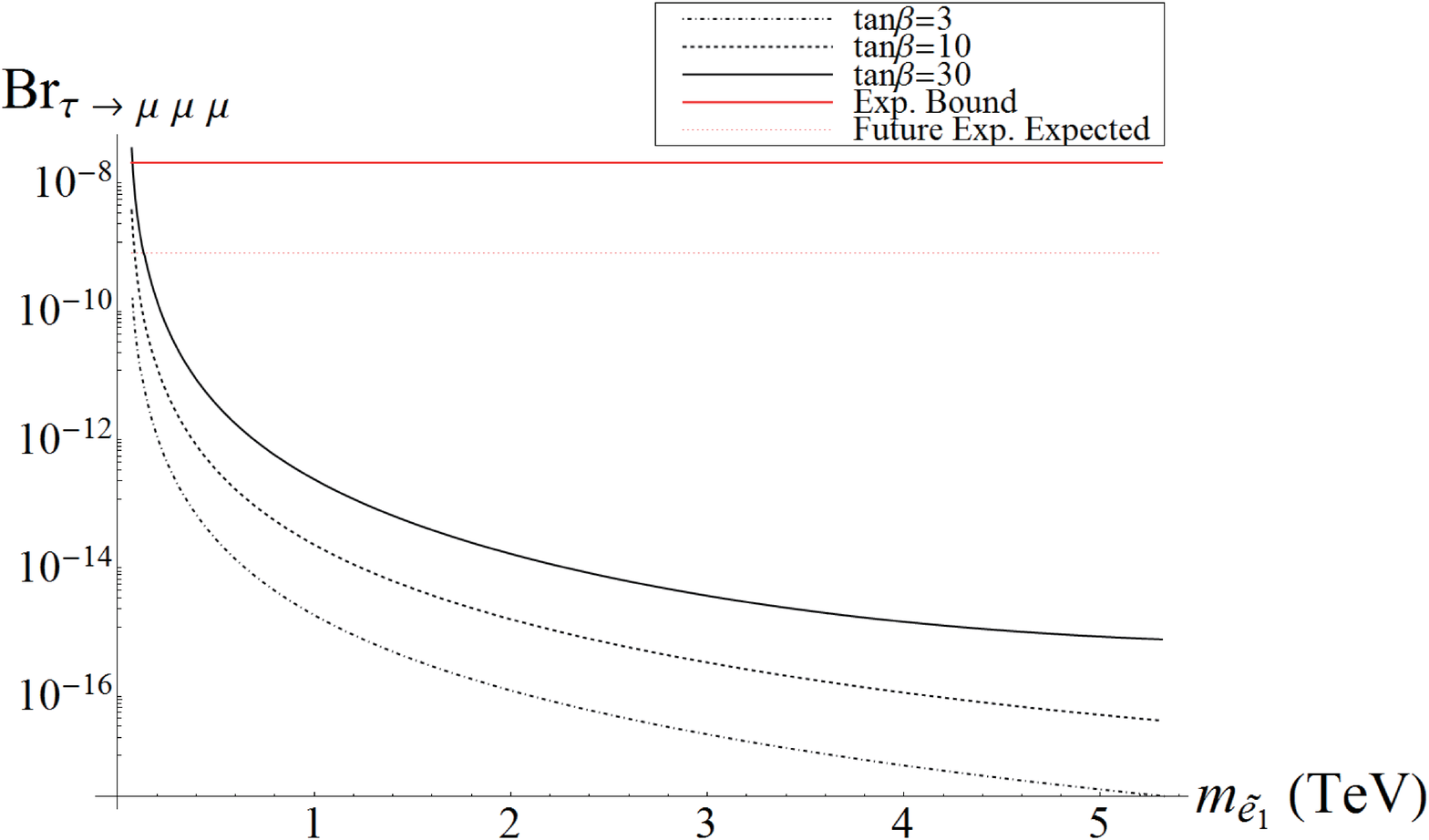}
  \end{center}
 \caption{ Branching Ratios of $\mu \to 3e$, $\tau \to 3e$ and $\tau \to 3\mu$ with respect to the lightest selectron mass for $\tan\beta=3, 10, 30$, $y_\nu=0.65$ and $\rho=0.1$.
  }
\label{fig:ltolll}
\end{figure}

In Fig. \ref{fig:ltolll}, we show branching ratios of various $l_j\to 3l_i $ processes for for Sec. \ref{Model I}. Fixed parameters are the same as $l_j\to l_i \gamma$ process. We see that ${\rm Br}(\mu \to 3e)$ is about 0.018 times suppressed than ${\rm Br}(\mu \to e \gamma)$ so Eq. (\ref{eq:l3lllg}) is satisfied. Photon penguin is a leading contribution for $\mu \to 3e$ process.
In the absence of special characteristic which can overcome the natural size of the branching ratio,
Br($l_j^- \to l_i^-l_i^-l_i^+$) is $\alpha/\pi$ suppressed compared to Br($l_j^- \to l_i^- \gamma$).

   %%%%%%%%%%%%%%%%%%%%%%%%%%%%%%%%%%%%%%%%%%%%%%%%%%%%%%%%%%%%%%%%%%%%%%%%%%%%%%%%%%%%%%%%
%%%%%%%%%%%%%%%%%%%%%%%%%%%%%%%%%%%%%%%%%%%%%%%%%%%%%%%%%%%%%%%%%%%%%%

\subsection{$\mu-e$ conversion}

\begin{figure}[!t]
  \begin{center}
      \includegraphics[width=0.70\textwidth]{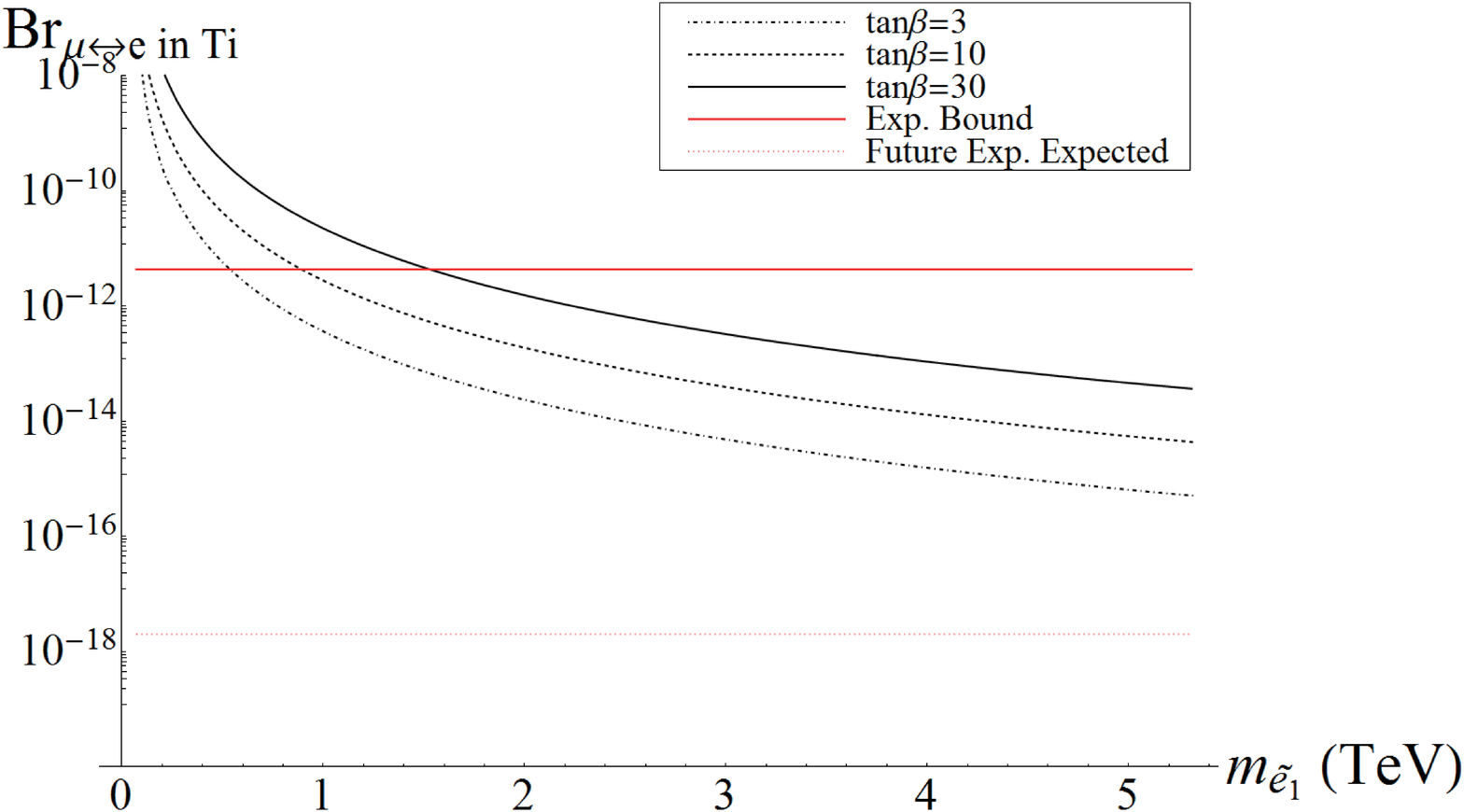}
  \end{center}
 \caption{ $\mu-e$ conversion rate with respect to the lightest selectron mass for $\tan\beta=3, 10, 30$, $y_\nu=0.65$ and $\rho=0.1$.
  }
\label{fig:ueconv}
\end{figure}

% REWRITE THIS PART
%
%
%
%
%
Conversion of the stopped muons in a nuclei to the electron is a promising channel to look for the charged lepton flavor violation. In principle there are many different operators including scalar, photon mediated vector, $Z$-boson mediated vector operators in addition to the dipole operator.
Muon to electron conversion rate is conventionally normalised by muon capture rate.
\dis{
B_{\mu \to e} (Z)  =  \frac{\Gamma_{\rm conv} (Z,A)}{\Gamma_{\rm capt}(Z,A)}.
}
Here $Z$ is the atomic number of the atom. Different target provide a different $B_{\mu \to e}(Z)$ and relative ratio of $B_{\mu \to e}(Z)$ of at least two different target can provide information on possible types of the operators as different operators predict different ratios.
In supersymmetric models\cite{Kitano:2002mt}, dominant contribution to $\mu-e$ conversion
comes from the dipole operator.
As a result, $B(\mu \to e)(Z)$ is predicted to be suppressed by $\alpha/\pi$
compared to $B(\mu \to e \gamma)$. For different choice of $Z$,
the conversion is suppressed by $10^{-3} \sim 5 \times 10^{-3}$.
Current limit on the conversion rate is comparable to $\mu \to e \gamma$ branching ratio,
but the future experiments on $\mu$ to e conversion will have better sensitivity.
We plot $\mu-e$ conversion rate with the expected future sensitivity of planned experiments in Fig.\ref{fig:ueconv}.

   %%%%%%%%%%%%%%%%%%%%%%%%%%%%%%%%%%%%%%%%%%%%%%%%%%%%%%%%%%%%%%%%%%%%%%%%%%%%%%%%%%%%%%%%
%%%%%%%%%%%%%%%%%%%%%%%%%%%%%%%%%%%%%%%%%%%%%%%%%%%%%%%%%%%%%%%%%%%%%%

%%%%%%%%%%%%%%%%%%%%%%%%%%%%%%%%%%%%%%%%%%%
\subsection{Correlation between Muon $g-2$, $\theta_{13}$, cLFV and the Higgs }\label{sec:mug-2}
%%%%%%%%%%%%%%%%%%%%%%%%%%%%%%%%%%%%%%%%%%%

%%%%%%%%%%%%%%%%%%%%%%%%%%%%%%%%%%%%%%%%%%%%%%%%%%%%%%%%%%%%%%%%%%%%%%%%%%%%%%%%%%%%%%%%
 \begin{figure}[ht]
  \begin{center}
      \includegraphics[width=0.70\textwidth]{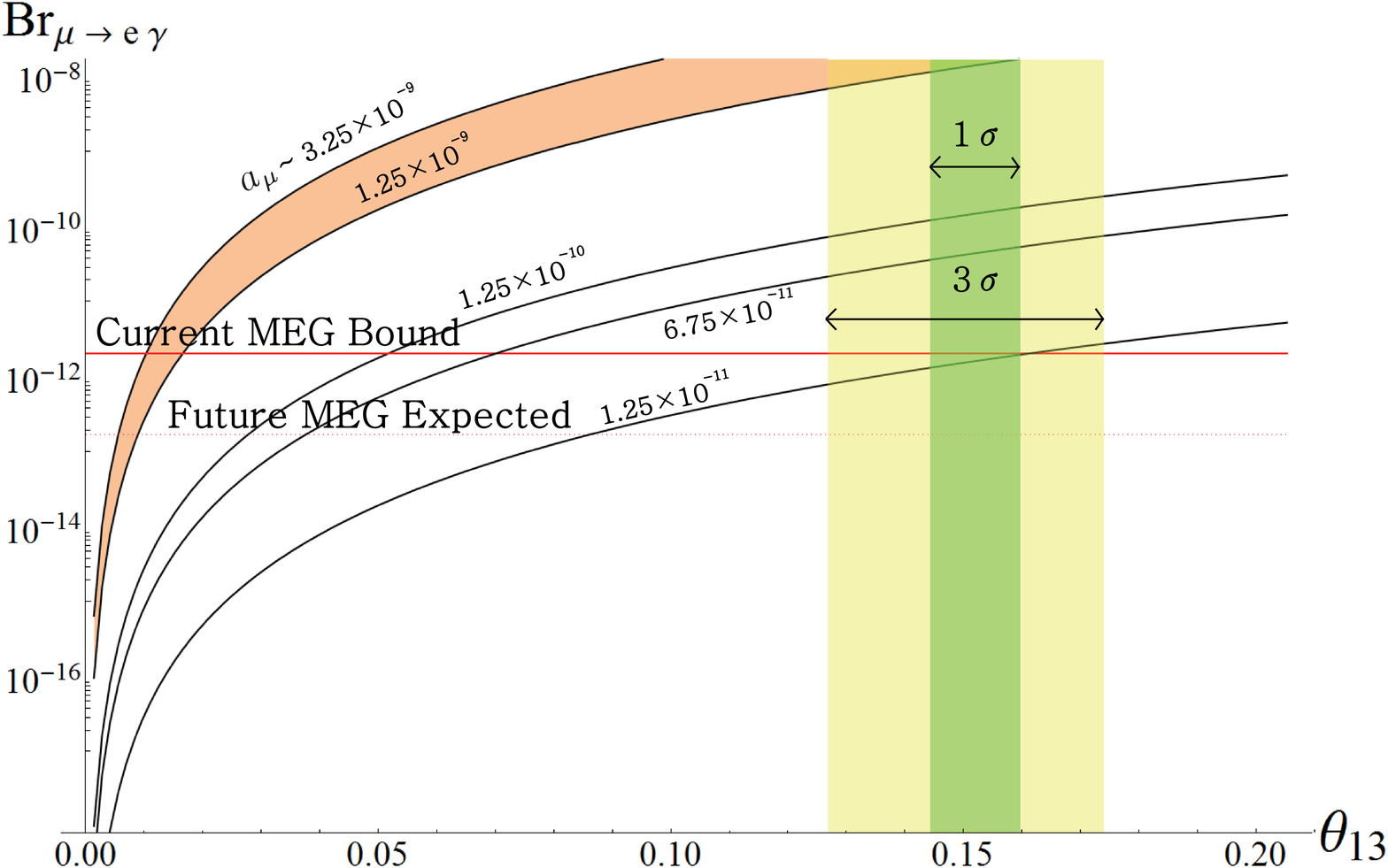}
      \includegraphics[width=0.70\textwidth]{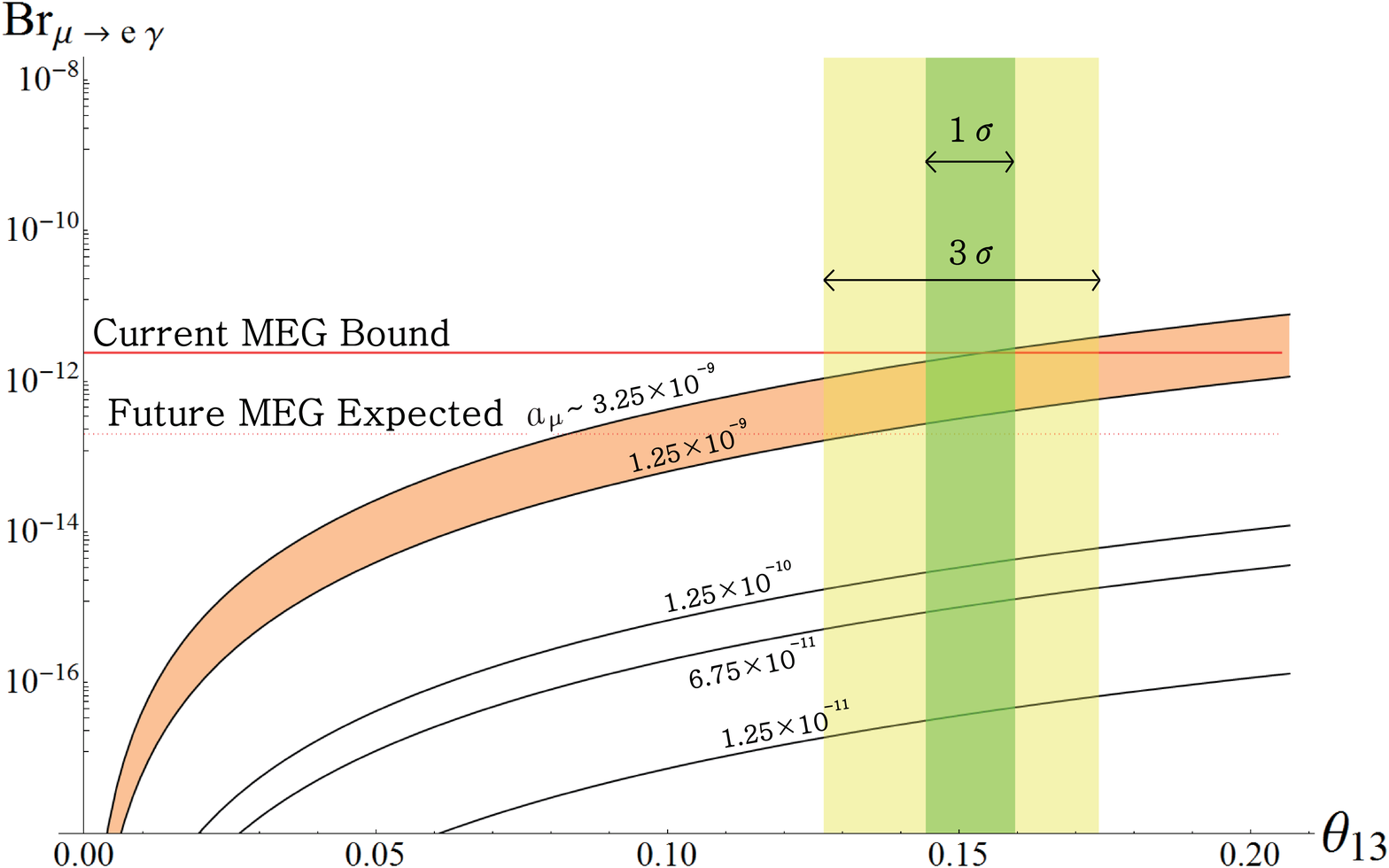}
  \end{center}
 \caption{ Branching ratio of $\mu \to e \gamma$ as a function of $\theta_{13}$ for $\tan \beta = 10$, $y_{\nu}= 0.62$, $\rho = 0.1$. Future MEG expected bound is $O(10^{-13})$, we set the value $2 \times 10^{-13}$. Observed muon $g-2$ discrepancy is about $(2.25 \pm 1) \times 10^{-9} $, we draw Br($\mu \to e \gamma$) at each muon $g-2$ contribution. Green and yellow band indicate $1 \sigma$, $3 \sigma$ level of neutrino $\theta_{13}$, respectively. In upper figure, $\theta_{13}$ is purely obtained from neutrino Dirac Yukawa splitting. In lower figure, only $1/15$ portion of $\theta_{13}$ is obtained from neutrino Dirac Yukawa.
  }
\label{fig:connection}
\end{figure}
%%%%%%%%%%%%%%%%%%%%%%%%%%%%%%%%%%%%%%%%%%%%%%%%%%%%%%%%%%%%%%%%%%%%%%

%%%%%%%%%%%%%%%%%%%%%%%%%%%%%%%%%%%%%%%%%%%%%%%%%%%%%%%%%%%%%%%%%%%%%%%%%%%%%%%%%%%%%%%%
 \begin{figure}[ht]
  \begin{center}
      \includegraphics[width=0.70\textwidth]{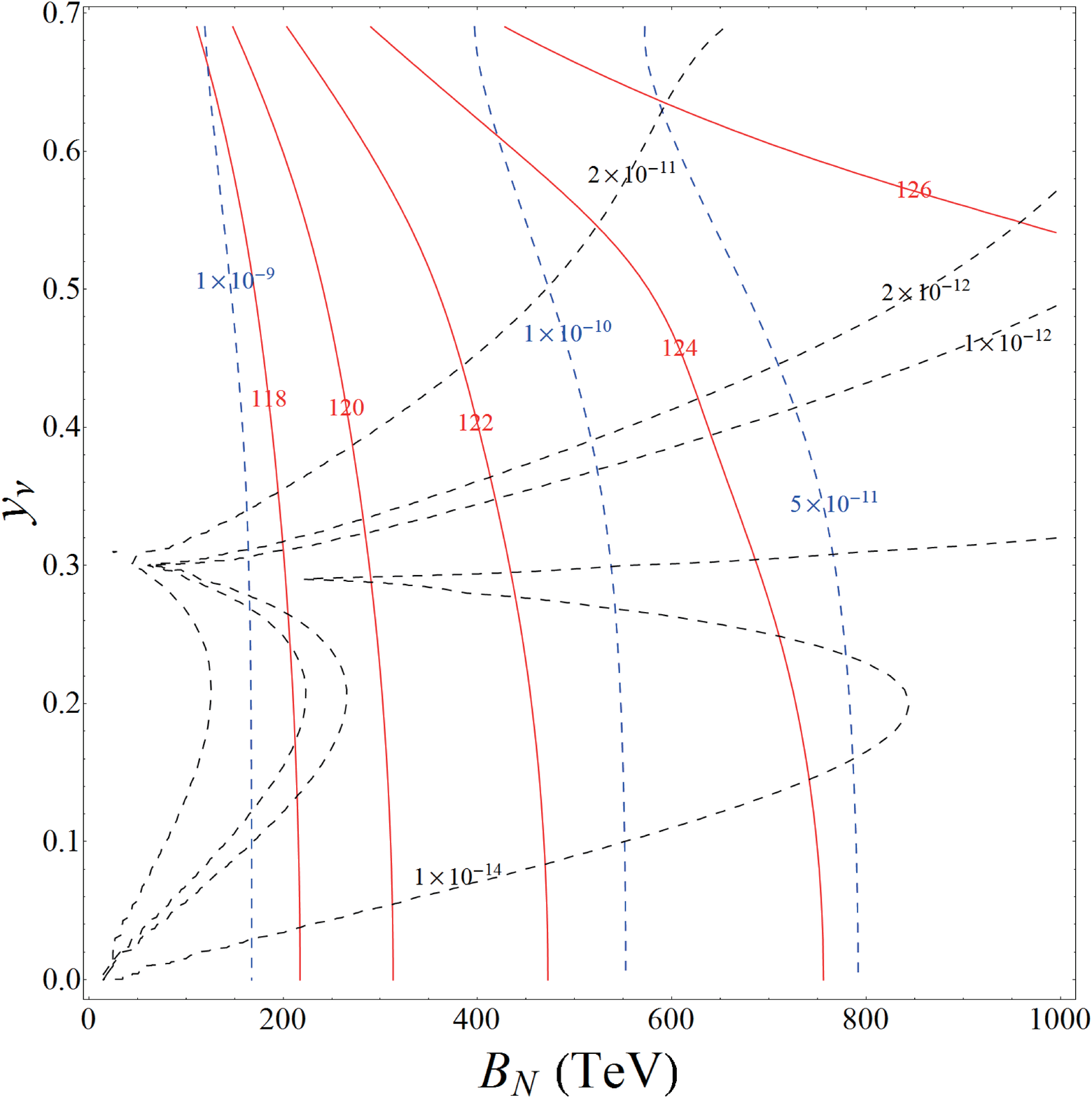}
  \end{center}
 \caption{ Contour plot of Higgs mass(red solid line), cLFV(black dashed line), and muon $g-2$(blue dashed line) in $B_N$ - $y_\nu$ plane for $\rho = 0.1$, $\tan \beta = 30$.
  }
\label{fig:BNYN}
\end{figure}
%%%%%%%%%%%%%%%%%%%%%%%%%%%%%%%%%%%%%%%%%%%%%%%%%%%%%%%%%%%%%%%%%%%%%%

The anomalous magnetic moment of muon (muon $g-2$) has a long standing
sizable deviation from the SM prediction. The observed value is \cite{Bennett:2006fi}
   \dis{a_\mu({\rm Exp})=11 659 208.9(6.3)\times10^{-10}, }
   whereas the SM prediction\cite{Hagiwara:2011af} is given by
   \dis{a_\mu({\rm SM})=11 659 182.8(4.9)\times10^{-10}}
   so we may have new physics contribution explaining the 3.3 $\sigma$ discrepancy,
   \dis{\delta a_\mu\equiv a_\mu({\rm Exp})-a_\mu({\rm SM})=(26.1\pm 8.0)\times10^{-10}.}
 In the context of SUSY \cite{Moroi:1995yh, Cho:2011rk}, muon $g-2$ has the same Feynman diagram structure as the cLFV process $\mu \to e\gamma$. The crucial difference is that muon $g-2$ is flavor-conserving process, while $\mu \to e\gamma$ violates lepton flavor, $L_\mu$ and $L_e$.
Therefore, $\delta a_\mu$ and ${\rm Br}(\mu \to e\gamma)$ have a strong correlation\cite{Hisano:2001qz},
 \dis{ {\rm Br}(\mu \to e\gamma)\simeq 3\times10^{-5}\Big(\frac{\delta a_\mu^{\rm SUSY}}{10^{-9}}\Big)^2\Big(\frac{(m_L^2)_{12}}{m_{\rm SUSY}^2}\Big)^2.}
Moreover, the neutrino Dirac Yukawa $Y_\nu$ contains information on the neutrino oscillation observables. Since we consider the model where parameters of $Y_\nu$ are related to $\theta_{13}$ and cLFV, we have a strong correlation between Br$(\mu \to e \gamma)$, $\theta_{13}$, and muon $g-2$ as discussed in  \cite{Hisano:2001qz}.

Fig. \ref{fig:connection} summarises the result.
Both muon anomalous magnetic moment and cLFV is a function
of $\tan \beta/M^2$ in which $M$ is the typical supersymmetry breaking scale.
The cLFV has extra suppression proportional to $(m_{L}^2)_{12}$.
The $S_4$ flavor model discussed here is constructed from the neutrino Dirac Yukawa matrix
which is proportional to the identity matrix and does not provide any off-diagonal entry in the slepton mass squared matrix if $\theta_{13}$ vanishes.
Recently measured $\theta_{13} \sim 0.15$ provides an extra information depending
on the origin of modification for nonzero $\theta_{13}$.

If the full $\theta_{13}$ is explained by the degeneracy lift of the neutrino Dirac Yukawa matrix
and if  the entire discrepancy of the muon anomalous magnetic moment should be explained by light slepton,
the current MEG bound tells that $\theta_{13}$ should be smaller than 0.01 which is incompatible with the observation recently made.
The parameter space which is consistent with $\mu \to e \gamma$ bound
and the $\theta_{13}$ predicts that muon anomalous magnetic moment is at least 1/20 times
smaller than what is needed.

If the nonzero $\theta_{13}$ is entirely generated by modifying the neutrino Majorana mass matrix,
there would be no cLFV even for the sizeable $\theta_{13}$.
In reality, the subleading corrections in the simple flavor model would appear in both sectors
and the observed $\theta_{13}$ would be a combined result of various sources.
The bottom plot of Fig. \ref{fig:connection} shows the hybrid case in which only $1/15$ of the $\theta_{13}$ is from the neutrino Dirac Yukawa modification. In this case $\theta_{13}$, muon anomalous magnetic moment can be explained at the same time. The $\mu \to e \gamma$ bound is satisfied and the the consistent region can be reached by the planned future MEG experiment as it predicts larger branching ratio of $\mu \to e \gamma$ than the planned expected sensitivity.

Fig. \ref{fig:BNYN} shows the tension between the muon $g-2$ and the Higgs mass.
Even if we take the model in which the neutrino Dirac Yukawa remains to be proportional to the identiy matrix such that no cLFV constraints apply,
125 GeV Higgs mass needs $B_N$ much larger thna 300 TeV. Then the slepton is too heavy and the muon $g-2$ is much smaller than $10^{-9}$.
The figure also shows an interesting feature that the off-diagonal elements of $m_L^2$ vanish at around $y_\nu = 0.3$ and the cLFV bounds are very weak at around $y_\nu = 0.3$.

%%%%%%%%%%%%%%%%%%%%%
%%%%%%%%%%%%%%%%%%%%%%%%%%%%%%%%%%%%%%%%%%%%%%%%%%%%%%%%%
\section{Conclusion}
%%%%%%%%%%%%%%%%%%%%%%%%%%%%%%%%%%%%%%%%%%%%%%%%%%%%%%%%%
%%%%%%%%%%%%%%%%%%%%%%

We considered the right-handed neutrinos as messengers of supersymmetry breaking
in the minimal gauge mediation. Direct coupling of neutrino messenger with the Higgs field $H_u$ and the lepton doublets $L_i$ provides soft-trilinear $A$ term for the top Yukawa
and can help increase the light Higgs mass by realising the maximal stop mixing scenario.
We call this setup as 'neutrino assisted gauge mediation'. The Yukawa mediation given by neutrino messengers also appear at soft scalar masses of the Higgs $H_u$, the lepton doublets $L_i$.
At the same time it affects the soft scalar masses of the fields which couple to $H_u$ and $L_i$
at two loop. Among those, the stop mass squared gets the largest correction
as the top Yukawa coupling is of order one. For $y_{\nu}$ slightly larger than $0.7$,
the correction is big enough to make stop tachyonic.
Therefore, this realises the natural supersymmetry spectrum.
At the same time maximal mixing is achieved by two effects, large $A_t$
and small $m_{\tilde{t}}^2$ at around $y_\nu \sim 0.7$.
In general this effect allows to explain the observed Higgs mass at around 125 GeV
using around 1 TeV stop mass.
Compared to the case when the neutrino assistance is turned off ($y_\nu=0$),
about 5 GeV of the Higgs mass is enhanced.

In general the off-diagonal entry of the slepton mass squared, $m_{L}^2$, appears at the messenger scale and can make the charged lepton flavor violating process to occur.
The detailed quantitative prediction of cLFV highly depends on flavor model building.
We provided a representative model based on $S_4$ flavor symmetry
in which the Dirac neutrino Yukawa can be proportional to the identity if $\theta_{13}=0$.
For nonzero $\theta_{13}$, two options were considered. Firstly, the total $\theta_{13}$
can be explained by the modification of the neutrino Dirac Yukawa matrix.
Secondly, the $\theta_{13}$ can be explained by modifying the Majorana mass matrix of neutrinos.
For the former, very stringent bound on the slepton mass comes from $\mu \to e \gamma$
and the slepton should be heavier than $2 \sim 4$ TeV, depending on $\tan \beta$.
Also for the slepton mass at around 2 TeV with $\tan \beta = 10$, the $\mu \to e \gamma$ is just below the current experimental bound and we expect to observe the $\mu \to e \gamma$ in the near future.

Even for the second case in which we can safely avoid cLFV constraints,
the neutrino assisted gauge mediation (in its minimal form with one copy of $5$ and $\bar{5}$ messenger) sets the lower bound on the slepton mass to explain the Higgs mass.
1 $\sim$ 2 TeV slepton mass at the same time sets an upper bound on the possible contribution
to muon anomalous magnetic moment and $a_\mu  \sim 10^{-10}$ is the upper bound.

In this paper we proposed the neutrino assisted gauge mediation and showed a possible way out to avoid the strong cLFV constraints. Even then the current scheme has a tension with the muon anomalous magnetic moment which needs a lighter slepton. The extension of the minimal neutrino assisted gauge mediation to multiple messengers might ameliorate the tension between the spectrum needed to explain the Higgs mass and the muon anomalous magnetic moment.

%%%%%%%%%%%%%%%%%%%%%%%%%%%%%%%%%%%%%%%%%%%%%%%%%%%%%%%%%%%%%%%%%%%%%%%%%%%%%
%%%%%%%%%%%%%%%%%%%%%%%%%%%%%%%%%%%%%%%%%%%%%%%%%%%%%%%%%%%%%%%%%%%%%%%%%%%%%

%%%%%%%%%%%%%%%%%%%%%%%%%%%%%%%%%%%%
\acknowledgements
This work is supported by the NRF of Korea No. 2011-0017051.
%%%%%%%%%%%%%%%%%%%%%%%%%%%%%%%%%%%%

%%%%%%%%%%%%%%%%
%%%%%%%%%%%%%%%%
\section*{Appendix 0: Sparticle Spectrum Sample Point}\label{sec:App0}
%%%%%%%%%%%%%%%%
  \begin{table}[h]
\begin{center}
\begin{tabular}{|c|c|c|}
\hline   & $(\tan \beta = 10 , B_N = 360 \tev)$ & $(\tan \beta = 30 , B_N = 300 \tev)$
\\[0.2em]
\hline  &&\\ [-1.1em]
$\tilde{\nu}_e, \tilde{\nu}_{\mu}, \tilde{\nu}_{\tau}$ & 2957, 2961, 3013  & 2429, 2465, 2502
\\[0.2em]
$\tilde{e}_1, \tilde{\mu}_1, \tilde{\tau}_1$&1364, 1364, 1333 &  1139, 1138, 880
\\[0.2em]
$\tilde{e}_2, \tilde{\mu}_2, \tilde{\tau}_2$&3013, 2962, 2954 &  2503, 2467, 2427
\\[0.2em]
$\tilde{u}_1 ,\tilde{c}_1, \tilde{t}_1$&2827, 2827, 634   &  2384,  2384, 637
\\[0.2em]
$\tilde{d}_1, \tilde{s}_1, \tilde{b}_1$&2853, 2853, 2820 &  2406, 2406, 2283
\\[0.2em]
$\tilde{u}_2, \tilde{c}_2, \tilde{t}_2$&3177, 3177, 2252 &  2675, 2675, 1868
\\[0.2em]
$\tilde{d}_2, \tilde{s}_2 ,\tilde{b}_2$&3178, 3178, 2297 &   2676, 2676, 1893
\\[0.2em]
$h_0, A, H_0 ,H_{\pm} $&  125, 1705, 1705, 1707  &125, 1031, 1031, 1034
\\[0.2em]
$\chi_1, \chi_2, \chi_3 ,\chi_4 $&487, 850, -892, 980 &  405, 713, -758, 829
\\[0.2em]
$\chi_{+}, \chi_{-}$& 849, 980 & 712, 829
\\[0.2em]
$\tilde{g} $&2514 & 2126
\\[0.2em]
\hline
\end{tabular}
\end{center}
\caption{Sparticle spectrum at the point giving $125 \gev$ Higgs mass with the lowest $B_N$} \label{table:Spectrum125}
\end{table}

%%%%%%%%%%%%%%%%%%%%%%%%%%%%%%%%%%%%%%%%%%%%%%%%%%%%%%%%%%%%%%%%%%%%%%
  \begin{table}[h]
\begin{center}
\begin{tabular}{|c|c|c|}
\hline   &    $(\tan \beta = 10, B_N = 240 \tev)$ & $(\tan \beta = 30 , B_N = 200 \tev)$
\\[0.2em]
\hline  &&\\ [-1.1em]
$\tilde{\nu}_e, \tilde{\nu}_{\mu}, \tilde{\nu}_{\tau}$ &1971,  1974,  2009 & 1657,  1682,  1707
\\[0.2em]
$\tilde{e}_1, \tilde{\mu}_1, \tilde{\tau}_1$ & 915,  915,  894  & 770,  769, 590
\\[0.2em]
$\tilde{e}_2, \tilde{\mu}_2, \tilde{\tau}_2$ &  2010,  1976,  1970  &  1709, 1684, 1657
\\[0.2em]
$\tilde{u}_1, \tilde{c}_1, \tilde{t}_1$& 1937,  1937,  521 &  1633,  1633,  404
\\[0.2em]
$\tilde{d}_1, \tilde{s}_1, \tilde{b}_1$ &  1954, 1954,  1931 &  1650,  1650,  1564
\\[0.2em]
$\tilde{u}_2, \tilde{c}_2, \tilde{t}_2$ &2169,  2169,  1569  &  1828,  1828,  1286
\\[0.2em]
$\tilde{d}_2, \tilde{s}_2, \tilde{b}_2$ &2170,  2170,  1586 &  1829, 1829, 1291
\\[0.2em]
$h_0, A, H_0, H_{\pm} $ &123, 1220, 1220, 1223&  123, 679, 679, 684
\\[0.2em]
$\chi_1, \chi_2, \chi_3, \chi_4 $& 322,  600, -729,  757   &   267, 466, -520,  579
\\[0.2em]
$\chi_{+}, \chi_{-}$&600,  757    &  465,  579
\\[0.2em]
$\tilde{g} $&1737     &  1470

\\[0.2em]
\hline
\end{tabular}
\end{center}
\caption{Sparticle spectrum at the point giving $123 \gev$ Higgs mass with the lowest $B_N$} \label{table:Spectrum123}
\end{table}

%%%%%%%%%%%%%%%%
%%%%%%%%%%%%%%%%
\section*{Appendix A: representations of $S_4$ symmetry and tensor products}\label{sec:AppA}
%%%%%%%%%%%%%%%%
  $S_4$ is a non-abelian discrete symmetry and consists of all permutations among four quantities. For a review, see \cite{Ishimori:2010au}. Irreducible representations of $S_4$ are two singlets ${\bf 1}, {\bf 1^\prime}$, one singlet ${\bf 2}$, and two triplets ${\bf 3}, {\bf 3^\prime}$. Tensor products among them are given as follows:
   \dis{\left(
\begin{array}{c}
x_1 \\
x_2 \\
x_3
  \end{array}\right)_{\bf 3} \times
  \left(
\begin{array}{c}
y_1 \\
y_2 \\
y_3
  \end{array}\right)_{\bf 3}&= (x_1y_1+x_2y_2+x_3y_3)_{\bf 1}+
  \left(
\begin{array}{c}
x_1 y_1+\omega x_2y_2 +\omega^2 x_3y_3\\
x_1 y_1+\omega^2 x_2y_2 +\omega x_3y_3
  \end{array}\right)_{\bf 2}
  \\
  &+
    \left(
\begin{array}{c}
 x_2y_3 + x_3y_2\\
x_3 y_1+ x_1y_3 \\
x_1y_2+x_2y_1
  \end{array}\right)_{\bf 3}
  +
    \left(
\begin{array}{c}
 x_2y_3 - x_3y_2\\
x_3 y_1- x_1y_3 \\
x_1y_2-x_2y_1
  \end{array}\right)_{\bf 3^\prime}    }

   \dis{\left(
\begin{array}{c}
x_1 \\
x_2 \\
x_3
  \end{array}\right)_{\bf 3^\prime} \times
  \left(
\begin{array}{c}
y_1 \\
y_2 \\
y_3
  \end{array}\right)_{\bf 3^\prime}&= (x_1y_1+x_2y_2+x_3y_3)_{\bf 1}+
  \left(
\begin{array}{c}
x_1 y_1+\omega x_2y_2 +\omega^2 x_3y_3\\
x_1 y_1+\omega^2 x_2y_2 +\omega x_3y_3
  \end{array}\right)_{\bf 2}
  \\
  &+
    \left(
\begin{array}{c}
 x_2y_3 + x_3y_2\\
x_3 y_1+ x_1y_3 \\
x_1y_2+x_2y_1
  \end{array}\right)_{\bf 3}
  +
    \left(
\begin{array}{c}
 x_2y_3 - x_3y_2\\
x_3 y_1- x_1y_3 \\
x_1y_2-x_2y_1
  \end{array}\right)_{\bf 3^\prime}    }

     \dis{\left(
\begin{array}{c}
x_1 \\
x_2 \\
x_3
  \end{array}\right)_{\bf 3} \times
  \left(
\begin{array}{c}
y_1 \\
y_2 \\
y_3
  \end{array}\right)_{\bf 3^\prime}&= (x_1y_1+x_2y_2+x_3y_3)_{\bf 1^\prime}+
  \left(
\begin{array}{c}
x_1 y_1+\omega x_2y_2 +\omega^2 x_3y_3\\
-(x_1 y_1+\omega^2 x_2y_2 +\omega x_3y_3)
  \end{array}\right)_{\bf 2}
  \\
  &+
    \left(
\begin{array}{c}
 x_2y_3 + x_3y_2\\
x_3 y_1+ x_1y_3 \\
x_1y_2+x_2y_1
  \end{array}\right)_{\bf 3^\prime}
  +
    \left(
\begin{array}{c}
 x_2y_3 - x_3y_2\\
x_3 y_1- x_1y_3 \\
x_1y_2-x_2y_1
  \end{array}\right)_{\bf 3}    }

     \dis{\left(
\begin{array}{c}
x_1 \\
x_2
  \end{array}\right)_{\bf 2} \times
  \left(
\begin{array}{c}
y_1 \\
y_2
  \end{array}\right)_{\bf 2}&= (x_1y_2+x_2y_1)_{\bf 1}+ (x_1y_2-x_2y_1)_{\bf 1^\prime}+
  \left(
\begin{array}{c}
x_2y_2 \\
x_1 y_1
  \end{array}\right)_{\bf 2}
   }

      \dis{\left(
\begin{array}{c}
x_1 \\
x_2
  \end{array}\right)_{\bf 2} \times
  \left(
\begin{array}{c}
y_1 \\
y_2 \\
y_3
  \end{array}\right)_{\bf 3}=
    \left(
\begin{array}{c}
 (x_1+x_2)y_1\\
(\omega^2 x_1+\omega x_2) y_2 \\
(\omega x_1 + \omega^2 x_2)y_3
  \end{array}\right)_{\bf 3} +
    \left(
\begin{array}{c}
  (x_1-x_2)y_1\\
(\omega^2 x_1-\omega x_2) y_2 \\
(\omega x_1 - \omega^2 x_2)y_3
  \end{array}\right)_{\bf 3^\prime}    }

        \dis{\left(
\begin{array}{c}
x_1 \\
x_2
  \end{array}\right)_{\bf 2} \times
  \left(
\begin{array}{c}
y_1 \\
y_2 \\
y_3
  \end{array}\right)_{\bf 3^\prime}=
    \left(
\begin{array}{c}
 (x_1+x_2)y_1\\
(\omega^2 x_1+\omega x_2) y_2 \\
(\omega x_1 + \omega^2 x_2)y_3
  \end{array}\right)_{\bf 3^\prime} +
    \left(
\begin{array}{c}
  (x_1-x_2)y_1\\
(\omega^2 x_1-\omega x_2) y_2 \\
(\omega x_1 - \omega^2 x_2)y_3
  \end{array}\right)_{\bf 3}    }
   and trivially, we have ${\bf 3} \times {\bf 1^\prime} ={\bf 3^\prime}$, ${\bf 3^\prime} \times {\bf 1^\prime} ={\bf 3^\prime}$, and ${\bf 2} \times {\bf 1^\prime} ={\bf 2}$.

%%%%%%%%%%%%%%%%
\section*{Appendix B: Remarks on the flavon vacuum stability}\label{sec:AppB}
%%%%%%%%%%%%%%%%
 In \cite{BenTov:2012tg}, it was shown that $A_4$ triplet flavon vacuum in the direction of $(1,1,1)$ and $(1,0,0)$, ($(0,1,0)$, $(0,0,1)$ are the same) is favored compared to other directions, such as $(1,1,0)$. Since $A_4$ symmetry is the subgroup of the $S_4$ composed of even permutations, similar arguments hold. In this Appendix, we argue that triplet flavon directions favored in $A_4$ model are also favored in the $S_4$ model and that $S_4$ doublet vacuum favors the $(1,1)$ direction.

   Rigid SUSY makes the discussion more simple, because the potential  $V$ is minimized at $\langle V \rangle=0$.  On the other hand, extra symmetries like $Z_4$ and U(1)$_L$ more restrict possible terms in the superpotential. Suppose that U(1)$_L$ symmetry is discretized to, for example, $Z_8$ symmetry. In this case, only quartic terms $\Phi^4$ and $\chi^4$ are allowed.
 Let us assume that breaking of extra symmetries introduces quadratic term, like $m_1\Phi^2$ or $m_2 \chi^2$. To achieve this, let us consider `$Z_4$ breaking singlets' $\psi_1$, $\bar{\psi}_1$ and `lepton number breaking singlets' $\psi_2$, $\bar{\psi}_2$ with $S_4 \times Z_4 \times{\rm U(1)}_L $ quantum numbers
 \dis{&\psi_1:  ({\bf 1},3,0),~~~\bar{\psi}_1 : ({\bf 1},1,0),
 \\
 & \psi_2:  ({\bf 1},0,2),~~~\bar{\psi}_2 : ({\bf 1},0,6).}
  They do not combine with $\bar{E}LH_d$, $NLH_u$ and $NN$ to make singlets under all symmetries imposed. Then, they can couple to $\Phi^2$ or $\chi^2$ such that a superpotential is given by
  \dis{W(\psi_1, \bar{\psi}_1, \psi_2, \bar{\psi}_2)=&\frac{1}{\Lambda}[\Phi^2\bar{\psi}_1\psi_1 + \chi^2\bar{\psi}_2\psi_2]
  \\
  &-M_1 \bar{\psi}_1\psi_1 +\frac{1}{\Lambda}[\kappa_1(\bar{\psi}_1\psi_1)^2+\kappa_2 (\psi_1)^4 + \kappa_3 (\bar{\psi}_1)^4]
  \\
    &-M_2 \bar{\psi}_2\psi_2 +\frac{1}{\Lambda}[\kappa_1^\prime(\bar{\psi}_2\psi_2)^2+\kappa_2^\prime (\psi_2)^4 + \kappa_3^\prime (\bar{\psi}_2)^4] .}
    In this superpotential, $\bar{\psi_1}{\psi}_1$ and $\bar{\psi_2}{\psi}_2$ pairs have VEVs and they provide $m_1 \Phi^2+m_2 \chi^2$ terms. With this setup, the triplet superpotential has the form of
   \dis{W&=mS^2+\frac{\lambda_1}{\Lambda} (x^2+y^2+z^2)^2+\frac{\lambda_2}{\Lambda}(x^2+\omega y^2+\omega^2 z^2)(x^2+\omega^2 y^2+\omega z^2)
   \\
   &+\frac{\lambda_3}{\Lambda}(xy+yz+zx)^2}
      where $S=(x, y, z)$ represents the generic $S_4$ triplet such as $\Phi$ or $\chi$.  Note also that the superpotential has an accidental $Z_2$ symmetry under which $\psi_{1,2}$ and $\bar{\psi}_{1,2}$ are odd whereas other fields are even. If this $Z_2$ symmetry is imposed,  $(\Phi^2\psi_1/\Lambda^3)\bar{E}LH_d$ and $(\Phi^2\psi_2/\Lambda^2)NN$ terms, which change the flavor structure in the subleading orders are forbidden. In this case, charged lepton Yukawa coupling structure in dimension-4 operator is preserved up to dimension-6 operator whereas Majorana mass structure in dimension-3 operator is preserved up to dimension-5 operator so corrections to them are highly suppressed.

 Each term of the F-term potential $V=|\partial W/\partial x|^2+|\partial W/\partial y|^2+|\partial W/\partial z|^2$ is given by
   \dis{&\frac{\partial W}{\partial x}=mx+\frac{4\lambda_1}{\Lambda}x(x^2+y^2+z^2)+\frac{2\lambda_2}{\Lambda}x(2x^2-y^2-z^2)+\frac{2\lambda_3}{\Lambda}(y+z)(xy+yz+zx)
   \\
  &\frac{\partial W}{\partial y}=my+\frac{4\lambda_1}{\Lambda}x(x^2+y^2+z^2)+\frac{2\lambda_2}{\Lambda}y(2y^2-z^2-x^2)+\frac{2\lambda_3}{\Lambda}(z+x)(xy+yz+zx)
  \\
  &\frac{\partial W}{\partial z}=mz+\frac{4\lambda_1}{\Lambda}z(x^2+y^2+z^2)+\frac{2\lambda_2}{\Lambda}z(2z^2-x^2-y^2)+\frac{2\lambda_3}{\Lambda}(x+y)(xy+yz+zx). }
  Stable vacuum requires that these three terms should be zero simultaneously. For vacuum $\langle S \rangle =v(1,1,1)$, three terms give the same condition,
  \dis{12(\lambda_1+\lambda_3)\Big(\frac{v^3}{\Lambda}\Big)+mv=0}
  so the vacuum is stabilized at $v^2=-m\Lambda/[12(\lambda_1+\lambda_3)]$. For vacuum $\langle S \rangle =v(1,0,0)$, the second and third terms vanish trivially and the first term gives
  \dis{4(\lambda_1+\lambda_3)\Big(\frac{v^3}{\Lambda}\Big)+mv=0}
  so the vacuum is stabilized at $v^2=-m\Lambda/[4(\lambda_1+\lambda_3)]$.
  The vacuum in the direction $(0,1,0)$ and $(0,0,1)$ gives the same result by permutational property of $S_4$. On the other hand, vacuum $\langle S \rangle =v(1,1,0)$ gives two conditions,
  \dis{&\frac{v^3}{\Lambda}(8\lambda_1+2\lambda_2+2\lambda_3)+mv=0
  \\
  &\lambda_3 v^3=0.}
If $\lambda_3$ is not forbidden by another symmetry, $v=0$ is the only solution
and nontrivial vacuum can not be developed.

   $S_4$ doublet stabilization can be discused in the same way. Renormalizable superpotential for doublet $(x,y)$ is written as
   \dis{W=m (xy)+\lambda (x^3+y^3)}
and stabilization condition
\dis{&\frac{\partial W}{\partial x}=2my+3\lambda x^2=0
\\
&\frac{\partial W}{\partial y}=2mx+3\lambda y^2=0}
requires that $x=y$. So the vacuum choice for Eq. (\ref{eq:modYnu}) is stable.

%%%%%%%%%%%%%%%%
\section*{Appendix C: Comment on K\"ahler potential corrections}\label{sec:AppC}
%%%%%%%%%%%%%%%%

In our setup, Yukawa couplings are constructed from non-renormalizable dimension-4 superpotential with several flavons. These flavons also appear in the non-renormalizable K\"ahler potential and kinetic terms are written in the form of
\dis{K_{i\bar{j}}\partial_\mu \phi^{\bar{j} \dagger} \partial^\mu \phi^i
-iK_{i\bar{j}}\bar{\psi}^{\bar j}\bar{\sigma}_\mu \partial^\mu \psi^i}
where $\phi$ and $\psi$ represent bosonic and fermionic fields, respectively. The K\"ahler potential of charged lepton supermultiplet $L$ is given by
\dis{K=\Big[1+a_1 \frac{\Phi^\dagger \Phi}{\Lambda^2} +a_2 \frac{\chi^\dagger \chi}{\Lambda^2} \Big] L^\dagger L \Big|_{S_4 ~{\rm singlets}}+\cdots}
and similar terms can be written for other fields, $\bar{E}^\dagger \bar{E}$, $N^\dagger N$, $H_{u,d}^\dagger H_{u,d}$, etc. Then we have quite complicate terms. For example, from $(\Phi_{\bf 3}^\dagger \Phi_{\bf 3}/\Lambda^2) L^\dagger L$ where $\Phi_{\bf 3}$ vacuum is given by $v_2(1,1,1)$, we have
\dis{a_1\frac{\Phi_{\bf 3}^\dagger \Phi_{\bf 3}}{\Lambda^2} L^\dagger L  \Big|_{S_4 ~{\rm singlets}}=& a_{1,1}\frac{v_2^2}{\Lambda^2} (L_1^\dagger L_1 + L_2^\dagger L_2 + L_3^\dagger L_3)
\\
&+a_{1,2}\frac{v_2^2}{\Lambda^2}\Big[ L_2^\dagger L_3 + L_3^\dagger L_2
+ L_3^\dagger L_1+ L_1^\dagger L_3 + L_1^\dagger L_2 + L_2^\dagger L_1  \Big].}
 Since $\langle \Phi \rangle/\Lambda=v_2/\Lambda$ is responsible for charged lepton Yukawa couplings, we see $4\pi v_2/\Lambda \gtrsim Y_\tau=m_\tau/[(v/\sqrt2)\cos\beta] \sim 0.1$ for $\tan\beta =10$. On the other hand, $\chi_{\bf 3}$ has another vacuum direction, $w_2(0,1,0)$. Then
 \dis{a_2\frac{\chi^\dagger \chi}{\Lambda^2}L^\dagger L\Big|_{S_4 ~{\rm singlets}}=&a_{2,1}\frac{w_2^2}{\Lambda^2}(L_1^\dagger L_1 + L_2^\dagger L_2 + L_3^\dagger L_3)
 \\
 &+a_{2,2}\frac{w_2^2}{\Lambda^2}(-L_1^\dagger L_1 + L_2^\dagger L_2 - L_3^\dagger L_3)}
  so it just rescales the fields.  Moreover, since See-Saw scale is about $10^{14}\gev$, we have suppressed effect,  $4\pi\chi/\Lambda \sim 0.01$ with $\Lambda$ is the GUT scale. In the same way, doublet and singlet flavons in the K\"ahler potential just contribute to the field rescalings.

 Physical fields are defined with canonical kinetic terms, so we should make field redefinitions and they affect flavor structures in principle. In our work, however, such effects are not considered by assuming small coeffecients $a_{1,2}$. For example, diagonalization of $Y_E$ demonstrated above is not affected if $a_1 ( v_2^2/\Lambda^2) \lesssim (m_e/m_\tau)  \sim 3 \times 10^{-4}$, {\it i.e.} $a_1 \lesssim 3 $.

  On the other hand, mixings in the K\"ahler potential between flavons can be dangerous.  For example, kinetic mixing between flavons such as $\bar{\psi}_1^\dagger \psi_2^\dagger \Phi_{\bf 3}^\dagger \chi_{\bf 3}/\Lambda^2$ can introduce small correction to $Y_E$ or $M_N$ with unwanted $S_4$ triplet vacuum direction.  Such effect is suppressed by $\bar{\psi}_1^\dagger \psi_2^\dagger/\Lambda^2$ and can be more suppressed with tiny coefficient.

%%%%%%%%%%%%%%%%%%%%%%%%%%%%%%%%%%%%
%\begin{thebibliography}{99}
%%%%%%%%%%%%%%%%%%%%%%%%%%%%%%%%%%%%%
%%%%%%%%%%%%

%%%%%%%%%
%%%%%%%%%%%%%%%%%%%%%%
%\end{thebibliography}
%%%%%%%%%%%%%%%%%%%%%%%%%%
%%%%%%%%%%%%%%%%%%%%%%%%%%%%%%%%%%%%%%%%%%%%%%%%

%%%%%%%%%%%%%%%%%%
%%%%%%%%%%%%%%%%%%
%%%%%%%%%%%%%%%%%%
%%%%%%%%%%%%%%%%%%

%%%%%%%%%%%%%%%%%%%%%%%%%%%%%%%%%%%%%%%%%%%%%%%%
%%%%%%%%%%%%%%%%%%%%%%%%%%%%%%%%%%%%%%%

\begin{thebibliography}{99}

\def\prp#1#2#3{Phys.\ Rep.\ {\bf #1} (#3) #2}
\def\rmp#1#2#3{Rev. Mod. Phys.\ {\bf #1} (#3) #2}
\def\npb#1#2#3{Nucl.\ Phys.\ {\bf B#1} (#3) #2}
\def\plb#1#2#3{Phys.\ Lett.\ {\bf B#1} (#3) #2}
\def\prd#1#2#3{Phys.\ Rev.\ {\bf D#1} (#3) #2}
\def\prl#1#2#3{Phys.\ Rev.\ Lett.\ {\bf #1} (#3) #2}
\def\err#1#2#3{\ {\bf #1} (#3) #2\,(E)}
\def\jhep#1#2#3{JHEP\ {\bf #1} (#3) #2}
\def\jcap#1#2#3{JCAP\ {\bf #1} (#3) #2}
\def\zp#1#2#3{Z.\ Phys.\ {\bf #1} (#3) #2}
\def\epjc#1#2#3{Euro. Phys. J.\ {\bf C#1} (#3) #2}
\def\ijmp#1#2#3{Int.\ J.\ Mod.\ Phys.\ {\bf #1} (#3) #2}
\def\mpl#1#2#3{Mod.\ Phys.\ Lett.\ {\bf A#1} (#3) #2 }
\def\apj#1#2#3{Astrophys.\ J.\ {\bf #1} (#3) #2}
\def\nat#1#2#3{Nature\ {\bf #1} (#3) #2}
\def\sjnp#1#2#3{Sov.\ J.\ Nucl.\ Phys.\ {\bf #1} (#3) #2}
\def\apj#1#2#3{Astrophys.\ J.\ {\bf #1} (#3) #2}
\def\ijmp#1#2#3{Int.\ J.\ Mod.\ Phys.\ {\bf #1} (#3) #2}
\def\mpl#1#2#3{Mod.\ Phys.\ Lett.\ {\bf A#1} (#3) #2 }
\def\nat#1#2#3{Nature\ {\bf #1} (#3) #2}
\def\npb#1#2#3{Nucl.\ Phys.\ {\bf B#1} (#3) #2}
\def\pthp#1#2#3{Prog.\ Theor.\ Phys.\ {\bf #1} (#3) #2}


%\cite{:2012gk}
\bibitem{:2012gk}
  G.~Aad {\it et al.}  [ATLAS Collaboration],
  %``Observation of a new particle in the search for the Standard Model Higgs boson with the ATLAS detector at the LHC,''
   Phys.\ Lett.\ B {\bf 716}, 1 (2012)  [arXiv:1207.7214 [hep-ex]].  %%CITATION = ARXIV:1207.7214;%%

%\cite{:2012gu}
\bibitem{:2012gu}
  S.~Chatrchyan {\it et al.}  [CMS Collaboration],
  %``Observation of a new boson at a mass of 125 GeV with the CMS experiment at the LHC,''
   Phys.\ Lett.\ B {\bf 716}, 30 (2012)  [arXiv:1207.7235 [hep-ex]].  %%CITATION = ARXIV:1207.7235;%%

  %\cite{Dine:1993yw}
\bibitem{Dine:1993yw}
  M.~Dine and A.~E.~Nelson,
  %``Dynamical supersymmetry breaking at low-energies,''
  Phys.\ Rev.\ D {\bf 48}, 1277 (1993)  [hep-ph/9303230].  %%CITATION = HEP-PH/9303230;%%


%\cite{Dine:1994vc}
\bibitem{Dine:1994vc}
  M.~Dine, A.~E.~Nelson and Y.~Shirman,
  %``Low-energy dynamical supersymmetry breaking simplified,''
  Phys.\ Rev.\ D {\bf 51}, 1362 (1995)  [hep-ph/9408384].  %%CITATION = HEP-PH/9408384;%%

%\cite{Dine:1995ag}
\bibitem{Dine:1995ag}
  M.~Dine, A.~E.~Nelson, Y.~Nir and Y.~Shirman,
  %``New tools for low-energy dynamical supersymmetry breaking,''
  Phys.\ Rev.\ D {\bf 53}, 2658 (1996)  [hep-ph/9507378].  %%CITATION = HEP-PH/9507378;%%

%\cite{Giudice:1998bp}
\bibitem{Giudice:1998bp}
  G.~F.~Giudice and R.~Rattazzi,
  %``Theories with gauge mediated supersymmetry breaking,''
  Phys.\ Rept.\  {\bf 322}, 419 (1999)  [hep-ph/9801271].  %%CITATION = HEP-PH/9801271;%%

 %\cite{Ajaib:2012vc}
\bibitem{Ajaib:2012vc}
  M.~A.~Ajaib, I.~Gogoladze, F.~Nasir and Q.~Shafi,
  %``Revisiting mGMSB in Light of a 125 GeV Higgs,''
  Phys.\ Lett.\ B {\bf 713}, 462 (2012)
  [arXiv:1204.2856 [hep-ph]].
  %%CITATION = ARXIV:1204.2856;%%

%\cite{Feng:2012rn}
\bibitem{Feng:2012rn}
  J.~L.~Feng, Z.~'e.~Surujon and H.~-B.~Yu,
  %``Confluence of Constraints in Gauge Mediation: The 125 GeV Higgs Boson and Goldilocks Cosmology,''
  Phys.\ Rev.\ D {\bf 86}, 035003 (2012)
  [arXiv:1205.6480 [hep-ph]].
  %%CITATION = ARXIV:1205.6480;%%

%\cite{Bae:2012am}
\bibitem{Bae:2012am}
  K.~J.~Bae, K.~Choi, E.~J.~Chun, S.~H.~Im, C.~B.~Park and C.~S.~Shin,
  %``Peccei-Quinn NMSSM in the light of 125 GeV Higgs,''
  arXiv:1208.2555 [hep-ph].
  %%CITATION = ARXIV:1208.2555;%%

%\cite{Martin:2009bg}
\bibitem{Martin:2009bg}
  S.~P.~Martin,
  %``Extra vector-like matter and the lightest Higgs scalar boson mass in low-energy supersymmetry,''
  Phys.\ Rev.\ D {\bf 81}, 035004 (2010)  [arXiv:0910.2732 [hep-ph]].  %%CITATION = ARXIV:0910.2732;%%


%\cite{Martin:2012dg}
\bibitem{Martin:2012dg}
  S.~P.~Martin and J.~D.~Wells,
  %``Implications of gauge-mediated supersymmetry breaking with vector-like quarks and a ~125 GeV Higgs boson,''
  Phys.\ Rev.\ D {\bf 86}, 035017 (2012)
  [arXiv:1206.2956 [hep-ph]].
  %%CITATION = ARXIV:1206.2956;%%

  %\cite{Bae:2012ir}
\bibitem{Bae:2012ir}
  K.~J.~Bae, T.~H.~Jung and H.~D.~Kim,
  %``125 GeV Higgs as a pseudo-Goldstone boson in supersymmetry with vector-like matters,''
   arXiv:1208.3748 [hep-ph].  %%CITATION = ARXIV:1208.3748;%%


%\cite{Kang:2012ra}
\bibitem{Kang:2012ra}
  Z.~Kang, T.~Li, T.~Liu, C.~Tong and J.~M.~Yang,
  %``A Heavy SM-like Higgs and a Light Stop from Yukawa-Deflected Gauge Mediation,''
   arXiv:1203.2336 [hep-ph].  %%CITATION = ARXIV:1203.2336;%%


%\cite{Craig:2012xp}
\bibitem{Craig:2012xp}
  N.~Craig, S.~Knapen, D.~Shih and Y.~Zhao,
  %``A Complete Model of Low-Scale Gauge Mediation,''
  arXiv:1206.4086 [hep-ph].  %%CITATION = ARXIV:1206.4086;%%

  %\cite{Shadmi:2011hs}
\bibitem{Shadmi:2011hs}
  Y.~Shadmi and P.~Z.~Szabo,
  %``Flavored Gauge-Mediation,''
  JHEP {\bf 1206}, 124 (2012)
  [arXiv:1103.0292 [hep-ph]].
  %%CITATION = ARXIV:1103.0292;%%

%\cite{Albaid:2012qk}
\bibitem{Albaid:2012qk}
  A.~Albaid, K.~S.~Babu and K.~S.~Babu,
  %``Higgs boson of mass 125 GeV in GMSB models with messenger-matter mixing,''
  arXiv:1207.1014 [hep-ph].
  %%CITATION = ARXIV:1207.1014;%%

%\cite{Abdullah:2012tq}
\bibitem{Abdullah:2012tq}
  M.~Abdullah, I.~Galon, Y.~Shadmi and Y.~Shirman,
  %``Flavored Gauge Mediation, A Heavy Higgs, and Supersymmetric Alignment,''
  arXiv:1209.4904 [hep-ph].  %%CITATION = ARXIV:1209.4904;%%

%\cite{Evans:2012hg}
\bibitem{Evans:2012hg}
  J.~L.~Evans, M.~Ibe, S.~Shirai and T.~T.~Yanagida,
  %``A 125GeV Higgs Boson and Muon g-2 in More Generic Gauge Mediation,''
  Phys.\ Rev.\ D {\bf 85}, 095004 (2012)
  [arXiv:1201.2611 [hep-ph]].
  %%CITATION = ARXIV:1201.2611;%%

  %\cite{Evans:2011bea}
\bibitem{Evans:2011bea}
  J.~L.~Evans, M.~Ibe and T.~T.~Yanagida,
  %``Relatively Heavy Higgs Boson in More Generic Gauge Mediation,''
  Phys.\ Lett.\ B {\bf 705}, 342 (2011)
  [arXiv:1107.3006 [hep-ph]].
  %%CITATION = ARXIV:1107.3006;%%

%\cite{Buican:2008ws}
\bibitem{Buican:2008ws}
  M.~Buican, P.~Meade, N.~Seiberg and D.~Shih,
  %``Exploring General Gauge Mediation,''
  JHEP {\bf 0903}, 016 (2009)
  [arXiv:0812.3668 [hep-ph]].
  %%CITATION = ARXIV:0812.3668;%%

%\cite{Dermisek:2006ey}
\bibitem{Dermisek:2006ey}
  R.~Dermisek and H.~D.~Kim,
  %``Radiatively generated maximal mixing scenario for the Higgs mass and the least fine tuned minimal supersymmetric standard model,''
  Phys.\ Rev.\ Lett.\  {\bf 96}, 211803 (2006)
  [hep-ph/0601036].
  %%CITATION = HEP-PH/0601036;%%

%\cite{Dermisek:2006qj}
\bibitem{Dermisek:2006qj}
  R.~Dermisek, H.~D.~Kim and I.~-W.~Kim,
  %``Mediation of supersymmetry breaking in gauge messenger models,''
  JHEP {\bf 0610}, 001 (2006)
  [hep-ph/0607169].
  %%CITATION = HEP-PH/0607169;%%


%\cite{Choi:2011rs}
\bibitem{Choi:2011rs}
  K.~Choi, E.~J.~Chun, H.~D.~Kim, W.~I.~Park and C.~S.~Shin,
  %``The $\mu$-problem and axion in gauge mediation,''
  Phys.\ Rev.\ D {\bf 83}, 123503 (2011)
  [arXiv:1102.2900 [hep-ph]].
  %%CITATION = ARXIV:1102.2900;%%

%\cite{Dvali:1996cu}
\bibitem{Dvali:1996cu}
  G.~R.~Dvali, G.~F.~Giudice and A.~Pomarol,
  %``The Mu problem in theories with gauge mediated supersymmetry breaking,''
  Nucl.\ Phys.\ B {\bf 478}, 31 (1996)
  [hep-ph/9603238].
  %%CITATION = HEP-PH/9603238;%%

%\cite{Giudice:2007ca}
\bibitem{Giudice:2007ca}
  G.~F.~Giudice, H.~D.~Kim and R.~Rattazzi,
  %``Natural mu and B mu in gauge mediation,''
  Phys.\ Lett.\ B {\bf 660}, 545 (2008)
  [arXiv:0711.4448 [hep-ph]].
  %%CITATION = ARXIV:0711.4448;%%

%\cite{Joaquim:2006uz}
\bibitem{Joaquim:2006uz}
  F.~R.~Joaquim and A.~Rossi,
  %``Gauge and Yukawa mediated supersymmetry breaking in the triplet seesaw scenario,''
  Phys.\ Rev.\ Lett.\  {\bf 97}, 181801 (2006)  [hep-ph/0604083].  %%CITATION = HEP-PH/0604083;%%

 %\cite{Mohapatra:2008wx}
\bibitem{Mohapatra:2008wx}
  R.~N.~Mohapatra, N.~Okada and H.~-B.~Yu,
  %``nu-GMSB with Type III Seesaw and Phenomenology,''
  Phys.\ Rev.\ D {\bf 78}, 075011 (2008)  [arXiv:0807.4524 [hep-ph]].  %%CITATION = ARXIV:0807.4524;%%

%\cite{FileviezPerez:2009im}
\bibitem{FileviezPerez:2009im}
  P.~Fileviez Perez, H.~Iminniyaz, G.~Rodrigo and S.~Spinner,
  %``Gauge Mediated SUSY Breaking via Seesaw,''
  Phys.\ Rev.\ D {\bf 81}, 095013 (2010)  [arXiv:0911.1360 [hep-ph]].  %%CITATION = ARXIV:0911.1360;%%


  %\cite{Borzumati:1986qx}
\bibitem{Borzumati:1986qx}
  F.~Borzumati and A.~Masiero,
  %``Large Muon and electron Number Violations in Supergravity Theories,''
  Phys.\ Rev.\ Lett.\  {\bf 57}, 961 (1986).
  %%CITATION = PRLTA,57,961;%%

  %\cite{Ciuchini:2007ha}
\bibitem{Ciuchini:2007ha}
  M.~Ciuchini, A.~Masiero, P.~Paradisi, L.~Silvestrini, S.~K.~Vempati and O.~Vives,
  %``Soft SUSY breaking grand unification: Leptons versus quarks on the flavor playground,''
  Nucl.\ Phys.\ B {\bf 783}, 112 (2007)  [hep-ph/0702144 [HEP-PH]].  %%CITATION = HEP-PH/0702144;%%


  %\cite{Harrison:2002er}
\bibitem{Harrison:2002er}
  P.~F.~Harrison, D.~H.~Perkins and W.~G.~Scott,
  %``Tri-bimaximal mixing and the neutrino oscillation data,''
  Phys.\ Lett.\ B {\bf 530}, 167 (2002)  [hep-ph/0202074].  %%CITATION = HEP-PH/0202074;%%

%\cite{Lin:2009bw}
\bibitem{Lin:2009bw}
  Y.~Lin,
  %``Tri-bimaximal Neutrino Mixing from A(4) and theta(13) ~ theta(C),''
  Nucl.\ Phys.\ B {\bf 824}, 95 (2010)  [arXiv:0905.3534 [hep-ph]].  %%CITATION = ARXIV:0905.3534;%%

%\cite{Ishimori:2012fg}
\bibitem{Ishimori:2012fg}
  H.~Ishimori and E.~Ma,
  %``New Simple $A_4$ Neutrino Model for Nonzero $\theta_{13}$ and Large $\delta_{CP}$,''
  Phys.\ Rev.\ D {\bf 86}, 045030 (2012)  [arXiv:1205.0075 [hep-ph]].  %%CITATION = ARXIV:1205.0075;%%

%\cite{Altarelli:2012bn}
\bibitem{Altarelli:2012bn}
  G.~Altarelli, F.~Feruglio, L.~Merlo and E.~Stamou,
  %``Discrete Flavour Groups, $theta_{13}$ and Lepton Flavour Violation,''
  JHEP {\bf 1208}, 021 (2012)  [arXiv:1205.4670 [hep-ph]].  %%CITATION = ARXIV:1205.4670;%%

%\cite{King:2012vj}
\bibitem{King:2012vj}
  S.~F.~King,
  %``Tri-bimaximal-Cabibbo Mixing,''
  Phys.\ Lett.\ B {\bf 718}, 136 (2012)  [arXiv:1205.0506 [hep-ph]].  %%CITATION = ARXIV:1205.0506;%%

%\cite{Minkowski:1977sc}
\bibitem{Minkowski:1977sc}
  P.~Minkowski,
  %``mu --> e gamma at a Rate of One Out of 1-Billion Muon Decays?,''
  Phys.\ Lett.\ B {\bf 67}, 421 (1977).  %%CITATION = PHLTA,B67,421;%%

%\cite{Yanagida:1979as}
\bibitem{Yanagida:1979as}
  T.~Yanagida,
  %``Horizontal Symmetry And Masses Of Neutrinos,''
  Conf.\ Proc.\ C {\bf 7902131}, 95 (1979).  %%CITATION = CONFP,C7902131,95;%%

 %\cite{Yanagida:1980xy}
\bibitem{Yanagida:1980xy}
  T.~Yanagida,
  %``Horizontal Symmetry and Masses of Neutrinos,''
  Prog.\ Theor.\ Phys.\  {\bf 64}, 1103 (1980).  %%CITATION = PTPKA,64,1103;%%

%\cite{GellMann:1980vs}
\bibitem{GellMann:1980vs}
  M.~Gell-Mann, P.~Ramond and R.~Slansky,
  %``Complex Spinors And Unified Theories,''
   Conf.\ Proc.\ C {\bf 790927}, 315 (1979).  %%CITATION = CONFP,C790927,315;%%

%\cite{Mohapatra:1979ia}
\bibitem{Mohapatra:1979ia}
  R.~N.~Mohapatra and G.~Senjanovic,
  %``Neutrino Mass and Spontaneous Parity Violation,''
  Phys.\ Rev.\ Lett.\  {\bf 44}, 912 (1980).  %%CITATION = PRLTA,44,912;%%

%\cite{Giudice:2010zn}
\bibitem{Giudice:2010zn}
  G.~F.~Giudice, P.~Paradisi and A.~Strumia,
  %``The electron and neutron EDM from supersymmetric see-saw thresholds,''
 Phys.\ Lett.\ B {\bf 694}, 26 (2010)  [arXiv:1003.2388 [hep-ph]].  %%CITATION = ARXIV:1003.2388;%%

 %\cite{Giudice:1997ni}
\bibitem{Giudice:1997ni}
  G.~F.~Giudice and R.~Rattazzi,
  %``Extracting supersymmetry breaking effects from wave function renormalization,''
  Nucl.\ Phys.\ B {\bf 511}, 25 (1998)  [hep-ph/9706540].  %%CITATION = HEP-PH/9706540;%%

%\cite{ArkaniHamed:1998kj}
\bibitem{ArkaniHamed:1998kj}
  N.~Arkani-Hamed, G.~F.~Giudice, M.~A.~Luty and R.~Rattazzi,
  %``Supersymmetry breaking loops from analytic continuation into superspace,''
  Phys.\ Rev.\ D {\bf 58}, 115005 (1998)  [hep-ph/9803290].  %%CITATION = HEP-PH/9803290;%%

%\cite{Chacko:2001km}
\bibitem{Chacko:2001km}
  Z.~Chacko and E.~Ponton,
  %``Yukawa deflected gauge mediation,''
  Phys.\ Rev.\ D {\bf 66}, 095004 (2002)  [hep-ph/0112190].  %%CITATION = HEP-PH/0112190;%%

%\cite{Grossman:2011fz}
\bibitem{Grossman:2011fz}
  D.~Grossman and Y.~Nir,
  %``Probing the Seesaw and Gauge Mediation Scales with BR(\mu\to e\gamma) and |U_{e3}|,''
  Phys.\ Rev.\ D {\bf 85}, 055004 (2012)  [arXiv:1111.5751 [hep-ph]].  %
  %CITATION = ARXIV:1111.5751;%%

%\cite{Giudice:2006sn}
\bibitem{Giudice:2006sn}
  G.~F.~Giudice and R.~Rattazzi,
  %``Living Dangerously with Low-Energy Supersymmetry,''
  Nucl.\ Phys.\ B {\bf 757}, 19 (2006)  [hep-ph/0606105].  %%CITATION = HEP-PH/0606105;%%

%\cite{Abe:2011fz}
\bibitem{Abe:2011fz}
  Y.~Abe {\it et al.}  [DOUBLE-CHOOZ Collaboration],
  %``Indication for the disappearance of reactor electron antineutrinos in the Double Chooz experiment,''
  Phys.\ Rev.\ Lett.\  {\bf 108}, 131801 (2012)  [arXiv:1112.6353 [hep-ex]].  %%CITATION = ARXIV:1112.6353;%%

%\cite{Hartz:2012np}
\bibitem{Hartz:2012np}
  M.~Hartz [T2K Collaboration],
  %``First Oscillation Results for the T2K Experiment,''
  arXiv:1201.1846 [hep-ex].  %%CITATION = ARXIV:1201.1846;%%

%\cite{Adamson:2012rm}
\bibitem{Adamson:2012rm}
  P.~Adamson {\it et al.}  [MINOS Collaboration],
  %``An improved measurement of muon antineutrino disappearance in MINOS,''
  Phys.\ Rev.\ Lett.\  {\bf 108}, 191801 (2012)  [arXiv:1202.2772 [hep-ex]].  %%CITATION = ARXIV:1202.2772;%%

%\cite{An:2012eh}
\bibitem{An:2012eh}
  F.~P.~An {\it et al.}  [DAYA-BAY Collaboration],
  %``Observation of electron-antineutrino disappearance at Daya Bay,''
  Phys.\ Rev.\ Lett.\  {\bf 108}, 171803 (2012)  [arXiv:1203.1669 [hep-ex]].  %%CITATION = ARXIV:1203.1669;%%

%\cite{Ahn:2012nd}
\bibitem{Ahn:2012nd}
  J.~K.~Ahn {\it et al.}  [RENO Collaboration],
  %``Observation of Reactor Electron Antineutrino Disappearance in the RENO Experiment,''
  Phys.\ Rev.\ Lett.\  {\bf 108}, 191802 (2012)  [arXiv:1204.0626 [hep-ex]].  %%CITATION = ARXIV:1204.0626;%%

%\cite{He:2006dk}
\bibitem{He:2006dk}
  X.~-G.~He, Y.~-Y.~Keum and R.~R.~Volkas,
  %``A(4) flavor symmetry breaking scheme for understanding quark and neutrino mixing angles,''
  JHEP {\bf 0604}, 039 (2006)  [hep-ph/0601001].  %%CITATION = HEP-PH/0601001;%%

%\cite{He:2006qd}
\bibitem{He:2006qd}
  X.~-G.~He and A.~Zee,
  %``Minimal modification to the tri-bimaximal neutrino mixing,''
  Phys.\ Lett.\ B {\bf 645}, 427 (2007)  [hep-ph/0607163].  %%CITATION = HEP-PH/0607163;%%

%\cite{He:2011gb}
\bibitem{He:2011gb}
  X.~-G.~He and A.~Zee,
  %``Minimal Modification to Tri-bimaximal Mixing,''
  Phys.\ Rev.\ D {\bf 84}, 053004 (2011)  [arXiv:1106.4359 [hep-ph]].  %%CITATION = ARXIV:1106.4359;%%

%\cite{BenTov:2012tg}
\bibitem{BenTov:2012tg}
  Y.~BenTov, X.~-G.~He and A.~Zee,
  %``An A4 x Z4 model for neutrino mixing,''
  arXiv:1208.1062 [hep-ph].  %%CITATION = ARXIV:1208.1062;%%


%\cite{Bazzocchi:2012st}
\bibitem{Bazzocchi:2012st}
  F.~Bazzocchi and L.~Merlo,
  %``Neutrino Mixings and the S4 Discrete Flavour Symmetry,''
   arXiv:1205.5135 [hep-ph].  %%CITATION = ARXIV:1205.5135;%%

%\cite{Beringer:1900zz}
\bibitem{Beringer:1900zz}
  J.~Beringer {\it et al.}  [Particle Data Group Collaboration],
  %``Review of Particle Physics (RPP),''
  Phys.\ Rev.\ D {\bf 86}, 010001 (2012).  %%CITATION = PHRVA,D86,010001;%%

%\cite{Fogli:2012ua}
\bibitem{Fogli:2012ua}
  G.~L.~Fogli, E.~Lisi, A.~Marrone, D.~Montanino, A.~Palazzo and A.~M.~Rotunno,
  %``Global analysis of neutrino masses, mixings and phases: entering the era of leptonic CP violation searches,''
  Phys.\ Rev.\ D {\bf 86}, 013012 (2012)  [arXiv:1205.5254 [hep-ph]].  %%CITATION = ARXIV:1205.5254;%%

%\cite{GonzalezGarcia:2012sz}
\bibitem{GonzalezGarcia:2012sz}
  M.~C.~Gonzalez-Garcia, M.~Maltoni, J.~Salvado and T.~Schwetz,
  %``Global fit to three neutrino mixing: critical look at present precision,''
  arXiv:1209.3023 [hep-ph].  %%CITATION = ARXIV:1209.3023;%%


 %\cite{Hisano:1995cp}
\bibitem{Hisano:1995cp}
  J.~Hisano, T.~Moroi, K.~Tobe and M.~Yamaguchi,
  %``Lepton flavor violation via right-handed neutrino Yukawa couplings in supersymmetric standard model,''
  Phys.\ Rev.\ D {\bf 53}, 2442 (1996)  [hep-ph/9510309].  %%CITATION = HEP-PH/9510309;%%

%\cite{Arganda:2005ji}
\bibitem{Arganda:2005ji}
  E.~Arganda and M.~J.~Herrero,
  %``Testing supersymmetry with lepton flavor violating tau and mu decays,''
  Phys.\ Rev.\ D {\bf 73}, 055003 (2006)  [hep-ph/0510405].  %%CITATION = HEP-PH/0510405;%%

%\cite{Hewett:2012ns}
\bibitem{Hewett:2012ns}
  J.~L.~Hewett, H.~Weerts, R.~Brock, J.~N.~Butler, B.~C.~K.~Casey, J.~Collar, A.~de Govea and R.~Essig {\it et al.},
  %``Fundamental Physics at the Intensity Frontier,''
  arXiv:1205.2671 [hep-ex].  %%CITATION = ARXIV:1205.2671;%%

%\cite{Adam:2011ch}
\bibitem{Adam:2011ch}
  J.~Adam {\it et al.}  [MEG Collaboration],
  %``New limit on the lepton-flavour violating decay $\mu^{+} \to e^{+} \gamma$,''
  Phys.\ Rev.\ Lett.\  {\bf 107}, 171801 (2011)  [arXiv:1107.5547 [hep-ex]].  %%CITATION = ARXIV:1107.5547;%%

%\cite{Aubert:2009ag}
\bibitem{Aubert:2009ag}
  B.~Aubert {\it et al.}  [BABAR Collaboration],
  %``Searches for Lepton Flavor Violation in the Decays tau+- ---> e+- gamma and tau+- ---> mu+- gamma,''
  Phys.\ Rev.\ Lett.\  {\bf 104}, 021802 (2010)  [arXiv:0908.2381 [hep-ex]].  %%CITATION = ARXIV:0908.2381;%%

%\cite{Bellgardt:1987du}
\bibitem{Bellgardt:1987du}
  U.~Bellgardt {\it et al.}  [SINDRUM Collaboration],
  %``Search for the Decay mu+ ---> e+ e+ e-,''
  Nucl.\ Phys.\ B {\bf 299}, 1 (1988).  %%CITATION = NUPHA,B299,1;%%
%\cite{Hayasaka:2010np}

\bibitem{Hayasaka:2010np}
  K.~Hayasaka, K.~Inami, Y.~Miyazaki, K.~Arinstein, V.~Aulchenko, T.~Aushev, A.~M.~Bakich and A.~Bay {\it et al.},
  %``Search for Lepton Flavor Violating Tau Decays into Three Leptons with 719 Million Produced Tau+Tau- Pairs,''
  Phys.\ Lett.\ B {\bf 687}, 139 (2010)  [arXiv:1001.3221 [hep-ex]].  %%CITATION = ARXIV:1001.3221;%%

%\cite{Dohmen:1993mp}
\bibitem{Dohmen:1993mp}
  C.~Dohmen {\it et al.}  [SINDRUM II. Collaboration],
  %``Test of lepton flavor conservation in mu ---> e conversion on titanium,''
  Phys.\ Lett.\ B {\bf 317}, 631 (1993).  %%CITATION = PHLTA,B317,631;%%

%\cite{Bertl:2006up}
\bibitem{Bertl:2006up}
  W.~H.~Bertl {\it et al.}  [SINDRUM II Collaboration],
  %``A Search for muon to electron conversion in muonic gold,''
  Eur.\ Phys.\ J.\ C {\bf 47}, 337 (2006).  %%CITATION = EPHJA,C47,337;%%

%\cite{O'Leary:2010af}
\bibitem{O'Leary:2010af}
  B.~O'Leary {\it et al.}  [SuperB Collaboration],
  %``SuperB Progress Reports -- Physics,''
  arXiv:1008.1541 [hep-ex].  %%CITATION = ARXIV:1008.1541;%%

\bibitem{Blondel} A. Blondel {\it et al.}, http://www.psi.ch/mu3e/DocumentsEN/LOI\_Mu3e\_PSI.pdf

\bibitem{PRIME} The PRIME working group, unpublished; LOI to J-PARC 50-GeV PS, LOI-25,
http://www-ps.kek.jp/jhf-np/LOIlist/pdf/L25.pdf

%\cite{Abada:2012cq}
\bibitem{Abada:2012cq}
  A.~Abada, D.~Das, A.~Vicente and C.~Weiland,
  %``Enhancing lepton flavour violation in the supersymmetric inverse seesaw beyond the dipole contribution,''
  JHEP {\bf 1209}, 015 (2012)  [arXiv:1206.6497 [hep-ph]].  %%CITATION = ARXIV:1206.6497;%%
%\cite{Ishimori:2010au}

%\cite{Lopez:1993vi}
\bibitem{Lopez:1993vi}
  J.~L.~Lopez, D.~V.~Nanopoulos and X.~Wang,
  %``Large (g-2)-mu in SU(5) x U(1) supergravity models,''
  Phys.\ Rev.\ D {\bf 49}, 366 (1994)  [hep-ph/9308336].  %%CITATION = HEP-PH/9308336;%%

%\cite{Chattopadhyay:1995ae}
\bibitem{Chattopadhyay:1995ae}
  U.~Chattopadhyay and P.~Nath,
  %``Probing supergravity grand unification in the Brookhaven g-2 experiment,''
  Phys.\ Rev.\ D {\bf 53}, 1648 (1996)  [hep-ph/9507386].  %%CITATION = HEP-PH/9507386;%%

%\cite{Hirsch:2012ax}
\bibitem{Hirsch:2012ax}
  M.~Hirsch, F.~Staub and A.~Vicente,
  %``Enhancing $l_i \to 3 l_j$ with the $Z^0$-penguin,''
  Phys.\ Rev.\ D {\bf 85}, 113013 (2012)  [arXiv:1202.1825 [hep-ph]].  %%CITATION = ARXIV:1202.1825;%%

 %\cite{Fok:2010vk}
\bibitem{Fok:2010vk}
  R.~Fok and G.~D.~Kribs,
  %``\mu to e in R-symmetric Supersymmetry,''
  Phys.\ Rev.\ D {\bf 82}, 035010 (2010)  [arXiv:1004.0556 [hep-ph]].  %%CITATION = ARXIV:1004.0556;%%


%\cite{Kitano:2002mt}
\bibitem{Kitano:2002mt}
  R.~Kitano, M.~Koike and Y.~Okada,
  %``Detailed calculation of lepton flavor violating muon electron conversion rate for various nuclei,''
  Phys.\ Rev.\ D {\bf 66}, 096002 (2002)  [Erratum-ibid.\ D {\bf 76}, 059902 (2007)]  [hep-ph/0203110].  %%CITATION = HEP-PH/0203110;%%

%\cite{Bennett:2006fi}
\bibitem{Bennett:2006fi}
  G.~W.~Bennett {\it et al.}  [Muon G-2 Collaboration],
  %``Final Report of the Muon E821 Anomalous Magnetic Moment Measurement at BNL,''
  Phys.\ Rev.\ D {\bf 73}, 072003 (2006)  [hep-ex/0602035].  %%CITATION = HEP-EX/0602035;%%


%\cite{Hagiwara:2011af}
\bibitem{Hagiwara:2011af}
  K.~Hagiwara, R.~Liao, A.~D.~Martin, D.~Nomura and T.~Teubner,
  %``(g-2)_mu and alpha(M_Z^2) re-evaluated using new precise data,''
  J.\ Phys.\ G {\bf 38}, 085003 (2011)  [arXiv:1105.3149 [hep-ph]].  %%CITATION = ARXIV:1105.3149;%%


%\cite{Moroi:1995yh}
\bibitem{Moroi:1995yh}
  T.~Moroi,
  %``The Muon anomalous magnetic dipole moment in the minimal supersymmetric standard model,''
  Phys.\ Rev.\ D {\bf 53}, 6565 (1996)  [Erratum-ibid.\ D {\bf 56}, 4424 (1997)]  [hep-ph/9512396].  %%CITATION = HEP-PH/9512396;%%

  %\cite{Cho:2011rk}
\bibitem{Cho:2011rk}
  G.~-C.~Cho, K.~Hagiwara, Y.~Matsumoto and D.~Nomura,
  %``The MSSM confronts the precision electroweak data and the muon g-2,''
  JHEP {\bf 1111}, 068 (2011)  [arXiv:1104.1769 [hep-ph]].  %%CITATION = ARXIV:1104.1769;%%

%\cite{Hisano:2001qz}
\bibitem{Hisano:2001qz}
  J.~Hisano and K.~Tobe,
  %``Neutrino masses, muon g-2, and lepton flavor violation in the supersymmetric seesaw model,''
  Phys.\ Lett.\ B {\bf 510}, 197 (2001)  [hep-ph/0102315].  %%CITATION = HEP-PH/0102315;%%


\bibitem{Ishimori:2010au}
  H.~Ishimori, T.~Kobayashi, H.~Ohki, Y.~Shimizu, H.~Okada and M.~Tanimoto,
  %``Non-Abelian Discrete Symmetries in Particle Physics,''
  Prog.\ Theor.\ Phys.\ Suppl.\  {\bf 183}, 1 (2010)  [arXiv:1003.3552 [hep-th]].  %%CITATION = ARXIV:1003.3552;%%

















\end{thebibliography}
\end{document}